\documentclass{aa}  
\usepackage{graphicx}
\usepackage{txfonts}
\usepackage{enumitem}
\usepackage{tabularx}

\usepackage{fixltx2e}
\usepackage{color}
\usepackage{lscape}
\usepackage{longtable}

\begin{document}

\title{The ALHAMBRA survey\thanks{Based on observations collected at the German-Spanish Astronomical Center, Calar Alto, jointly operated by the Max-Planck-Institut f\"ur Astronomie (MPIA) at Heidelberg and the Instituto de Astrof\'{\i}sica de Andaluc\'{\i}a (CSIC)}: 2-D analysis of the stellar populations in  massive early-type galaxies at $z$ < 0.3}

\author{I. San Roman\inst{\ref{a1}}\fnmsep\thanks{e-mail: isanroman@cefca.es}, A. J. Cenarro\inst{\ref{a1}}, L. A. D\'{i}az-Garc\'{i}a\inst{\ref{a1}}, C. L\'{o}pez-Sanjuan\inst{\ref{a1}}, J. Varela\inst{\ref{a1}}, R. M. Gonz\'{a}lez Delgado\inst{\ref{a2}}, P. S\'{a}nchez-Bl\'{a}zquez\inst{\ref{a3}},  E. J. Alfaro\inst{\ref{a2}}, B. Ascaso\inst{\ref{a4}}, S. Bonoli\inst{\ref{a1}}, A. Borlaff\inst{\ref{a6}}, F. J. Castander\inst{\ref{a5}}, M. Cervi\~{n}o\inst{\ref{a2},\ref{a6},\ref{a7}}, A. Fern\'{a}ndez-Soto\inst{\ref{a8},\ref{a9}}, I. M\'{a}rquez\inst{\ref{a2}}, J. Masegosa\inst{\ref{a2}}, D. Muniesa\inst{\ref{a1}}, M. Povic\inst{\ref{a2}}, K. Viironen\inst{\ref{a1}}, J.~A.~L.~Aguerri\inst{\ref{a6},\ref{a7}}, N.~Ben\'itez\inst{\ref{a2}},  T.~Broadhurst\inst{\ref{a12},\ref{a13}}, J.~Cabrera-Ca\~no\inst{\ref{a14}}, J.~Cepa\inst{\ref{a6},\ref{a7}}, D.~Crist\'obal-Hornillos\inst{\ref{a1}}, L.~Infante\inst{\ref{a15},\ref{a16}}, V.~J.~Mart\'inez\inst{\ref{a9},\ref{a10},\ref{a11}}, M.~Moles\inst{\ref{a1},\ref{a2}}, A.~del~Olmo\inst{\ref{a2}}, J.~Perea\inst{\ref{a2}}, F.~Prada\inst{\ref{a2},\ref{a3},\ref{a17}}, J.~M.~Quintana\inst{\ref{a2}}}

    \institute{Centro de Estudios de F\'{i}sica del Cosmos de Arag\'{o}n (CEFCA), Unidad Asociada al CSIC, Plaza San Juan 1, E-44001 Teruel, Spain\label{a1}
\and
Instituto de Astrof\'{i}sica de Andaluc\'{i}a (IAA-CSIC), Glorieta de la Astronom\'{i}a s/n, 18080 Granada, Spain\label{a2}
\and
Departamento de F\'{i}sica Te\'{o}rica, Universidad Autonoma de Madrid (UAM-CSIC), 28049 Cantoblanco, Madrid, Spain\label{a3}
\and
APC, AstroParticule et Cosmologie, Universit\'e Paris Diderot, CNRS/IN2P3, CEA/lrfu, Observatoire de Paris, Sorbonne Paris Cit\'e, 10, rue Alice Domon et L\'eonie Duquet, 75205 Paris Cedex 13, France\label{a4}
\and
Institut de Ci\`encies de l'Espai (IEEC-CSIC), Facultat de Ci\`encies, Campus UAB, 08193 Bellaterra, Barcelona, Spain\label{a5}
\and
Instituto de Astrof\'{i}sica de Canarias (IAC), V\'{i}a L\'{a}ctea s/n, 38205 La Laguna, Tenerife, Spain\label{a6}
\and
Departamento de Astrof\'{i}sica, Universidad de La Laguna (ULL), 38205 La Laguna, Tenerife, Spain\label{a7}
\and
Instituto de F\'{i}sica de Cantabria (CSIC-UC), E-39005 Santander, Spain\label{a8}
\and
Unidad Asociada Observatorio Astron\'omico (IFCA-UV), E-46980, Paterna, Valencia, Spain\label{a9}
\and
Observatori Astron\`omic, Universitat de Val\`encia, C/ Catedr\`atic Jos\'e Beltr\'an 2, E-46980, Paterna, Spain\label{a10}
\and
Departament d'Astronomia i Astrof\'isica, Universitat de Val\`encia, E-46100, Burjassot, Spain\label{a11}
\and
Department of Theoretical Physics, University of the Basque Country UPV/EHU, 48080 Bilbao, Spain\label{a12}
 \and
IKERBASQUE, Basque Foundation for Science, Bilbao, Spain\label{a13}
\and
Departamento de F\'isica At\'omica, Molecular y Nuclear, Facultad de F\'isica, Universidad de Sevilla, 41012 Sevilla, Spain\label{a14}
\and
Instituto de Astrof\'{\i}sica, Universidad Cat\'olica de Chile, Av. Vicuna Mackenna 4860, 782-0436 Macul, Santiago, Chile\label{a15}
\and
Centro de Astro-Ingenier\'{\i}a, Universidad Cat\'olica de Chile, Av. Vicuna Mackenna 4860, 782-0436 Macul, Santiago, Chile\label{a16}
\and
Campus of International Excellence UAM+CSIC, Cantoblanco, E-28049 Madrid, Spain \label{a17}
}

\titlerunning {2D study of ALHAMBRA galaxies} 
\authorrunning {San Roman et al.}

  \abstract{We present a technique that permits the analysis of stellar population gradients in a relatively low cost way compared to IFU surveys analyzing a vastly larger samples as well as out to larger radii. We developed a technique to analyze unresolved stellar populations of spatially resolved galaxies based on photometric multi-filter surveys.  We derived spatially resolved stellar population properties and radial gradients by applying a Centroidal Voronoi Tesselation and performing a multi-color photometry SED fitting. This technique has been successfully applied to a sample of 29 massive (M$_{\star}$ > 10$^{10.5}$ M$_{\sun}$), early-type galaxies at $z$ < 0.3 from the ALHAMBRA survey. We produced detailed 2D maps of stellar population properties (age, metallicity and extinction) which allow us to identify galactic features. Radial structures have been studied and luminosity-weighted and mass-weighted gradients have been derived out to 2 -- 3.5 R$\textsubscript{eff}$. We find that the spatially resolved stellar population mass, age and metallicity are well represented by their integrated values. We find the gradients of early-type galaxies to be on average flat in age ($\nabla$log Age$\textsubscript{L}$ = 0.02 $\pm$ 0.06 dex/R$\textsubscript{eff}$) and negative in metallicity ($\nabla$[Fe/H]$\textsubscript{L}$ = --0.09 $\pm$ 0.06 dex/R$\textsubscript{eff}$). Overall, the extinction gradients are flat ($\nabla$A$\textsubscript{v}$ = --0.03 $\pm$ 0.09 mag/R$\textsubscript{eff}$ ) with a wide spread. These results are in agreement with previous studies that used standard long-slit spectroscopy as well as with the most recent integral field unit (IFU) studies. According to recent simulations, these results are consistent with a scenario where early-type galaxies were formed through major mergers and where their final gradients are driven by the older ages and higher metallicity of the accreted systems. We demonstrate the scientific potential of multi-filter photometry to explore the spatially resolved stellar populations of local galaxies and confirm previous spectroscopic trends from a complementary technique. }


\maketitle

\section{Introduction}
Large spectroscopic surveys such as  the Sloan Digital Sky Survey \citep[SDSS;][]{Yorketal2000}, the Galaxy and Mass Assembly project \citep[GAMA;][]{Driveretal2011}  or the 2dF Galaxy Redshift Survey \citep[2dFGRS;][]{Collessetal2001} dramatically improved our understanding of galaxy formation in the local universe. Through the analysis of sensitive absorption lines or via full spectral fitting techniques, one can derive galaxy properties such as the star formation history, chemical content or stellar mass. However, these large galaxy surveys are restricted to single aperture observations per galaxy usually limited to their central regions. Large photometric surveys are in general also restricted to global galaxy information since they mostly rely on integrated photometry. Internal inhomogeneities of a galaxy, as radial age and metallicity gradients, are the results of its star formation and enrichment history. Therefore, spatially resolved studies of galaxies are essential to uncover the formation and assembly of local galaxies. 

Early attempts to study radially resolved stellar populations are based on multi-wavelength broadband photometry \citep[e.g.][]{MacArthuretal2004, Wuetal2005,LaBarberaetal2005, LaBarberaetal2010,Tortoraetal2010, MunozMateosetal2011} as well as long-slit spectroscopy for nearby galaxies \citep[e.g.][]{Gorgasetal1990, Davidge1992, SanchezBlazquezetal2006, SanchezBlazquezetal2007, MacArthuretal2009, SanchezBlazquezetal2011}.  Several line-strength gradient studies on early-type galaxies \citep[e.g.][]{Gonzalezetal1995, Tantaloetal1998, Kolevaetal2011,SanchezBlazquezetal2006} show mean flat or slightly positive age gradients.

Theoretically, shallow metallicity gradients are expected if major mergers are a key factor in the formation of elliptical galaxies \citep{Kobayashi2004}. On the other hand, minor mergers of low mass and metal poor galaxies can change the age and metallicity radial structure of the already formed galaxy as it increases in size. Different ingredients as galactic winds, metal cooling or AGN feedback can have important effect modifying the initial formation scenario producing a different behavior of the galaxy assembly and the resultant radial gradients of the present stellar population \citep{Gibsonetal2013, Hopkinsetal2013, Hirschmannetal2013,Hirschmannetal2015}. These theoretical works show that more detailed observational data containing spatial information are required to further constrain the formation history and the different mechanisms involved in the assembly of galaxies. 

The arrival of integral field spectroscopy (IFS) has brought a significant breakthrough in the field. First generation of IFS surveys, which targeted one galaxy at a time have been completed (\citealp[SAURON,][]{deZeeuwetal2002}; \citealp[VENGA,][]{Blancetal2010};  \citealp[PINGS,][]{RosalesOrtegaetal2010}; \citealp[ATLAS$^{3D}$,][]{Cappellarietal2011}; \citealp[CALIFA,][]{Sanchezetal2012}; \citealp[DiskMass,][]{Bershadyetal2010}). Currently, a new generation of multiplexed IFS surveys, which can observe multiple galaxies simultaneously, have started (\citealp[SAMI,][]{Bryantetal2015}; \citealp[MaNGA,][]{Bundyetal2015}). These IFS surveys allow detailed internal analysis through multiple spectra of each galaxy performing a 2D map of the object. Recent studies using data from CALIFA provided the most comprehensive results so far regarding the radial variations of the stellar population and star formation history of nearby galaxies. Results from this survey support an inside-out scenario of galaxy formation through different studies \citep[e.g.][]{Perezetal2013, Sanchezetal2014, SanchezBlazquezetal2014}. Using a sample of 107 galaxies, \cite{GonzalezDelgadoetal2014a} study the radial structure of the stellar mass surface density as well as the age distribution as a function of morphology and mass. Negative radial gradients of the stellar population ages (inner regions older than outer ones) are present in most of the galaxies, supporting an inside-out formation. However they find a clear trend with galaxy mass when galaxies are separated in early-type and late-type systems. In a more extended galaxy sample, \cite{GonzalezDelgadoetal2015} confirm this inside-out scenario but find that age gradients at larger distances (R > 2 R$_\mathrm{eff}$, i.e half-light radius) are only mildly negative or flat indicating that star formation is more uniformly distributed or that stellar migration is important at these distances. They also find mildly negative metallicity gradients, shallower than predicted from models of galaxy evolution in isolation. 

Using a small sample of 12 galaxies produced during the MaNGA prototype (P-MaNGA) observations, \cite{Lietal2015} obtain maps and radial profiles for different age-sensitive spectral indices that suggest galaxy growth is a smooth process.  \cite{Wilkinsonetal2015} find, by performing full spectral fitting of P-MaNGA data, that the gradients for galaxies identified as early-type are on average flat in age, and negative in metallicity. Most recently MaNGA survey \citep{Goddardetal2016} find, using a large representative sample of  $\sim$ 500  galaxies, that early-type galaxies generally exhibit shallow luminosity-weighted age gradients and slightly positive mass-weighted median age gradients pointing to an outside-in scenario of star formation. On the other hand, \citet{Zhengetal2016} using also MaNGA data but different spectral fitting routines and different stellar population models, find mean ages and metallicity gradients slightly negative, consistent with the inside-out formation scenario. These MaNGA studies are restricted to R < 1.5 R$_\mathrm{eff}$  and a redshift range of 0.01 <  z < 0.15 with a median redshift at z $\sim$ 0.03. The conclusions of these diversity of IFU and long-slit spectroscopy studies are not fully consistent so a substantially larger  sample size, analyzing larger galactocentric distances and higher redshifts will help us to restrict the radial gradients as well as restrict the formation scenario of these objects.

Currently, the number of multi-filter surveys is significantly increasing (e.g. \citealp[COMBO-17,][]{Wolfetal2003}; \citealp[ALHAMBRA,][]{Molesetal2008}; \citealp[PAU,][]{Castanderetal2012}; \citealp[SHARDS,][]{PerezGonzalezetal2013}: \citealp[J-PAS,][]{Benitezetal2014}; J-PLUS, Cenarro et al. in prep). Half-way between classical photometry and spectroscopy, these surveys will build a formidable legacy data set by delivering low resolution spectroscopy for every pixel over a large area of the sky.  Although multi-filter observing techniques suffer from the lack of high spectral resolution, their advantages over standard spectroscopy are worth noticing: 1) a narrow/medium-band filter system provides low resolution spectra (e.g. resolving power $\sim$ 50 for J-PAS) that results in an adequate sampling of galaxy SEDs; 2) No sample selection criteria other than the photometric depth in the detection band resulting in a uniform and non-biased spatial sampling that allows environmental studies; 3) IFU-like character, allowing a pixel-by-pixel investigation of extended galaxies; 4) large survey areas leading to much larger galaxy samples than multi-object spectroscopic surveys; and 5) a much greater multiplexing advantage, in terms of galaxy-pixels per night, than multiplexed IFU surveys.Furthermore, direct imaging is more efficient than spectroscopy so that multi-filter surveys are generally deeper than traditional spectroscopic studies (i.e., better access to galaxy outskirts). 

It is therefore clear that multi-filter surveys open a way to improve our knowledge of galaxy formation and evolution complementing standard multi-object spectroscopic surveys.  During the last few years, several SED fitting codes have been developed to deal with the peculiarities of multi-filter surveys \citep[see][]{Molinoetal2014,DiazGarciaetal2015}. These codes are specifically designed to analyze the stellar content of galaxies with available multi-filter data. To fully exploit the capabilities of multi-filter surveys, we have implemented a technique that combines these multi-filter SED fitting codes with an adequate spatial binning (e.g. Voronoi Tessellation). This approach allows us to analyze unresolved stellar populations of spatially resolved galaxies based on large sky multi-filter surveys. In this paper we present the technique used. In order to prove and test the reliability of our method we present a 2D analysis of the stellar populations for a sample of early-type galaxies observed by the ALHAMBRA survey \citep{Molesetal2008}. 

This paper is organized as follows. Section 2 gives  a quick overview of the ALHAMBRA survey as well as the photometric properties of our sample. In Sect. 3, we describe the technical aspect of the implemented method and present the 2D maps of age, [Fe/H], and A$_\mathrm{v}$.  Section 4 describes the integrated properties for our sample galaxies and analyzes the potential degeneracies. Section 5 presents the radial profiles and gradients. We discuss the results in Sect. 6 and Sect. 7 presents the conclusions. Throughout this paper we assume a $\Lambda$CDM cosmology with $H_0 = 70 $~km s$^{-1}$, $\Omega_\mathrm{M}=0.30$, and $\Omega_\mathrm{\Lambda}=0.70$.

\section{The sample}
The ALHAMBRA survey is a multi-filter survey carried out with the 3.5m telescope in the Calar Alto Observatory (CAHA) using the wide-field optical camera LAICA  and the near-infrared (NIR) instrument Omega-2000. ALHAMBRA uses a specially designed filter system that covers  the  optical  range  from  3500  $\AA$  to  9700  $\AA$  with  20  contiguous,  equal  width  (FWHM $\sim$ 300 $\AA$),  medium  band  filters, plus the three standard broad-bands, \textit{J}, \textit{H}, and \textit{K$_{s}$}, in the NIR. The survey spans a total area of  4 deg$^{2}$ over 8 non-contiguous regions of the northern hemisphere. This characteristic filter set provides a low-resolution 23 band photo-spectrum corresponding to a resolving power of $\sim$20. The final survey parameters and scientific goals, as well as the technical requirements of the filter set, were described by \citet{Molesetal2008}.  The  full  characterization,  description,  and performance  of  the  ALHAMBRA  optical  photometric  system were presented in \citet{AparicioVillegasetal2010}. For  details  about  the NIR data reduction see \citet{CristobalHornillosetal2009}, while the  optical reduction is  described  in  Crist\'{o}bal-Hornillos et al. (in prep). Note that from the total ALHAMBRA survey area, only 2.8 deg$^{2}$ have been analyzed and publicly released. We make use of these observations to test and prove the reliability of our method.  These observations are available through the ALHAMBRA web page\footnote{http://alhambrasurvey.com/}. 
 
The selection criteria of our initial sample is based on visual morphology and apparent size. In this work, we focus only on the analysis of early-type galaxies (spheroidal galaxies) for simplicity. For the purposes of this early paper, we are interested in a high quality sample of the biggest and brightest galaxies. Morphological catalogs based on objective algorithms \citep[e.g.][]{Povicetal2013} are very successful but are incomplete for very extended objects. This effect produces that some of the obvious ideal objects to test our method are not included in those catalogs. For this reason a visual inspection has been chosen to identify our sample. The standard separation between early-type galaxies and spiral galaxies is entirely based on the presence of spiral arms or extended dust lanes in edge-on galaxies \citep[e.g.][]{Devaucouleursetal1991}. This nearly universal definition of early-type galaxies is adopted in this paper. In order to properly bin each galaxy, we also required a minimum apparent size of every object. As a size indicator we determine the equivalent radius as R$\textsubscript{eq}$ = $\sqrt{(A/\pi)}$ where \textit{A} is the isophotal aperture area in the  ALHAMBRA catalog \citep{Molinoetal2014}. Early-type galaxies that occupy an  equivalent radius of R$\textsubscript{eq}$ >  20 pixels have been selected.  This selection criteria restrict this work to the study of early-type galaxies located at $z$ < 0.3. This selection provides an initial sample of 58 early-type galaxies. After the Voronoi Tessellation (see Sect. \ref{sec:voronoi}), galaxies with an insufficient signal-to-noise (S/N) to provide a reasonable 2D map were rejected. This last cut leads a final sample of 29 early-type galaxies with enough quality for our analysis.  This very restricted quality cut is based exclusively in the inspection of the final radial profiles and maps to limit the sample to well mapped objects. Although our final sample is not complete in mass or redshift, these selection criteria identify ideal objects to test the method at the same time that preserve a reasonable number of objects.  We note, however, that without applying a volume correction or completeness study, this sample does not represent the local galaxy population. Table \ref{tab:A1} of Appendix \ref{ap:tables} presents the stellar properties of the final sample. 

\begin{figure*}[h]
\begin{center}

\includegraphics[width=0.65\textwidth]{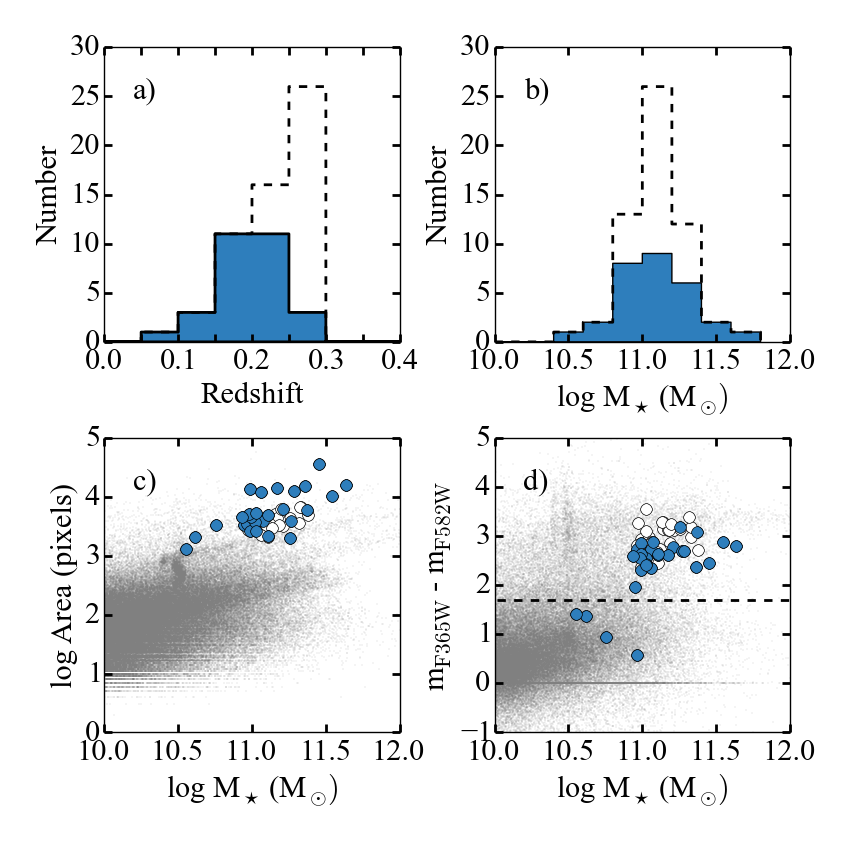}
\caption{Properties of the analyzed sample from the ALHAMBRA catalog.   a) Best photometric redshift distribution; b) Stellar mass distribution; c) Mass  versus isophotal area relation; and d) Color (m$\textsubscript{F365W}$ -- m$\textsubscript{F582W}$) versus stellar mass diagram.  Gray symbols correspond to the total ALHAMBRA sample. Dashed histograms and open circles correspond to our initial sample and blue histograms/symbols correspond to our final sample of 29 early-type galaxies after quality control cuts. The dashed black line represents the arbitrary red sequence-blue cloud division at m$\textsubscript{F365W}$ -- m$\textsubscript{F582W}$  < 1.7.}
\label{fig:1}
\end{center}
\end{figure*}

\begin{figure*}
\begin{center}
\includegraphics[width=0.95\textwidth]{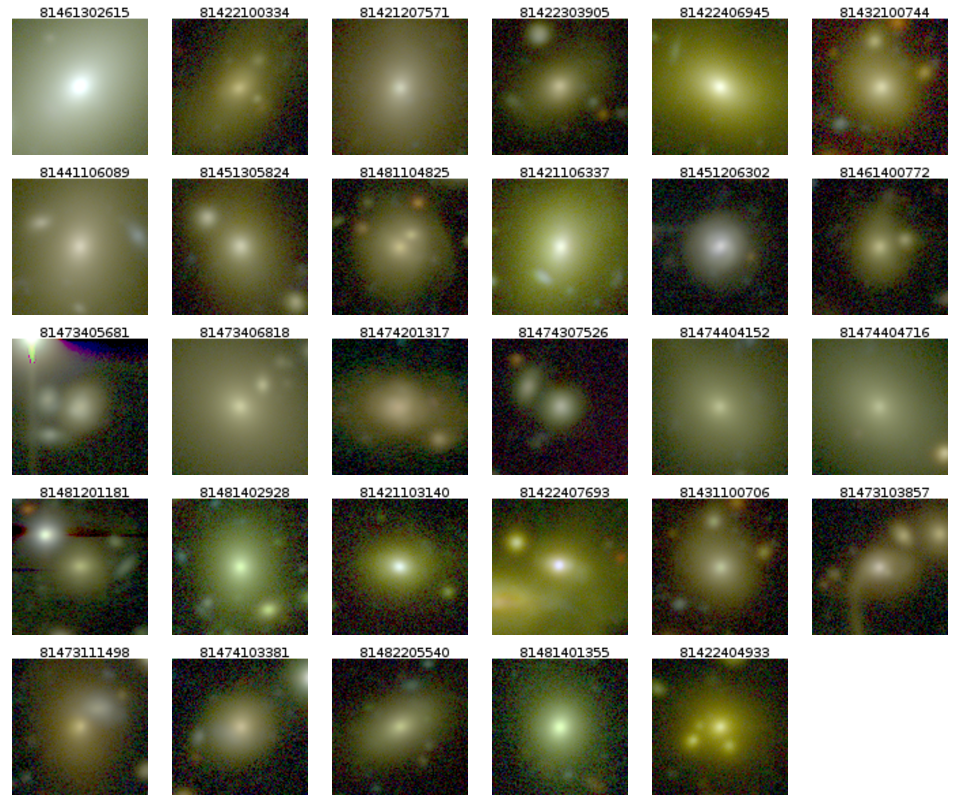}
\caption{Color images of the 29 massive, early-type galaxies analyzed in this paper. Each image corresponds to 22"x 22" centered in each target. The number of each image corresponds to the ID from Table \ref{tab:A1}} 
\label{fig:stamps}
\end{center}
\end{figure*}

Figure \ref{fig:1} shows redshifts, masses, isophotal aperture areas and colors (m$\textsubscript{F365W}$ -- m$\textsubscript{F582W}$) of our final sample derived from the ALHAMBRA catalog \citep{Molinoetal2014}. To characterize the final sample, we over-plotted the total ALHAMBRA sample as well as our initial sample of 58 early-type galaxies.  Panel a) in Fig. \ref{fig:1} shows that galaxies removed during the S/N quality check correspond to galaxies beyond $z$ $\sim$ 0.25. For comparison, CALIFA parent sample is selected in a redshift range $z$ = 0.005 -- 0.03. P-MaNGA galaxies are located at $z$ < 0.06 although the final MaNGA survey expands a range $z$ = 0.01 -- 0.15. Panel d) revels that 4 galaxies of our final sample have a blue color (m$\textsubscript{F365W}$ -- m$\textsubscript{F582W}$  < 1.7) not populating the so-called red sequence. This is indicative of current star formation and we should expect younger stellar populations in those 4 galaxies. Figure \ref{fig:stamps} presents the color images of the 29 massive, early-type galaxies analyzed in this study. From Fig. \ref{fig:1}, we conclude that our sample of galaxies comprises massive, early-type galaxies at 0.05 $\lesssim$ $z$ $\lesssim$ 0.3.

\section{The method}\label{sec:method}
The method used in the analysis can be resumed in four main steps: the homogenization of the point spread function (PSF), the spatial binning of each object, the determination of the different stellar populations through the SED fitting of each bin, and finally the representation of the 2D maps. The homogenization of the PSF ensures a homogeneous photometry. The spatial binning technique used in the analysis is the Centroidal Voronoi Tesselation. For the SED fitting, we used the \textrm{MUFFIT} code \citep[MUlti-Filter FITting in photometric surveys; ][]{DiazGarciaetal2015}. In this section we describe each step in detail.

\subsection{PSF homogenization}
One of the challenges faced by data management of current large surveys is to provide homogeneous photometry and morphology for a large number of objects over large areas of the sky. 
 To perform good quality multi-color photometry, it is necessary to sample the same physical region of the galaxy taking into account the smearing produced by the different PSFs of each filter. These PSF variations may produce artificial structure that could bias our results \citep{Bertin2011}. To avoid this problem, we have developed a method to analyze, characterize and homogenize the PSF using the code PSFEx \citep{Bertinetal2013}.

We first analyze the seeing of every image and choose a target PSF for the homogenization. The algorithm allows the user to choose the target PSF of the analyzed images. For this specific case, the worst PSF value of the image set was chosen for the homogenization. Once the target PSF has been selected, we generate a homogenization kernel that depends on the image position by running  SExtractor \citep{Bertinetal1996}  and PSFEx \citep{Bertinetal2013} in every image. To generate the homogenization kernel PSFEx performs a  $\chi^{2}$ minimization to fit the target PSF. A two-dimensional Moffat model is used as an homogenization kernel. Once the variable PSF models have been determined, we convolve each image with its corresponding kernel using a Fast Fourier transform (FFT).  Although the method penalizes computer resources, it increases the speed of the process in large images. The homogenization process allows image combinations and photometry measurements more consistent at the expense of degrading some images. Therefore it is important to apply the method to a set of images where quality is comparable. Finally, we need to take into account that the homogenization process has consequences in the image noise producing pixel-by-pixel correlations. To correct from this, the algorithm recalculates the noise model of the images following the procedure described in \citet{Labbeetal2003} and \citet{Molinoetal2014}. For square apertures of different areas, the algorithm make a high number of  measurements of the sky flux in each aperture. The side of each square aperture varies from 2 to 30 pixels. A gaussian fit is made to the resulting histogram and the sky noise in an aperture of N pixels is modeled. Using apertures at different locations on the images, but also different aperture sizes, we account for the pixel-by-pixel correlation. This recalculated noise model has been used for computing the photometric errors.

\subsection{The Centroidal Voronoi Tessellation} \label{sec:voronoi}

Spatial binning technique is a widely used technique to reach a required minimum S/N for a reliable and unbiased extraction of information. Given the large variations in the S/N across the detector, ordinary binning and smoothing techniques are not suitable to capture detailed structure. \cite{Cappellarietal2003} test different methods of adaptively binning IFS data and conclude that the Centroidal Voronoi Tessellation (CVT) is the optimal algorithm that solves the 2D-binning problems. We use this binning scheme adapting the size of the bin to the local S/N (e.g., bigger bins are applied in the low-S/N regions, while a higher resolution is retained in the higher-S/N parts). Following their prescription, we have extended the adaptive spatial binning of IFS data proposed to photometric imaging data.

Previous to the CVT, a mask was applied to determine the area of the object to be analyzed. If the area is too small, we will lose information of the outskirt. If the area is too big, it will include regions with low S/N and the tessellation will not satisfy the morphological and uniformity requirements in the outer parts. Neither situation will produce an optimal tessellation so we need to reach a compromise between the resolution and the S/N achieved. All pixels inside the Kron radius of filter F365W, R$\textsubscript{Kron}$, were selected for the CVT\footnote{R$\textsubscript{Kron}$ is defined by SExtractor as  a flexible  elliptical aperture that confines most of the flux from an object and has been empirically tested to enclose > 90$\%$ of the object's light.}. If secondary objects were present in the image, they were masked to avoid contaminating light. An area of 1.5 R$\textsubscript{eff}$ of each secondary objects was masked during the analysis\footnote{R$\textsubscript{eff}$ is defined as the half light radius, i.e. the radius containing 50$\%$ of the object's light. R$\textsubscript{eff}$ was always defined from the synthetic F814W images created to emulate the analogous band of the Advanced Camera for Surveys (ACS) onboard of the Hubble Space Telescope (HST). These images are created as a linear combination of individual filter images and are used in ALHAMBRA for detection and completeness purposes \citep{Molinoetal2014}.}.

The CVT algorithm is divided into 3 different phases:
\begin{itemize}
\item  Phase I: Accretion phase
\begin{enumerate}[label=(\roman*)]
\item  Start the first bin from the highest S/N pixel of the image.
\item Evaluate the mass centroid of the current bin. 
\item Select a next pixel as candidate to be included in the current bin and analyze the following conditions:
\begin{enumerate}
\item The candidate pixel is adjacent to the current bin.
\item The roundness ($R_{c}$) of the bin would remain under 0.3 where $R_{c}=\frac{r_\mathrm{max}}{r_\mathrm{ref}} -1$, $r_\mathrm{max}$ is the maximum distance between the centroid of the bin and any of the bin pixels, and $r_\mathrm{ref}$ is the radius of a disk of same area as the whole bin.
\item The potential new bin S/N would not deviate from the target S/N more than the current bin.
\end{enumerate}
If the candidate pixel fulfilled the previous criteria, it will be binned to the current bin. Go back to step (ii).
\item Finish the addition of pixels to the current bin and start a new bin from the unbinned pixel closest to the centroid of the last bin. Go back to step (ii) until all pixels have been processed as successful and added to a bin or as unsuccessful and classified as unbinned.\\
\end{enumerate}

\item  Phase II: Reassigning phase
\begin{enumerate}[label=(\roman*), resume]
\item Evaluate the mass centroid of all the successful identified bins and reassign the unsuccessfully binned pixels to the closest of these bins.
\item Recompute the centroid of each final bin.\\ 
\end{enumerate}

\item  Phase III: Equi-mass CVT
\begin{enumerate}[label=(\roman*), resume]
\item Use the previous centroids as initial generators to perform a Voronoi Tessellation.
\item Determine the mass centroids of the Voronoi regions according to $\rho=(S/N)^{2}$ to force generate bins that enclose equal masses according to that density. Use these new set of centroids as new generators.
\item Iterate over step (vii) until the old generators and the new ones converge. 
\end{enumerate}
\end{itemize}

This optimal binning scheme satisfies the following requirements: a) topological requirement: a partition without overlapping or holes, b) morphological requirement: the bins have to be as compact as possible so the best spatial resolution is obtained along all directions, c) uniformity requirement: small S/N scatter of the bins around a target value, to not compromise spatial resolution in order to increase S/N of each bin,  d) equal-mass bins according to a density distribution of $\rho=(S/N)^{2}$.

Considering the photometric depth of the ALHAMBRA images as well as the characteristic SED of typical early type galaxies, the S/N ratio will vary significantly from filter to filter  and even along different redshifts. Based on \citet[][see their Fig. 7]{DiazGarciaetal2015}, the bluest filter, F365W, is on average the band with the minimum S/N  ratio at redshift $z$ < 0.4. Therefore, F365W will put constraints in our ability to determine reliable stellar population parameters. Enforcing a minimum S/N in filter F365W will ensure a proper S/N ratio in the rest of the bands as well as a reliable determination of stellar parameters.  In order to obtain typical uncertainties of  $\Delta$log Age$\textsubscript{L}$ =0.14 and $\Delta$[Fe/H]$\textsubscript{L}$=0.20 for $z$ < 0.4 in ALHAMBRA red sequence galaxies, the SED fitting code used, \textrm{MUFFIT}, requires a S/N $\sim$ 6 in the bluest filter, F365W. For a more detailed discussion about the impact of different S/N ratios on the derived stellar population parameters the reader is referred to Sec. 4.2 of \citet{DiazGarciaetal2015}. To be conservative, CVT has been performed in the filter F365W where  the target S/N was set to 10, defined as the average S/N of the bins. The tessellation was then applied to the images in all the filters and finally the photometry of every region in all the filters was determined.

Although ALHAMBRA images are already background subtracted, this subtraction was done globally over the entire image. We perform a local sky subtraction considering an area of 100 x 100 pixels (22" x 22") around each target galaxy.  Assuming the small area subtended in the sky by each object, we consider as a first approximation a constant sky subtraction. A second approach was tested applying CHEF functions \citep{JimenezTejaetal2012}. This tool models the light distribution of the galaxy through  an orthonormal polar base formed by a combination of Chebyshev rational functions and Fourier polynomials. We note that no significant improvement was detected so a constant sky subtraction was finally applied.  

\begin{figure*}
\begin{center}
\includegraphics[width=0.95\columnwidth]{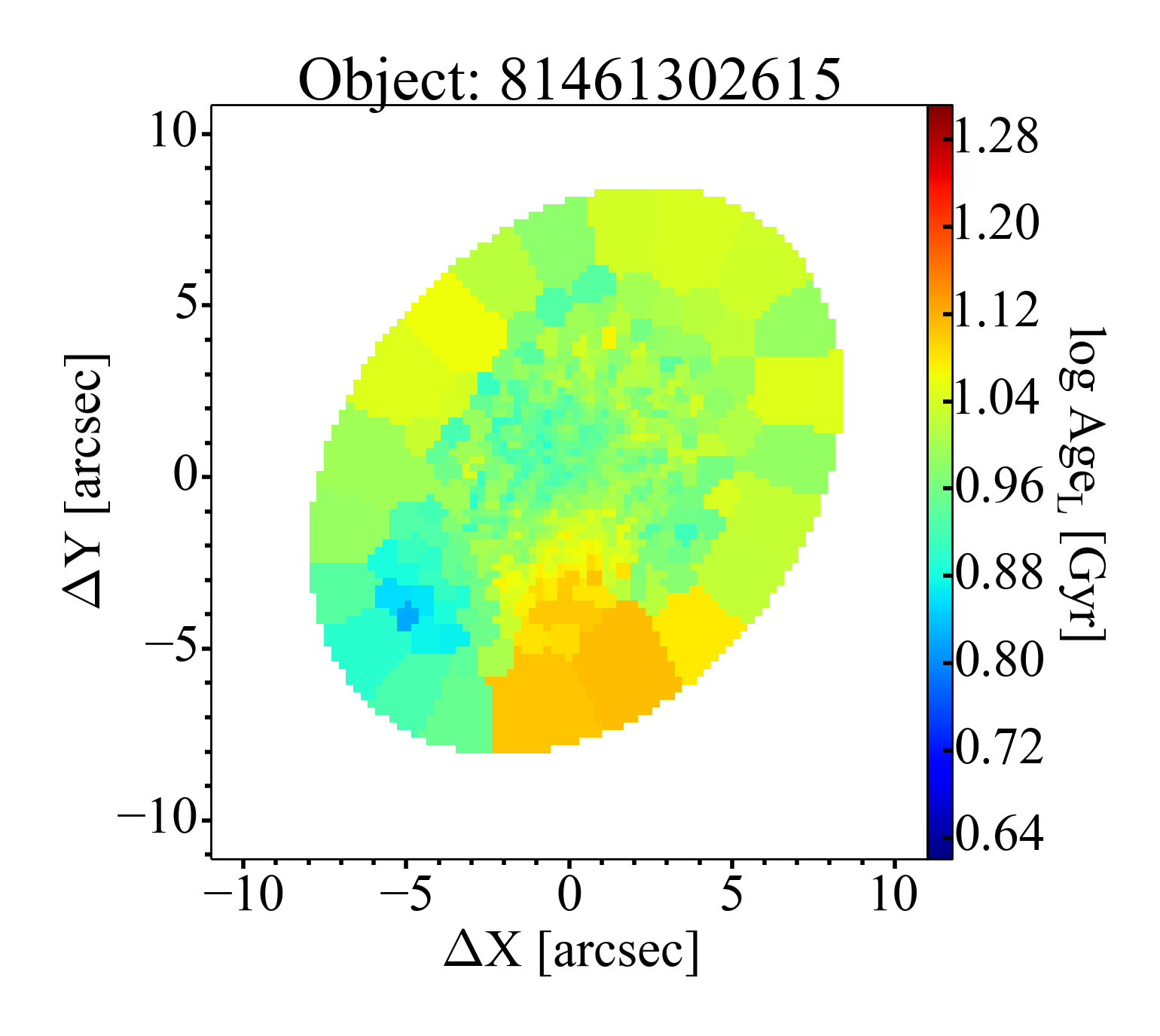}
\includegraphics[width=0.95\columnwidth]{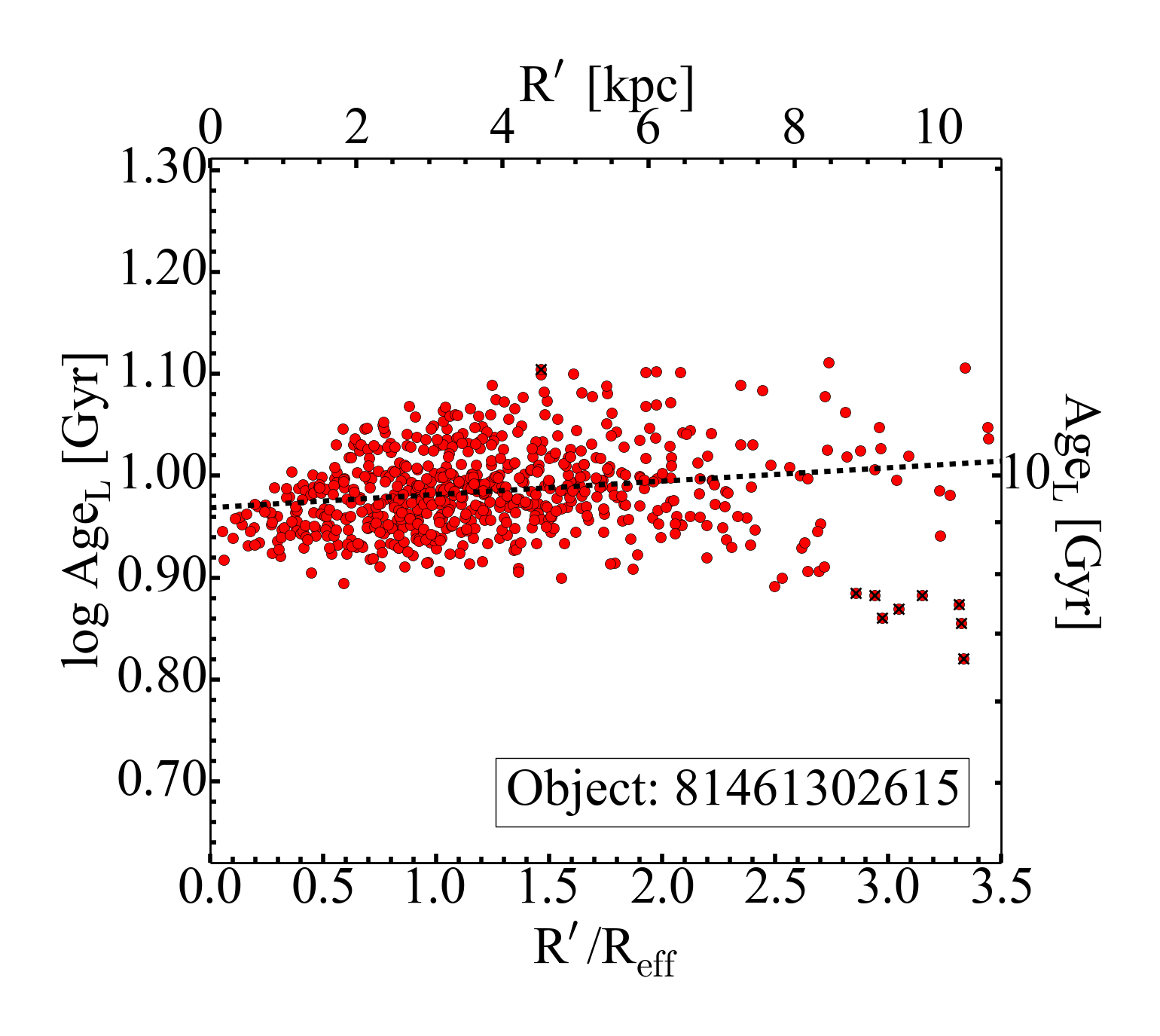}

\includegraphics[width=0.95\columnwidth]{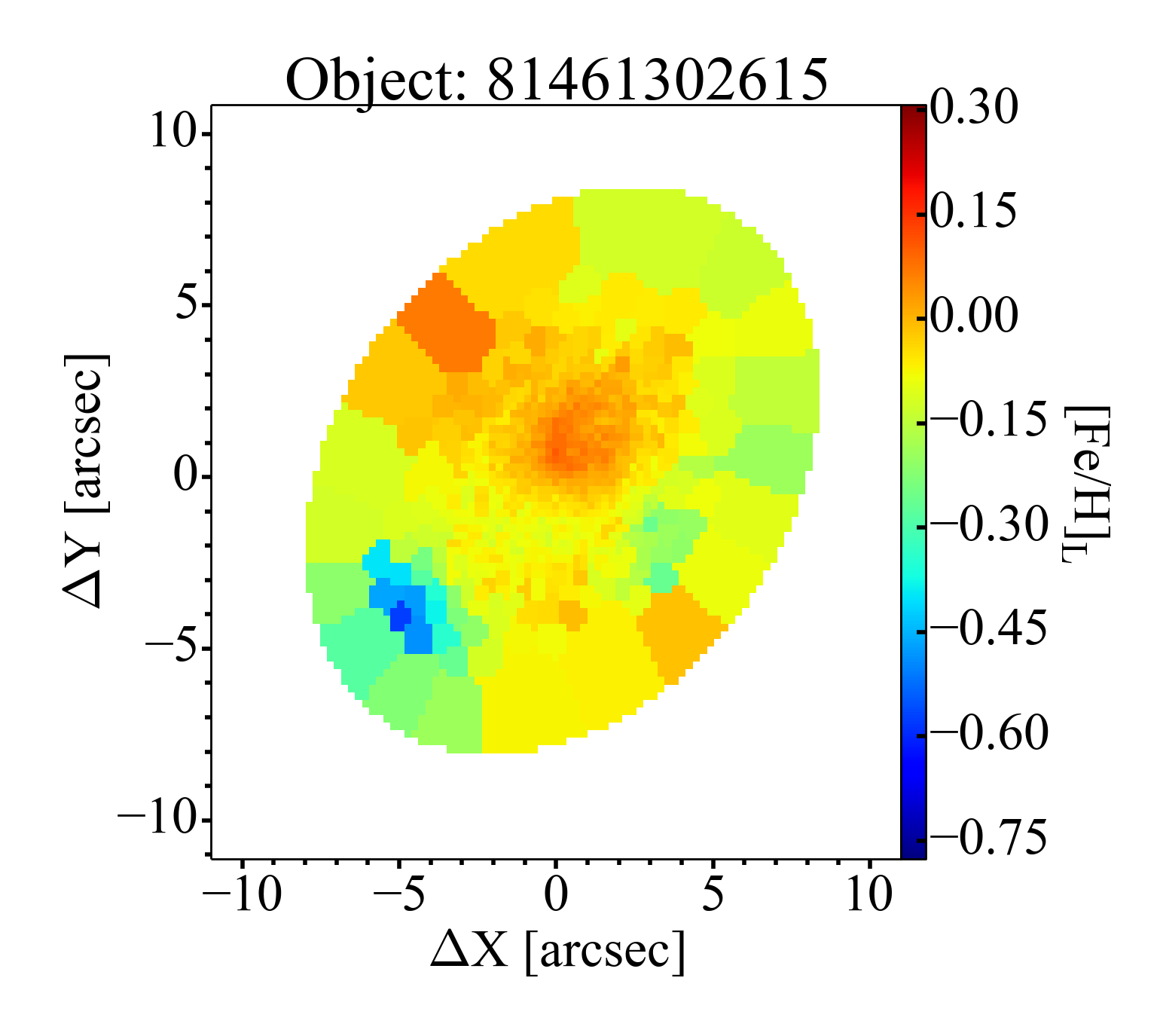}
\includegraphics[width=0.95\columnwidth]{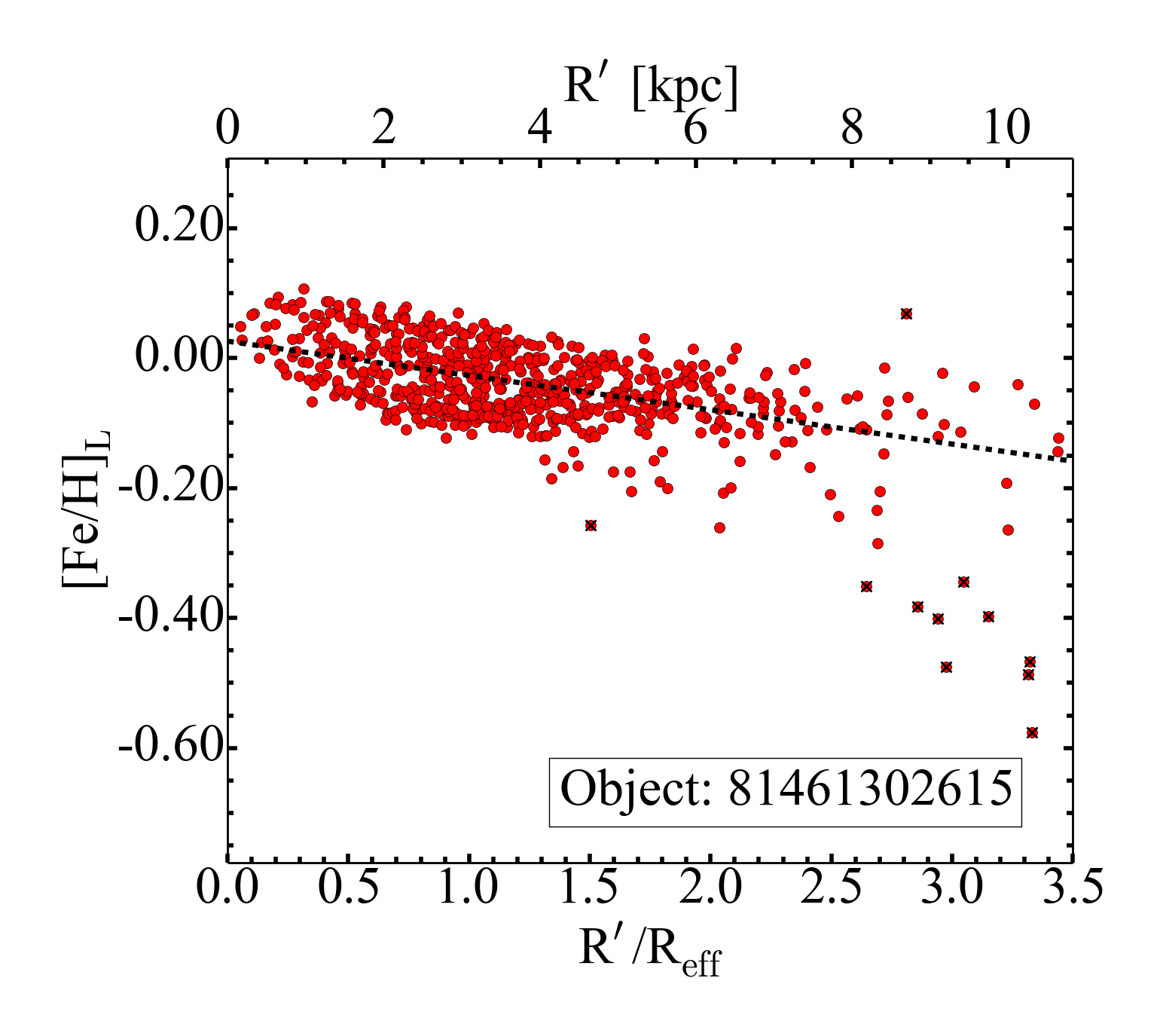}

\includegraphics[width=0.95\columnwidth]{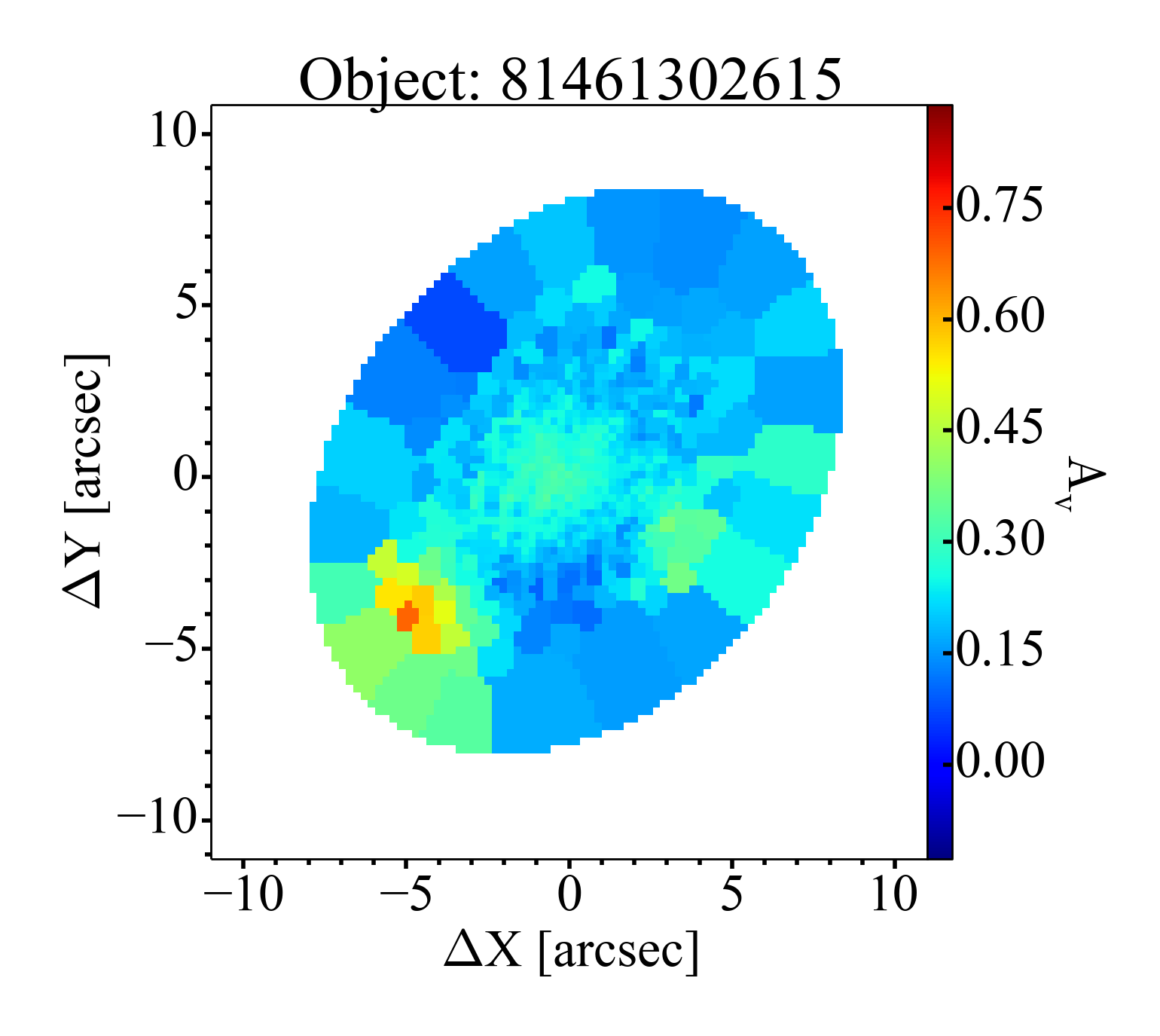}
\includegraphics[width=0.95\columnwidth]{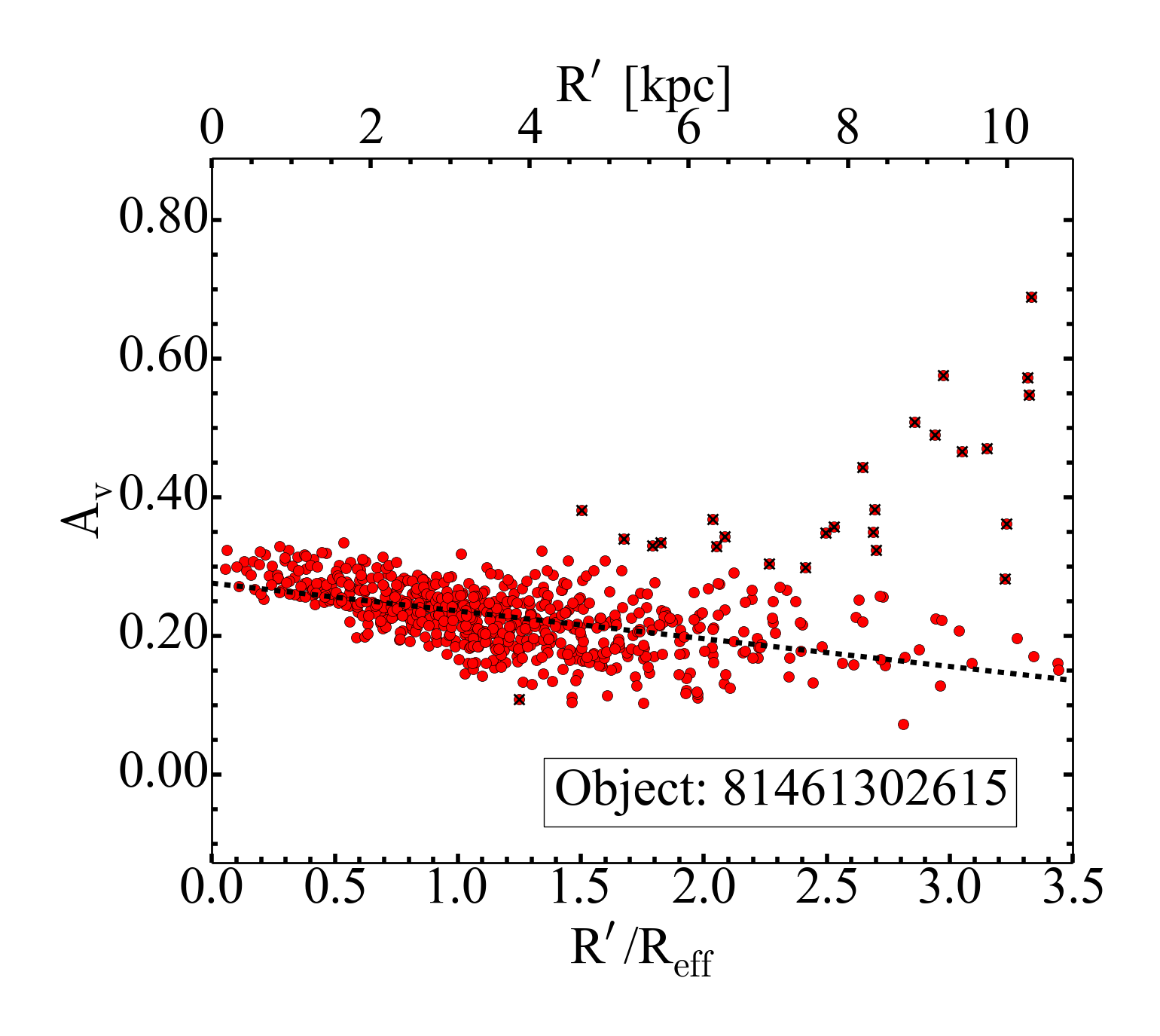}

\caption{Luminosity-weighted stellar population properties maps and radial profiles analyzed using E-MILES-based models for the example object 81461302615. \textit{Left column}: 2D maps for log Age, [Fe/H] and A$_\mathrm{v}$ covering an area of 22" x 22" around the target galaxy. The color range has been chosen to highlight inhomogenities. \textit{Right column}: Log Age, [Fe/H] and A$_\mathrm{v}$ as a function of the circularized galactocentric distance R$^\prime$ (see Sect. \ref{radial_profiles}). Each red circle symbol corresponds to a bin in the 2D map. Black dashed line corresponds to the error-weighted linear fitting. Black crosses correspond to outliers identified during the sigma clipping process and were not considered during the fit.} 
\label{fig:2}
\end{center}
\end{figure*}

\subsection{Stellar population parameters}
After the CVT has been performed, and the photometry of every region in every filter was determined, we run the code \textrm{MUFFIT} to obtain 2D maps of different stellar populations properties. \textrm{MUFFIT} is a generic code optimized to retrieve the main stellar population parameters of galaxies in photometric multi-filter surveys.  The code compares the multi-filter fluxes of galaxies with the synthetic photometry of mixtures of two single stellar populations (SSP) for a range of redshifts and extinctions through an error-weighted $\chi^{2}$ approach. In addition, the code removes during the fitting process those bands affected by emission lines in order to improve the quality of the fit. The determination of the best solution space is based on a Monte Carlo method. This approach assumes an independent Gaussian distribution in each filter, centered on the band flux/magnitude, with a standard deviation equal to its photometric error. Each filter is observed and calibrated independently of the remaining ones, so the errors of different filters are not expected to correlate. This approach not only provides the most likely range of ages, metallicities (both luminosity- and mass-weighted), extinctions, redshifts, and stellar masses but set constraints on the confidence intervals of the parameters provided.

Several studies have shown that the mixture of  two SSPs is a reasonable compromise that significantly improves the reliability of determining the stellar population parameters of multi-filter galaxy data \citep{Ferrerasetal2000, Kavirajetal2007, Lonoceetal2014}. In fact, \citet{Rogersetal2010} have shown that the mixture of two SSPs is the most reliable approach to describe the stellar population of early-type galaxies. Most recently, \citet{LopezCorredoiraetal2017} fit a set of 20 red galaxies with models of a single-burst SSP, combinations of two SSPs as well as an extended star formation history. They conclude that exponentially decaying extended star formation models ($\tau$-models) improve slightly the fits with respect to the single burst model, but they are considerably worse than the two SSPs based fits, further supporting the residual star formation scenario. Based on these studies, we consider the 2-SSP model fitting approach the best method for our study. That being said, \textrm{MUFFIT} is currently being tested with later-type galaxies and future versions of the code will also account for the use of different sets of SSPs or $\tau$-models for the best choose of the user. \textrm{MUFFIT}  has been tested and verified using data from the ALHAMBRA survey \citep{Molesetal2008}. In particular, \textrm{MUFFIT} accuracy and reliability tests are performed considering ALHAMBRA sensitivity and filter set, thus, analysis and results present in \citet{DiazGarciaetal2015} are directly applicable to the present study.

Single stellar population models are a key ingredient to disentangle physical properties of stellar populations. We provide \textrm{MUFFIT} with two different sets of SSP models: BC03 \citep{BruzualCharlot2003} and E-MILES \citep{Vazdekisetal2016}. To explore any potential dependence of the SSP model used, the analysis of this study was performed using both sets of models.  For BC03 with a spectral coverage from 91 $\AA$ to 160 $\mu$m, we selected ages up to 14 Gyr, metallicities [Fe/H]= --1.65, --0.64, --0.33, 0.09, and 0.55, Padova 1994 tracks \citep{Bressan1993, Fagotto1994a,Fagotto1994b,Girardi1996}, and a \cite{Chabrier2003} initial mass function (IMF). E-MILES provides a  spectral range of $1680 - 50000$ $\AA$ at moderately high resolution for BaSTI isochrones \citep{Pietrinfernietal2004}. E-MILES models include the Next Generation Spectral Library \citep[NGSL,][]{Heapetal2007}  for computing spectra of single-age and single-metallicity stellar populations in the wavelength range 1680--3540 $\AA$, and the NIR predictions from MIUSCATIR \citep{Rocketal2015}. For BaSTI isochrones, we chose 21 ages in the same age range as BC03, but with metallicities [Fe/H]= --1.26, --0.96, --0.66, --0.35, 0.06, 0.26, and 0.4. We assume a Kroupa Universal-like  IMF \citep{Kroupa2001}. \textrm{MUFFIT} allows the user to choose amongst several extinction laws. Throughout this work the Fitzpatrick reddening law has been used \citep{Fitzpatrick1999} with extinctions values A$\textsubscript{v}$ in the range of 0 and 3.1. This extinction law is suitable for dereddening any photospectroscopic data, such as ALHAMBRA \citep[further details in ][]{Fitzpatrick1999}.

To minimize the free fitting parameters we provide \textrm{MUFFIT} with an initial constraint of redshift values. We have used the spectroscopic redshifts determined by SDSS \citep{Gallazzietal2005} as input parameter. When spectroscopic redshifts were not available, photometric redshifts determined by ALHAMBRA \citep{Molinoetal2014} were used. We note that \textrm{MUFFIT} linearly averages the ages and metallicities (their eq. 16 and eq. 17) (log < Age> and log <[Fe/H]> versus < log Age > and <log[Fe/H] >)\footnote{While < log x > corresponds to the logarithm of the geometric mean, log< x >  corresponds to the logarithm of the arithmetic mean. Although the geometric mean is smaller than the arithmetic mean and less sensitive to extreme values, the geometric mean is not necessarily a good estimator of the arithmetic mean.}. One finds both types of averaging in the literature (e.g. MaNGA versus CALIFA). We note that a logarithmic weighting formalism may give more weight to younger and metal poorer stellar populations. Due to the small range in Age and [Fe/H] that we cover in our analysis, no significant consequences are introduced in the analysis because of differences in the mathematical treatment. Therefore both approaches would lead to similar results.

\begin{figure}
\begin{center}
\includegraphics[width=0.95\columnwidth]{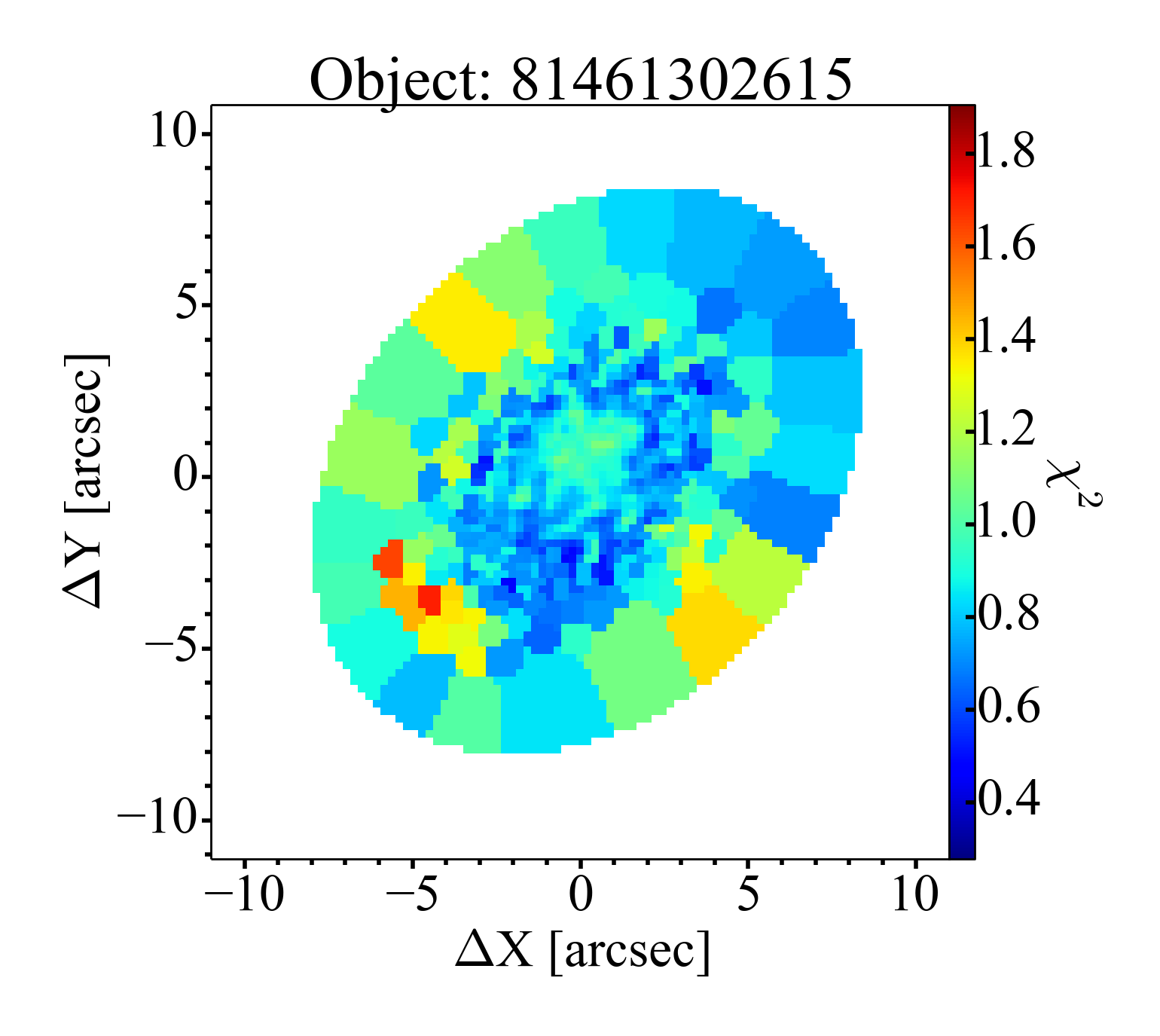}
\caption{Goodness-of-fit map, $\chi^{2}$, for the example object 81461302615.} 
\label{fig:2b}
\end{center}
\end{figure}

\subsection{ 2D Maps: Age, [Fe/H], and A\textsubscript{v}}
\textrm{MUFFIT} provides luminosity- and mass-weighted ages, metallicities, and extinctions. Mass-weighted properties better trace the whole evolutionary history of the galaxy since they give insight into its mass assembly history. On the other hand, luminosity-weighted properties are more sensitive to the most recent periods of star formation in the galaxy. Throughout this study, we show both, mass-weighted  (log Age$\textsubscript{M} $ and [Fe/H]$\textsubscript{M}$) and luminosity-weighted  (log Age$\textsubscript{L}$ and [Fe/H]$\textsubscript{L}$) properties, as a way to complement each other and identify more clearly certain processes. As an illustrative example, Fig. \ref{fig:2} presents the age, metallicity and extinction maps and radial profiles (see Sect. 5) of a galaxy using E-MILES models and luminosity-weighted properties. We note a very distinct region around ($\Delta$X, $\Delta$Y) = (--5, --5)  with a younger, more metal-poor and more extinct stellar population than the rest of the galaxy. To further probe the quality of the SED fits,  the $\chi^{2}$ map of every object is inspected as a goodness-of-fit quality check. A detailed definition of the error-weighted $\chi^{2}$ minimization process can be found in Sec. 3.2.1 of \cite{DiazGarciaetal2015}. We note that although generally speaking, a value of $\chi^{2}$ $\sim$ 1 is a mean value for a good fit, $\chi^{2}$ serves only as an indicator. One must also examine other parameters as the distribution of residuals. A good model fit should yield residuals normally distributed around zero with no systematic trends. Figure \ref{fig:2b} shows the $\chi^{2}$ map of the example object. Visual inspection of the SED does not show evidence of higher photometric errors that could suggest an artificial feature although the $\chi^{2}$ of the fits in that area are slightly higher than the median value. A detailed light decomposition of our sample will reveal the nature of these substructures and will be present in future studies. The 2D maps as well as the radial profiles of all the sample objects can be found in the online version of the paper (Appendix C).

\section{Integrated versus resolved properties of the sample}\label{sec:integrated}
Besides the multi-filter photometry of every bin in a galaxy tessellation, we have also determined the integrated stellar properties of each galaxy. This integrated photometry was obtained using exactly the same method described in Sect. \ref{sec:method} for the galaxy as a whole, with no tessellation, emulating integrated aperture photometry. Several studies point out that besides the uncertainties in the stellar properties inferred from all uncertainties in model assumptions, there are additional biases related with the fitting procedure that can affect the determination of integrated stellar population parameter. In particular the outshining bias, i.e., young stellar populations obscuring older stellar components behind their bright flux, could produce a missing mass effect \citep{Zibettietal2009, Sorbaetal2015}. This problem is less severe when a resolved analysis is done, as those young components are localized in specific areas. Although we do not expect a strong contribution from young components in our sample of massive early-type galaxies, we analyze here the differences obtained with both methods.

\begin{figure} [h]
\begin{center}
\includegraphics[width=\columnwidth]{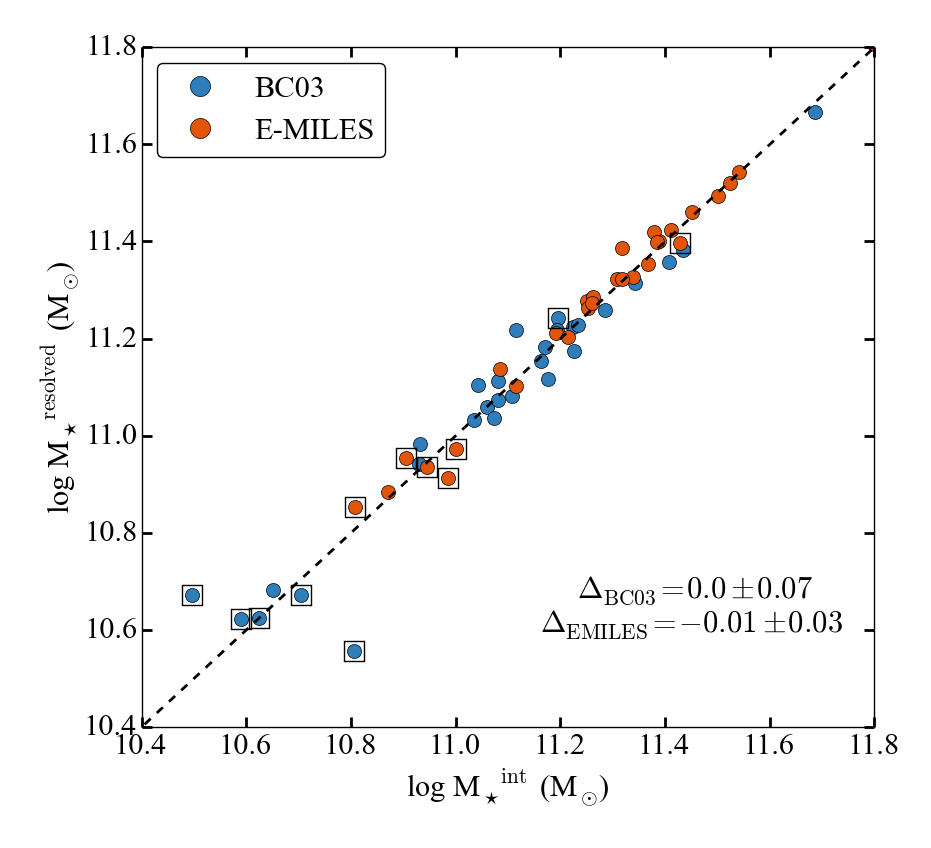}
\caption{Integrated versus spatially resolved stellar masses derived from \textrm{MUFFIT} for the 29 massive, early-type galaxies analyzed in this paper. Both BC03 and E-MILES SSP models were used. Symbols enclosed in squares indicate the 6 peculiar cases analyzed in Sect. \ref{Sec:peculiar}. A normal distribution has been fitted to the difference between the x-axis and the y-axis. The mean and standard deviation is labeled as $\Delta$. The black dashed line shows the one-to-one relation.} 
\label{fig:3}
\end{center}
\end{figure}

Figure \ref{fig:3} compares, for each galaxy,  the total stellar mass derived from the integrated method (log M$_{\star}$$^{\mathrm{int}}$) with the one obtained by adding the estimated masses in each tessellation bin (log M$_{\star}$$^{\mathrm{resolved}}$). The two values agree very well with a mean difference of --0.0 $\pm$ 0.07 dex for BC03 and --0.01 $\pm$ 0.03 dex for E-MILES.  

\begin{figure} [h]
\begin{center}
\includegraphics[width=\columnwidth]{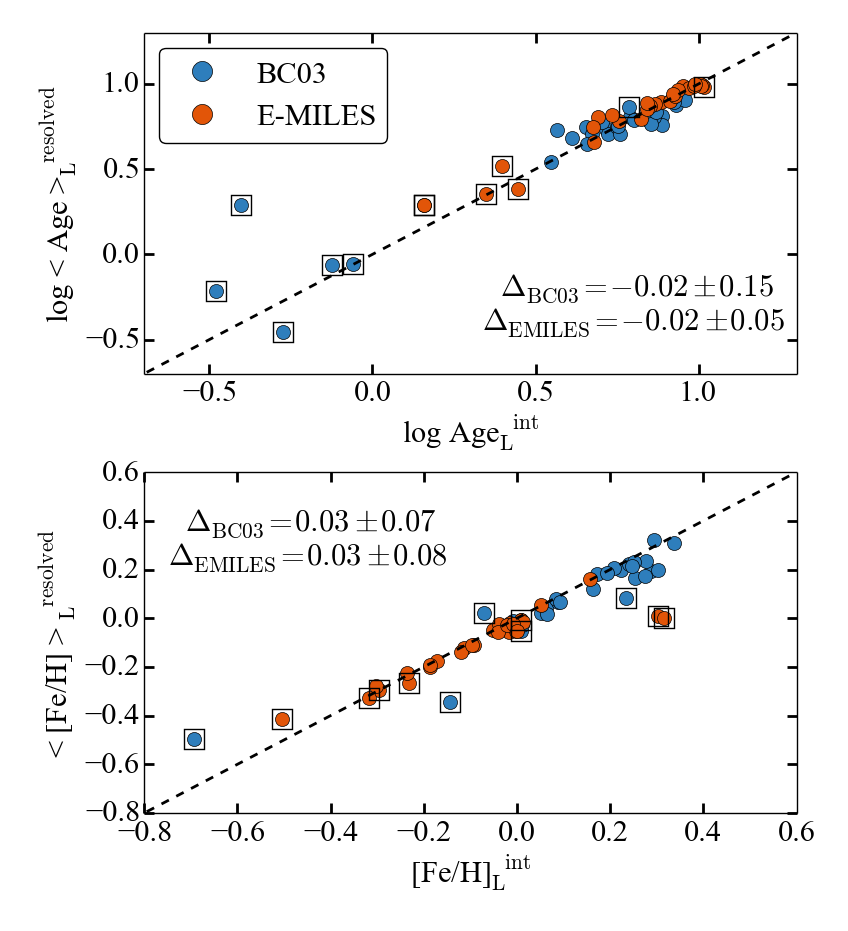}
\caption{Integrated versus spatially resolved ages and metallicities derived from \textrm{MUFFIT} for the 29 massive, early-type galaxies analyzed in this paper. Both BC03 and E-MILES SSP models were used. Symbols enclosed in squares indicate the 6 peculiar cases analyzed in Sect. \ref{Sec:peculiar}. A normal distribution has been fitted to the difference between the x-axis and the y-axis. The mean and standard dispersion is labeled as $\Delta$ in each panel. The black dashed line shows the one-to-one relation.} 
\label{fig:4}
\end{center}
\end{figure}

Figure \ref{fig:4} compares the Age$_\textsubscript{L}^\textsubscript{int}$ and [Fe/H]$_\textsubscript{L}^\textsubscript{int}$ from the integrated photometry to the weighted mean properties (< Age >$_\textsubscript{L}^\textsubscript{resolved}$ and < [Fe/H] >$_\textsubscript{L}^\textsubscript{resolved}$) from the spatially resolved analysis. The spatially resolved weighted mean properties are defined as:

\begin{equation}\label{eq:1}
< \mathrm{Age} >_\textsubscript{L}^\textsubscript{resolved} = \frac{{\sum_{i} L_{\mathrm{5515},i}\cdot \mathrm{Age}_{\mathrm{L},i}}}{{\sum_{i} L_{5515,i}}} \,,  \text{and} 
\end{equation}

\begin{equation}\label{eq:2}
< \mathrm{[Fe/H]}> _\textsubscript{L}^\textsubscript{resolved} = \frac{{\sum_{i} L_{5515,i}\cdot \mathrm{[Fe/H]}_{\mathrm{L},i}}}{{\sum_{i} L_{5515,i}}} \,, 
\end{equation}

\noindent where $L_{5515,i}$ is the luminosity of bin $i$ evaluated at a reference rest-frame wavelength of 5515$\AA$ and $\mathrm{Age}_{\mathrm{L},i}$ and $\mathrm{[Fe/H]}_{\mathrm{L},i}$ the luminosity-weighted age and metallicity for that $i$ bin.

Once again the values are in agreement with no significant offset ($\Delta$log Age$_\textsubscript{L, BC03}$ = --0.02 $\pm$ 0.15, $\Delta$log Age$_\textsubscript{L,EMILES}$= --0.02 $\pm$ 0.05, $\Delta$[Fe/H]$_\textsubscript{L,BC03}$=0.03 $\pm$ 0.07, $\Delta$[Fe/H]$_\textsubscript{L,EMILES}$= 0.03 $\pm$ 0.08). Results are shown for both SSP models. Overall, the total stellar mass, age and metallicity estimated from integrated photometry are remarkably robust when compared with those from a spatially resolved analysis. Similar conclusions (see Fig. \ref{fig:4a}) are reached when mass-weighted mean properties are evaluated where the mass-weighted version of equations \ref{eq:1} and \ref{eq:2} are applied (See Appendix \ref{ap:resolved}). These results agree with recent conclusion obtained using CALIFA galaxies \citep{GonzalezDelgadoetal2014a, GonzalezDelgadoetal2015} where galaxy-averaged stellar ages, metallicities, mass surface density and extinction are well matched by the corresponding values obtained from the analysis of the integrated spectrum. Table \ref{tab:A1} of Appendix \ref{ap:tables} summarizes the spatially resolved stellar properties of the sample as well as the median $\chi^{2}$ of the SED fits.

\subsection{Spatially resolved mass-age and mass-metallicity relations}
\begin{figure} [h]
\begin{center}
\includegraphics[width=\columnwidth]{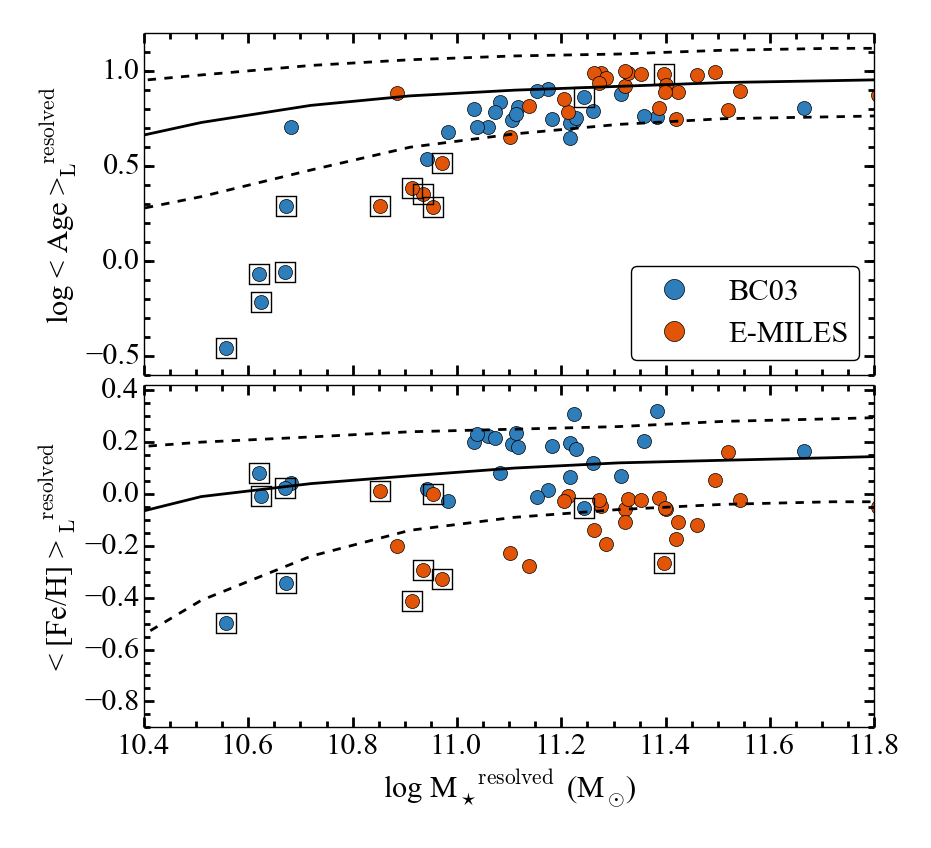}
\caption{Spatially resolved ages and metallicities versus mass. The lines indicate the relation obtained in \citet{Gallazzietal2005} where the solid line corresponds to the median distribution and the dashed lines to the 16$^\mathrm{th}$ and 84$^\mathrm{th}$ percentiles. Symbols enclosed in squares indicate the 6 peculiar cases analyzed in Sect. \ref{Sec:peculiar}.}
\label{fig:correlation}
\end{center}
\end{figure}

Galaxies in the local Universe exhibit a clear correlation between stellar mass and their properties in the well-known mass-age and mass-metallicity (MZR) relations. Rather than characterize the nature of these relations, we want to find out how well our spatially resolved mass-age and mass-metallicity relations follow previous trends to guarantee that our technique is able to reproduce previously observational evidences.

In Fig. \ref{fig:correlation}, we show the relation between the spatially resolved age and metallicity with the mass, assuming both SSP models. For comparison purposes, we have overplotted the relation obtained by \citet{Gallazzietal2005} that includes all type of galaxies. \citet{Gallazzietal2005} use a Bayesian approach to derive ages and metallicities by a simultaneous fit of five spectral absorption features. Their analysis uses absorption line indices with different sensitivities for age and metallicity focusing on Lick indices and the 4000 $\AA$ break \citep{Gorgasetal1993, Wortheyetal1994}. The population properties were determined using BC03 SSP models. The ages of our galaxies show a clear increase with the galaxy mass, thus, more massive galaxies tend to be older which represent the well know mass-age relation.  Although overall this trend follows the mass-age relation obtained by \citet{Gallazzietal2005} for massive galaxies, there are 5  low massive galaxies for each SSP (log M$_{\star}$ < 11.0 ) that significantly departure for the general trend. These outliers are identified as either star forming galaxies or AGNs candidates (see Sect. \ref{Sec:peculiar} for further details). \citet{Pengetal2015} split Gallazzi et al. population of galaxies into quiescent and star forming and found a different mass-age relation between both populations. The observational evidences found by \citet{Pengetal2015} would explain the position of our star forming galaxies outside the general trend.  Figure \ref{fig:correlation} shows that the spatially resolved stellar population properties derived from this method reproduces the well known downsizing effect \citep{Cowieetal1996}.

On average, the global metallicities of the galaxies increase with their mass (bottom panel in Fig. \ref{fig:correlation}) and agree with the general trend found by \citet{Gallazzietal2005}. \citet{Pengetal2015} also found a different metallicity-mass relation between quiescent and star forming galaxies. At a given stellar mass, the stellar metallicity of quiescent galaxies is higher than for star forming galaxies for log M$_{\star}$ < 11.  According with their Fig. 2 and for our mass range, the metallicity difference for the two galaxy populations would be  [Fe/H] $\sim$ --0.2 dex in agreement with our BC03 results. In general, the behavior of our galaxy sample reproduces well the mass-age and mass-metallicity relations of the local Universe based on SDSS observations although a clear offset is present based on the SSP models used.

\subsection{Comparison with previous spectroscopic studies}
No common objects were found between our sample and any IFU survey as SAURON, CALIFA, or MaNGA. This inconvenience hinders any one-to-one galaxy  comparison of our method. However, there is a subsample of 9 galaxies in the SDSS catalog for which individual integrated spectroscopy is available. Estimations of ages and metallicities of the central area of these galaxies are provided by \citet{Gallazzietal2005}.  SDSS spectra were taken using fiber aperture of 3" while our photo-spectra are not restricted to a fixed aperture. For a fair comparison and to avoid any aperture effects, we consider only the spatially resolved properties of our maps in a 3" circular aperture placed at the galaxy center (< Age >$_\textsubscript{L}^\textsubscript{resolved}$ (3") and < [Fe/H] >$_\textsubscript{L}^\textsubscript{resolved}$ (3")) 

\begin{figure} [h]
\begin{center}
\includegraphics[width=\columnwidth]{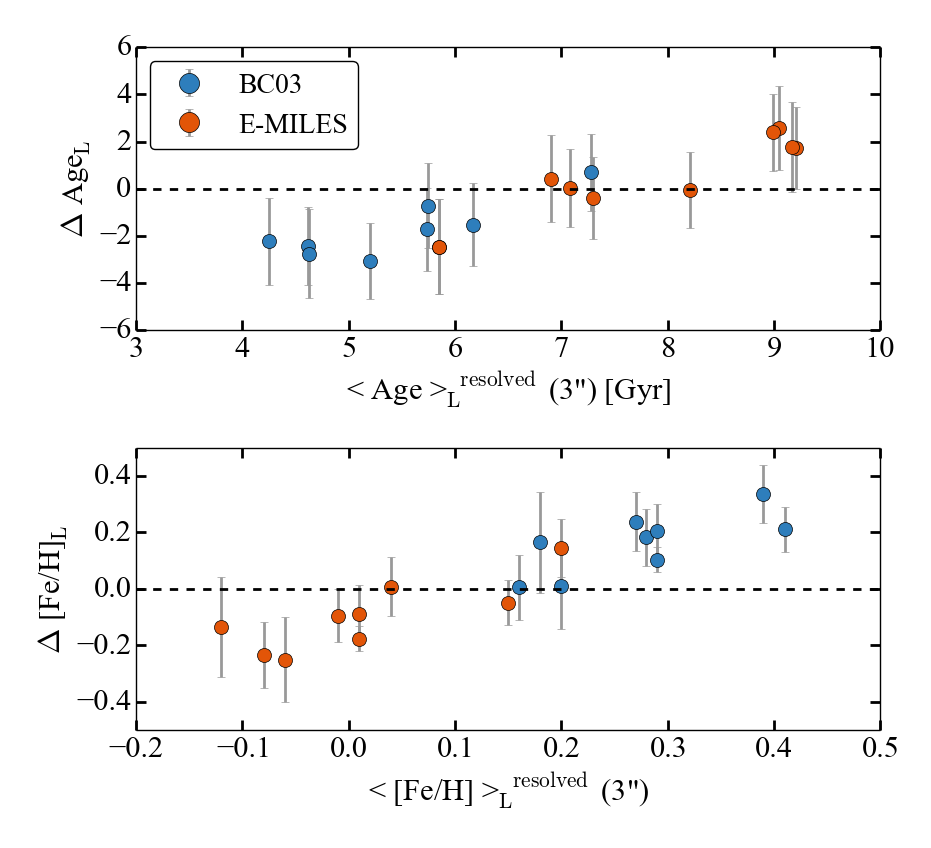}
\caption{Difference stellar population values between our study and SDSS \citep{Gallazzietal2005}  as a function of Age$\textsubscript{L}$ (top panel) and [Fe/H]$\textsubscript{L}$ (bottom panel) from this study. Photometric values of our analysis were determined in the same aperture size than SDSS (3"). The black dashed line shows the one-to-one relation.} 
\label{fig:5}
\end{center}
\end{figure}

Figure \ref{fig:5} presents a one-to-one comparison of the spectroscopic luminosity-weighted ages and metallicities given by \citet{Gallazzietal2005} and our photometric values analyzed in a 3" aperture of the sub-sample of 9 common objects  ($\Delta$Age$_\mathrm{L}$ = < Age >$_\mathrm{L}^\mathrm{resolved}$ (3") -- Age$_\mathrm{L}$(SDSS) and $\Delta$[Fe/H] = < [Fe/H] >$_\mathrm{L}^\mathrm{resolved}$(3") -- [Fe/H]$_\mathrm{L}$ (SDSS)) . The offsets between both studies are plotted against the luminosity-weighted age and metallicity to explore any dependency. A careful look at the top panel reveals that, except for one single object that drive a misleading trend, the sample shows, for BC03,  an offset in agreement with the expected $\sim$ 2 Gyr difference determined by \cite{DiazGarciaetal2015}. As the authors conclude, this discrepancy is due to intrinsic systematic differences between the two methods. In the bottom panel, an apparent trend is shown between the metallicity offsets as a function of the luminosity-weighted metallicities suggesting a possible degeneracy. In order to study the accuracy and reliability of the stellar population parameters retrieved, we quantify the expected degeneracies among the derived parameters in the next section.

\subsection{Degeneracies}
Understanding the degeneracies between the different parameters is a crucial step in order to avoid any misinterpretation that could mislead the analysis. Although some of them are unavoidable, in this section we analyze their extension and potential effects. As in any color-dependent diagnostic technique, any parameter that can modify the color of the object needs to be evaluated. This means that we need to include, in the analysis of the degeneracies, not only the age and metallicity, but also the extinction. To address the degeneracy problem we used the stellar population values recovered by \textrm{MUFFIT} during the Monte Carlo approach for every single object in every bin of the tessellation.  Then we stack each distribution to build a final distribution per object and obtain distributions among pairs of parameters (age, metallicity and extinction).  To quantify the final degeneracy between each pair of distributions, we follow the method used in \citet{DiazGarciaetal2015}.  Each distribution of potential values is characterized by setting confidence ellipses (2D confidence intervals) that enclose the results provided during the Monte Carlo process.  These ellipses are obtained from the covariance matrix of each distribution. The ellipticity, $e$, and the position angle, $\theta$ of these ellipses allow us to quantify the degeneracies.  A value of $e$  close to zero implies no degeneracy between the two parameters. Also, if $\theta$ lies on any of the two axes ($\theta$ is multiple of $\pi$/2) both parameters are not correlated and no degeneracy is found. In addition, \citet{DiazGarciaetal2015} quantify the level of degeneracy between parameters via the Pearson's correlation coefficient \citep[see Eq. 25 in][]{DiazGarciaetal2015}. If the Pearson's coefficient, $r_{xy}$, is close to 1 (or -1), the correlation (anti-correlation) is large, i.e., there is a degeneracy between the parameters. On the contrary, the closer the coefficient is to 0 (e.g., -0.1 $\leq r_{xy} \leq 0.1$) the more uncorrelated are the parameters. Thus, we end up with an \textit{ellipse of degeneracy} ($e$, $\theta$ and $r_{xy}$) per object for every pair of parameters where the ellipse represents the covariance error ellipse that enclose, at the 95$\%$ confidence level, the distribution of the parameters during the Monte Carlo approach. 

\begin{figure}
\begin{center}
\includegraphics[width=\columnwidth]{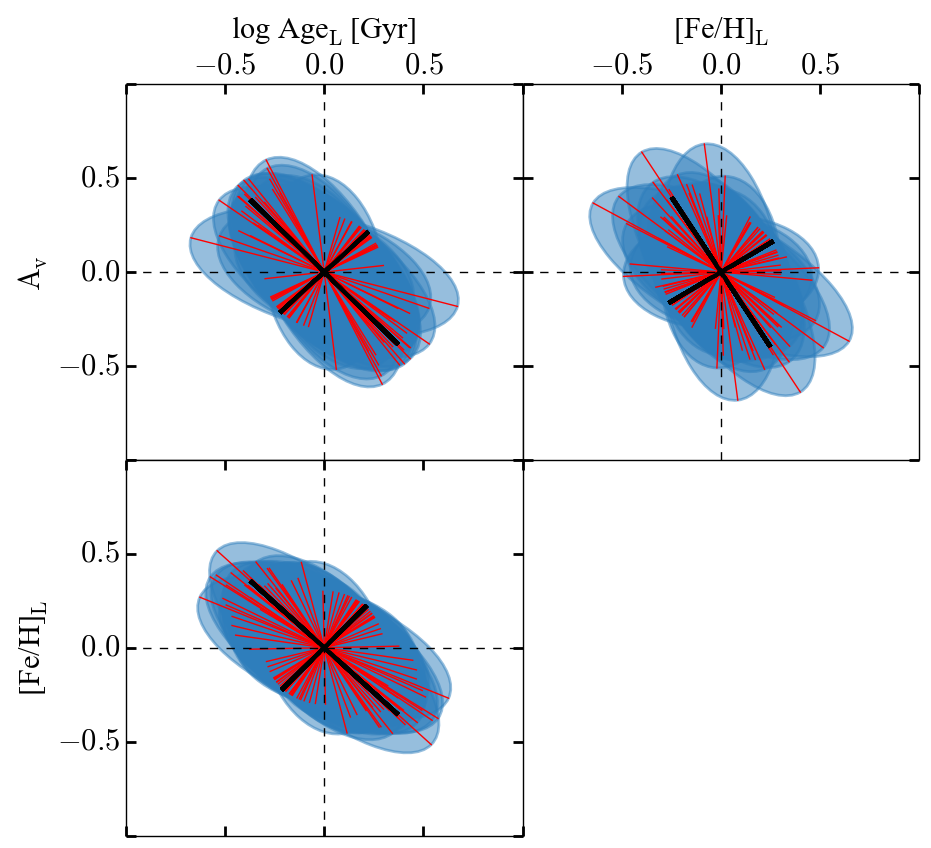}
\caption{Stacked covariance error ellipses, at the 95$\%$ confidence level, of the stellar population parameters for all the objects in the sample. The red lines illustrate the minor and major axis of each ellipse while the black lines correspond to the median values for the entire sample for E-MILES models.} 
\label{fig:6b}
\end{center}
\end{figure}

\begin{figure} [h]
\begin{center}
\includegraphics[width=\columnwidth]{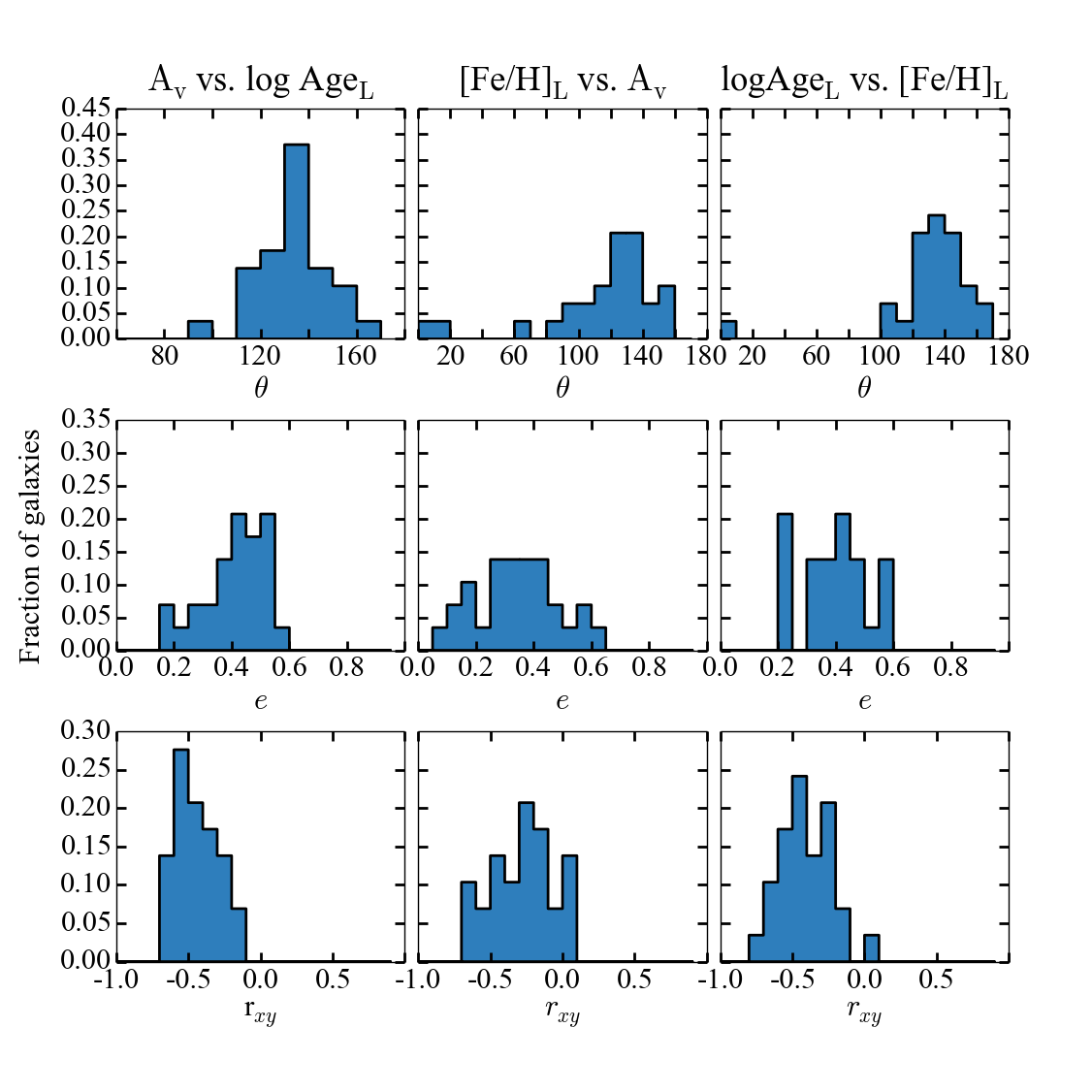}
\caption{Histogram of the position angle ($\theta$), ellipticity ($e$), and Pearson's coefficient (r$_{xy}$) for E-MILES models. The related variables are shown at the top of each column.} 
\label{fig:7}
\end{center}
\end{figure}

Figure \ref{fig:6b} corresponds to the stacking of the ellipses of degeneracy for all the objects where the red lines illustrate the minor and major axis of each ellipse and the black lines the median values for the entire sample. Figure \ref{fig:7} provides the histogram of $\theta$, $e$ and $r_{xy}$ in all regimes. Table \ref{tab:deg} summarizes the median values of the covariance error ellipses. An old, metal-rich sample as ours, exhibits anti-correlation in all cases. This means that, for example, a stellar population reddened by extinction can imitate the behavior of a population reddened because of the metal-rich content or viceversa.  We also checked that the general degeneracy trends do not significantly vary on the basis of the input model set (BC03 vs E-MILES) except for the case of log Age$_\mathrm{M}$ vs. [Fe/H]$\mathrm{M}$ where E-MILES SSP can break or reduce the degeneracy ($r_{xy}$ = --0.10). We emphasize that these results are restricted to our sample, i.e. early-type galaxies, and that the covariance error ellipses for more complicated stellar population histories could be different.

\begin{table}
\caption{Median values of the covariance error ellipses for all the objects in the sample: position angle ($\theta$), ellipticity ($e$), and Pearson's coefficient (r$_{xy}$).}
\label{tab:deg}
\centering
\small
\begin{tabular}{ccccc}

\hline 
\hline \\[-1ex]
& Models  &  $\theta$ &  $e$   & r$_{xy}$ \\[0.5ex]
 
\hline \\[-1ex]

A$\textsubscript{v}$ vs. log Age$\textsubscript{L}$ & BC03        & 134   & 0.43  & --0.44\\
                                & E-MILES & 132   & 0.44  & --0.50\\

  [Fe/H]$\textsubscript{L}$ vs. A$\textsubscript{v}$   & BC03        & 122   & 0.35 &  --0.27 \\
                                   & E-MILES & 135  & 0.46  & --0.51 \\
                                   
log Age$\textsubscript{L}$ vs. [Fe/H]$\textsubscript{L}$   & BC03        &136  & 0.40 & --0.43\\
                              & E-MILES  & 120 & 0.29 & --0.29\\[0.5ex]

\hline \\[-1ex]

A$\textsubscript{v}$ vs. log Age$\textsubscript{M}$ & BC03        & 143   & 0.34  & --0.28\\
                                & E-MILES & 133   & 0.42  & --0.38\\

  [Fe/H]$\textsubscript{M}$ vs. A$\textsubscript{v}$   & BC03        & 138   & 0.37 &  --0.36 \\
                                   & E-MILES & 142  & 0.48  & --0.53 \\
                                   
log Age$\textsubscript{M}$ vs. [Fe/H]$\textsubscript{M}$   & BC03        &129  & 0.29 & --0.23\\
                              & E-MILES  & 69 & 0.26 & --0.10\\

\hline
\hline
\end{tabular}
\end{table}

\section{Radial profiles and gradients} \label{radial_profiles}
To show and quantify radial variations of the galaxies, we obtain the radial profiles of the three parameters (age, metallicity and extinction) for all 29 galaxies. It is common practice to derive radial profiles of stellar population properties by binning the output values into elliptical annuli that are scaled in  along the major axis such that the bins are constant in effective radius.  These azimuthally averaged radial profiles assume a priori symmetry in the stellar population of the galaxies and remove the important 2D information imprint in the maps by directly collapsing the information to a 1D plot.  We take a different approach by plotting the stellar population values of each bin in each object as a function of the circularized galactocentric distance, $R^{\prime}$: 

\begin{equation}\label{eq:3}
R^{\prime}=R \cdot \sqrt{\mathrm{cos}^2\phi + (a/b) \cdot \mathrm{sin}^2 \phi}  \,,
\end{equation}

\noindent where $R$ is the galactocentric distance of the bin, $\phi$ is the angle of the bin with respect the semi-major axis, and $a/b$ is the semi-axis ratio of the galaxy as obtained by IRAF/Ellipse in the synthetic F814W band. Along the semi-major axis ($\phi$ = 0) $R^{\prime}=R$,  whereas in a more general case $R^{\prime} \geq R $. For the specific case of an apparent circular galaxy ($a/b$ = 1), $R^{\prime} = R$ at all $\phi$. This parameter, $R^{\prime}$, allows us to take into account the information in all the bins considering the morphology of the object. Right panels in Fig. \ref{fig:2} present the radial profiles for the age (log Age), metallicity ([Fe/H]) and extinction (A$_\mathrm{v}$) obtained from the 2D maps of an example object. Each red symbol corresponds to a bin in the 2D map. For a given $R^{\prime}$, the spread of the values gives us an idea of the departure of the stellar population from symmetry. It is important to note that the majority of previous studies can not go beyond R = 1.5 R$\textsubscript{eff}$ except CALIFA \citep{GonzalezDelgadoetal2015} that reaches 2.7 R$\textsubscript{eff}$. In contrast, our analysis derives gradients out to 2 -- 3.5 R$\textsubscript{eff}$.

Although the radial profiles of our sample seem to follow, in general, linear relations as a function of galactocentric distance, several authors assume power-law gradients. For comparison purposes, radial gradients were fitted assuming linear relation in linear space and logarithmic space as:

\begin{equation}\label{eq:4}
X= X_{0} + \nabla X \cdot R/R_\mathrm{eff} \,, \text{and}
\end{equation}
\vspace{-7pt}
\begin{equation}\label{eq:5}
X= X_{0} + \nabla X \cdot \mathrm{log}(R/R_\mathrm{eff}) \,,
\end{equation}

\noindent where $\nabla X$ corresponds to the radial gradient of the different stellar population analyzed (in linear scale $\nabla$log Age (dex/R$\textsubscript{eff}$), $\nabla$[Fe/H] (dex/R$\textsubscript{eff}$) and $\nabla$A$_\mathrm{v}$ (mag/R$\textsubscript{eff}$) and in log scale $\nabla$log Age (dex/dex), $\nabla$[Fe/H] (dex/dex) and $\nabla$A$_\mathrm{v}$ (mag/dex) and $X_{0}$ the fitted values at the central region (log Age$_{0}$, [Fe/H]$_{0}$ and A$_\mathrm{v0}$) for the case of luminosity-weighted values and mass-weighted values. Radial gradients were determined through an error-weighted linear fitting of the profiles out to 2 - 3.5 R$\textsubscript{eff}$ (depending on the photometric depth of each object). The process was done iteratively via a sigma clipping on the residuals to remove outliers. Black crosses in Fig. \ref{fig:2} represent values not considered during the fit. The black dashed line corresponds to the final gradient.  The individual uncertainties of each stellar population value of each region are derived by \textrm{MUFFIT} through a Monte Carlo method. For clarity we have not included individual errors in the figure but typical errors are $\Delta$log Age = 0.15 dex, $\Delta$[Fe/H] = 0.15 dex and $\Delta$A$_\mathrm{v}$ = 0.10 mag. The gradients determined for each object are summarized in Table \ref{tab:A2} of Appendix \ref{ap:tables}. The radial profiles as well as the 2D maps of all the sample objects can be found in the online version (Appendix C).  

\subsection{Peculiar cases} \label{Sec:peculiar}

Although it is not the scope of this paper to enter in the peculiarities of individual objects and maps, it is worth mentioning that the level of resolution is so high in some cases that a simple visual inspection of the maps reveals fine structures. In this section, we describe the peculiar cases that stand up from the analyzed sample. The 2D maps as well as the radial profiles of all the sample objects can be found in the online version of the paper (Appendix C).

\begin{itemize}
\item AGN profiles:\\
Galaxies 81422406945 (Fig. C.5), 81451206302 (Fig. C.11), and 81473103857 (Fig. C.24) show very distinct profile behaviors than the rest of the sample. With a very well defined core, their radial profiles are not well represented by a linear fitting. The objects 81451206302 and 81422406945 were previously identified as AGN candidates based on their X-ray detection \citep{Flesch2016, Pageetal2012}. 81473103857 seems to be an obscured AGN \citep{cutri2002, cutri2003}. Similar core profiles have been previously observed in several studies \citep{Kolevaetal2011, Lietal2015}. Some of the CALIFA spheroidal galaxies also show positive central gradients in age \citep{GonzalezDelgadoetal2014a, GonzalezDelgadoetal2015}. It is likely that these very young areas with high extinction levels are actually due to the AGN rather than the properties of the underlying stellar population. \\

\item Disk/Ring substructures: \\
Galaxy 81422406945 (Fig. C.5) shows an apparent disk specially evident in metallicity along the major axis. As mentioned before, this object has been identified as an AGN candidate. Galaxies 81473405681 (Fig. C.13) and 81422407693 (Fig. C.22) show apparent rings more prominent in age. A detailed light decomposition of these objects will reveal the nature of these substructures. \\

\item Very young profiles: \\
Besides 2 of the AGN profiles (81451206302 and 81473103857) and the objects with apparent rings (81473405681 and 81422407693) already mentioned, there is another galaxy with a very young profile (81474307526, Fig. C.16). These five galaxies correspond to the blue objects outside the red sequence in Fig. \ref{fig:1} (panel d). These objects are actually star forming galaxies or AGNs based on their location in panel d) of Fig. \ref{fig:1}. 
\end{itemize}

In order to not compromise the results, we have excluded from the following analysis the 6 objects with peculiar profiles described above. However, considering that the main goal of this study is to present and test our methodology, we have included the profiles, maps and spatially resolved properties in the final tables and appendices as a probe of the potential of the method.

\begin{table}
\caption{Median light-weighted and mass-weighted gradients in linear and logarithmic space. Errors correspond to the 1-$\sigma$ value from the distribution.}
\label{tab:4}
\centering
\tiny
\begin{tabular}{cccc}
\hline 
\hline \\[-1ex]
 &  $\mu_\mathrm{BC03}$ & $\mu_\mathrm{EMILES}$  & Units\\[0.5ex]
 \hline \\[-1ex]

$\nabla$log Age$_\mathrm{L}$    &    0.02 $\pm$ 0.08  &    0.02 $\pm$ 0.06    &    dex/R$_\mathrm{eff}$     \\
$\nabla$log Age$_\mathrm{M}$  &    0.02 $\pm$ 0.06  &    0.02 $\pm$ 0.04    &    dex/R$_\mathrm{eff}$       \\
 $\nabla$$\mathrm{[Fe/H]}$$_\mathrm{L}$   &    --0.11 $\pm$ 0.07  &    --0.09 $\pm$ 0.06    &    dex/R$_\mathrm{eff}$        \\
 $\nabla$$\mathrm{[Fe/H]}$$_\mathrm{M}$     &    --0.11 $\pm$ 0.08  &    --0.09 $\pm$ 0.07    &    dex/R$_\mathrm{eff}$     \\
 $\nabla$A$_\mathrm{v}$  &    --0.01 $\pm$ 0.15  &    --0.03 $\pm$ 0.09    &     mag/R$_\mathrm{eff}$     \\
 
 \hline \\[-1ex]
  
&  $\mu_\mathrm{BC03}$ & $\mu_\mathrm{EMILES}$  & Units\\[0.5ex]
 
 \hline \\[-1ex]

$\nabla$log Age$_\mathrm{L}$  &    0.04 $\pm$ 0.31  &    0.03 $\pm$ 0.21    &       dex/dex  \\
$\nabla$log Age$_\mathrm{M}$   &    0.01 $\pm$ 0.16  &    0.02 $\pm$ 0.07    &    dex/dex      \\
 $\nabla$$\mathrm{[Fe/H]}$$_\mathrm{L}$    &    --0.20 $\pm$ 0.27  &    --0.17 $\pm$ 0.13    &    dex/dex      \\
$\nabla$$\mathrm{[Fe/H]}$$_\mathrm{M}$     &    --0.22 $\pm$ 0.25  &    --0.15 $\pm$ 0.14    &     dex/dex     \\
  $\nabla$A$_\mathrm{v}$   &    0.01 $\pm$ 0.39  &    --0.04 $\pm$ 0.15    &   mag/dex      \\

\hline
\hline
\end{tabular}
\end{table}

 \begin{figure} [h]
\begin{center}
\includegraphics[width=\columnwidth]{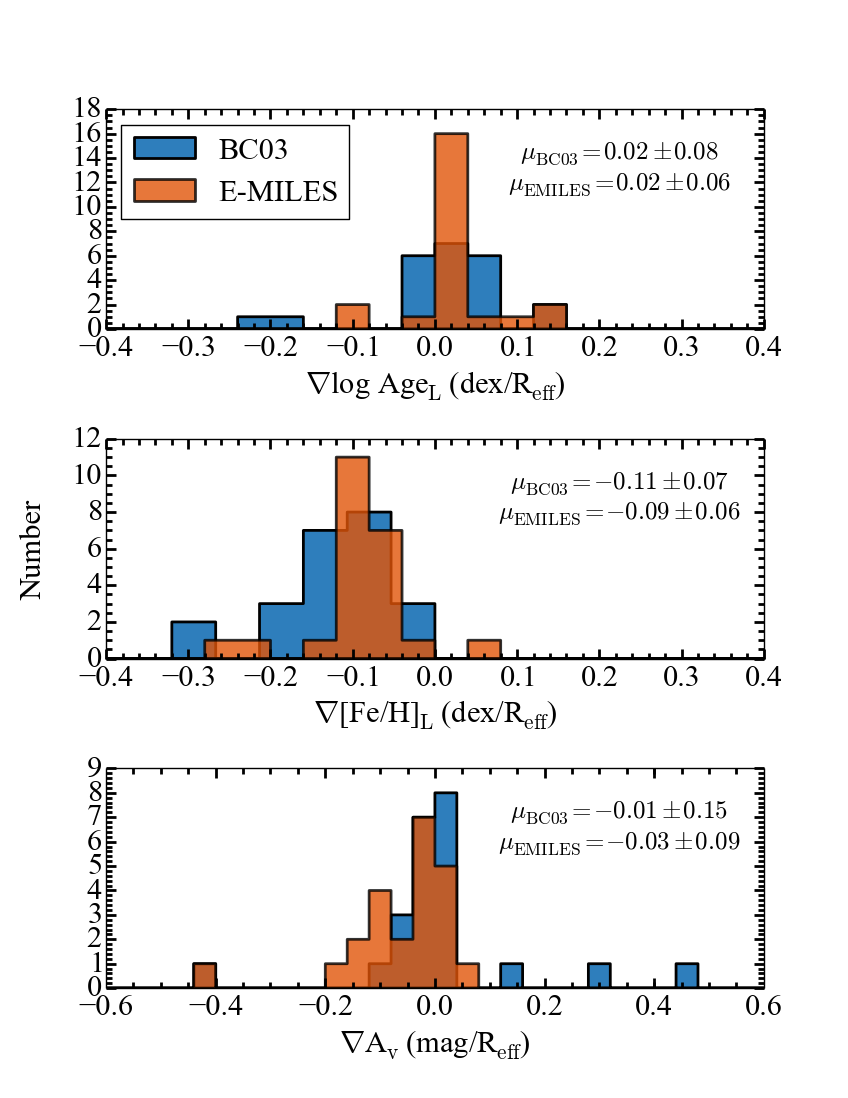}
\caption{Histograms of the linear space gradients for  log Age$\textsubscript{L}$ (upper panel), [Fe/H]$\textsubscript{L}$ (middle panel) and A$_\textsubscript{v}$ (bottom panel) for different SSP models. The median $\mu$ values are labeled in each panel. The 6 objects with peculiar profiles (Sect. \ref{Sec:peculiar})  have been removed from the analysis.} 
\label{fig:8}
\end{center}
\end{figure}

\subsection{Radial gradients}
 The age, metallicity and extinction luminosity-weighted gradients, in linear space, for our sample are summarized in the histograms of Fig. \ref{fig:8}. Mass-weighted stellar population gradients are shown in Fig. \ref{fig:8b}. In order to not compromise the results we have excluded from the analysis the peculiar cases (See Sect. \ref{Sec:peculiar}) for a final sub-sample of 23 objects. Table \ref{tab:4} presents the median light- and mass-weighted values of the linear and logarithmic space radial gradients. Although the median values do not seem to be affected by the SSP model used, BC03 produces a higher scatter driven by a few relatively large outliers. This evidence indicates systematic errors associated with the different SSP models used.

 \subsubsection{Age gradients}
 Our sample of early-type galaxies has, on average, flat age gradients with a median of $\nabla$log Age$\textsubscript{L}$ = 0.02 $\pm$ 0.08 dex/R$\textsubscript{eff}$  and $\nabla$log Age$\textsubscript{L}$ = 0.02 $\pm$ 0.06 dex/R$\textsubscript{eff}$ for BC03 and E-MILES, respectively. As seen from Table \ref{tab:4} and Fig. \ref{fig:8}, none of the input SSP models used during the analysis make significant differences in the median gradients, although BC03 produces some outliers. It is also remarkable how similar luminosity- and mass-weighted values are. This suggests that galaxies, on average, behave similarly on mass and on light and that the half-mass radius (R$_\mathrm{MLR}$) and half-light radius (R$_\mathrm{eff}$) on early-type galaxies should be similar. This result suggests that the contribution of the second SSP (the younger component) is small and the star formation history of our early-type population is well represented by mainly an old SSP component. This result agrees with a previous CALIFA analysis \citep{GonzalezDelgadoetal2014a} that finds the ratio  for early-type galaxies R$_\mathrm{MLR}$/R$_\mathrm{eff}$ $\sim$ 0.9. However our results differ from recent MaNGA analysis. \citet{Goddardetal2016} find a marked difference between luminosity- and mass-weighted age gradients where the mass-weighted median age does show some radial dependence with positive gradients at all galaxy masses. A more detailed study of mass-to-light (M/L) maps and gradients of different galaxy type will be performed in future works.

\begin{figure}
\begin{center}
\includegraphics[width=\columnwidth]{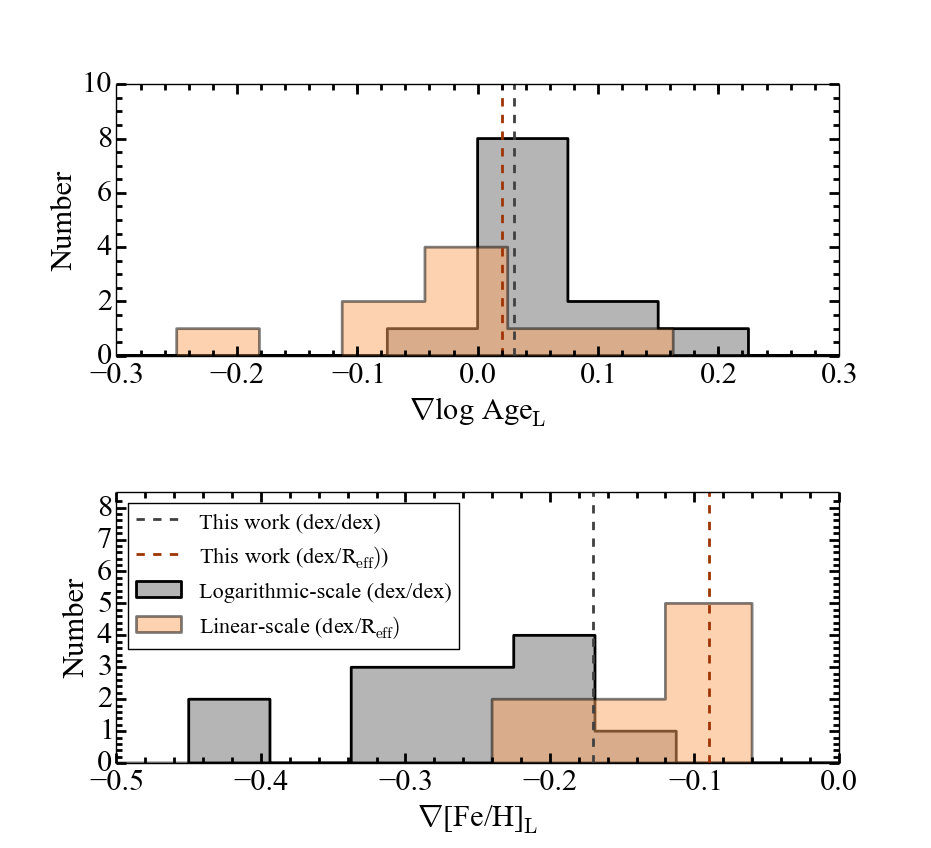}
\caption{Distribution of age and metallicity gradients for early-type galaxies obtained in previous studies. All values have been taken from Table \ref{tab:grad}. The dashed-lines represent the stellar population gradients obtained in this work.} 
\label{fig:11tab4}
\end{center}
\end{figure}

 To set our results in context with the literature and provide  a quantitative comparison, Table \ref{tab:grad} summarizes the most relevant previous studies on stellar population gradients in early-type galaxies. The list combines a large variety of different observational approaches, sample sizes and radial coverage. The radial coverage varies between 1 and 2 R$\textsubscript{eff}$ except for the photometric studies that reach further galactocentric distances. The distribution of the gradients derived in these studies are visualized in Fig. \ref{fig:11tab4}.  Most studies in the literature have found either a flat or slightly positive age gradients except for CALIFA inner gradient result.  \citet{GonzalezDelgadoetal2015}, using CALIFA galaxies ($\sim$ 41 early-type), found very negative inner ( < 1 R$_\mathrm{eff}$) gradients ($\sim$ -- 0.25) that become flat at larger galactocentric distances. Our results agree with previous long-slit analysis of early-type galaxies \citep[e.g.][]{Mehlertetal2003, Wuetal2005, Redaetal2007, Spolaoretal2010} as well as with the most recent IFU studies \citep[e.g.][]{Rawleetal2008, Rawleetal2010, Kuntschneretal2010, Wilkinsonetal2015, Goddardetal2016}, in particular regarding the lack of a significant age gradient. With a median age gradient value of  0.0 dex/R$_\mathrm{eff}$ and 0.03 dex/dex,  our results agree very well with previous studies.

  \begin{table*}
\caption{Comparison table providing the most relevant previous studies on stellar population gradients. Only early-type galaxies and luminosity-weighted age and metallicity gradients are considered.}
\label{tab:grad}
\centering
\tiny
\begin{tabular}{ccccccc}

\hline 
\hline \\[-1ex]
Reference &  Sample Size  &  Observation type & Radial range & $\nabla$log Age$_\mathrm{L}$    &   $\nabla$[Fe/H]$_\mathrm{L}$   & Units  \\[0.5ex]
 \hline \\[-1ex]

\citet{Daviesetal1993} &   13  &    Long-slit    &      1 -- 1.3 R$_\mathrm{eff}$   &   --  &   --0.2 $\pm$ 0.10  & dex/dex\\  
\citet{Mehlertetal2003} &   35  &    Long-slit   &      1 R$_\mathrm{eff}$   &  $\sim$ 0 &    -- 0.16 $\pm$ 0.12  & dex/dex\\
\citet{Wuetal2005} &   36  &    Photometry   &      < 5 R$_\mathrm{eff}$   &  0.02 $\pm$ 0.04  &   --0.25 $\pm$ 0.03 &  dex/dex\\
\citet{SanchezBlazquezetal2006} &   61$\textsuperscript{a}$  &    Long-slit   &      < 1 R$_\mathrm{eff}$    &  0.08 $\pm$ 0.01  &   --0.21 $\pm$ 0.02 &  dex/dex\\
\citet{SanchezBlazquezetal2006} &   21$\textsuperscript{b}$  &    Long-slit   &      < 1 R$_\mathrm{eff}$    &  0.03 $\pm$ 0.07  &   --0.33 $\pm$ 0.07 & dex/dex\\
\citet{Broughetal2007} &   6  &    Long-slit    &      < 1 R$_\mathrm{eff}$   &  0.01 $\pm$ 0.04 &   --0.31 $\pm$ 0.05 & dex/dex\\  
\citet{Redaetal2007}  & 12 & Long-slit &  1 R$_\mathrm{eff}$   &  0.04 $\pm$ 0.08  &   --0.25 $\pm$ 0.05 & dex/dex\\
\citet{SanchezBlazquezetal2007} &   11  &    Long-slit   &      < 1 R$_\mathrm{eff}$   &  0.16 $\pm$ 0.05  &   --0.33 $\pm$ 0.07 & dex/dex\\
\citet{Rawleetal2008} &   12  &    IFU   &      1 R$_\mathrm{eff}$   &  0.08 $\pm$ 0.08  &   --0.20 $\pm$ 0.05 & dex/R$\textsubscript{eff}$\\
\citet{Spolaoretal2008} &   2  &    Long-slit    &      < 1.5 R$_\mathrm{eff}$   &  --0.01 $\pm$ 0.04  &   --0.42 $\pm$ 0.06 & dex/dex\\  
\citet{Kuntschneretal2010} &   48  &    IFU    &      < 1 R$_\mathrm{eff}$   &  0.02 $\pm$ 0.13  &   --0.28 $\pm$ 0.12  & dex/dex\\
\citet{Rawleetal2010} &   19  &    IFU   &      1 R$_\mathrm{eff}$   &  --0.02 $\pm$ 0.06  &   --0.13 $\pm$ 0.04  & dex/R$\textsubscript{eff}$\\
\citet{Spolaoretal2010} &   14  &    Long-slit   &      1 -- 3 R$_\mathrm{eff}$   &  0.03 $\pm$ 0.17  &  -- 0.22 $\pm$ 0.14  & dex/dex\\
\citet{Kolevaetal2011} &   40  &    Long-slit    &      < 2 R$_\mathrm{eff}$   &  $\sim$ 0.1  &   $\sim$ --0.2  & dex/R$\textsubscript{eff}$\\
\citet{LaBarberaetal2012} &   674  &    Photometry   &    < 8 R$_\mathrm{eff}$   &  $\sim$ 0.1  &   $\sim$ -- 0.4 & dex/dex\\
\citet{GonzalezDelgadoetal2015} &   41  &    IFU    &      0 -- 1 R$_\mathrm{eff}$   &  $\sim$ --0.25  &   $\sim$ -- 0.1 & dex/R$\textsubscript{eff}$\\
\citet{GonzalezDelgadoetal2015} &   41  &    IFU    &      1 -- 2 R$_\mathrm{eff}$   &  $\sim$ --0.05  &   $\sim$ -- 0.1  & dex/R$\textsubscript{eff}$ \\
\citet{Wilkinsonetal2015} &   5  &    IFU    &      < 1.5 R$_\mathrm{eff}$   &  $\sim$ 0  &   --0.15 $\pm$ 0.14  & dex/R$\textsubscript{eff}$\\
\citet{Goddardetal2016} &   505  &    IFU    &      < 1.5 R$_\mathrm{eff}$   &  0.004 $\pm$ 0.06  &   --0.12 $\pm$ 0.05 & dex/R$\textsubscript{eff}$ \\
\citet{Zhengetal2016} &   463  &    IFU    &      < 1.5 R$_\mathrm{eff}$   &  --0.05 $\pm$ 0.01  &   --0.09  $\pm$ 0.01 & dex/R$\textsubscript{eff}$\\
 This work (E-MILES)    &    23  &    Photometry    &      < 2 -- 3.5 R$_\mathrm{eff}$   &  0.02 $\pm$ 0.06  &   --0.09 $\pm$ 0.06 & dex/R$\textsubscript{eff}$\\
  This work (E-MILES)   &    23  &    Photometry    &      < 2 -- 3.5 R$_\mathrm{eff}$   &  0.03 $\pm$ 0.21  &   --0.17 $\pm$ 0.13 & dex/dex\\
 \hline \\[-1.5ex]
  
Median Values &      &      &      &  0.00 $\pm$ 0.05    &   --0.12 $\pm$ 0.03 &  dex/R$\textsubscript{eff}$\\
                     &      &      &      &  0.03 $\pm$ 0.01   &   --0.25 $\pm$ 0.06  & dex/dex\\[0.5ex]

\hline
\hline \\[-1ex]
\multicolumn{6}{l}{$\textsuperscript{a}$Low-density environment galaxies.}\\
\multicolumn{6}{l}{$\textsuperscript{b}$High-density environment galaxies.}\\
\end{tabular}
\end{table*}

\subsubsection{Metallicity gradients} 
 The vast majority of our galaxies have negative metallicity\footnote{Although different notations are used to indicate metallicity ([Fe/H], [M/H] or [Z/H]), for solar-scaled models, all the notations are directly comparable. Through this paper, we maintain the metallicity and gradient metallicity notation of the original study.} gradients as expected in most of the galaxy formation scenarios. The median gradients of our sample are $\nabla$[Fe/H]$_\mathrm{L}$/R$\textsubscript{eff}$= --0.11 $\pm$ 0.07 dex/R$\textsubscript{eff}$ for BC03 and $\nabla$[Fe/H]$_\mathrm{L}$/R$\textsubscript{eff}$ = --0.09 $\pm$ 0.06 dex/R$\textsubscript{eff}$ for E-MILES.  
 When we use logarithmic scale values, for comparison purposes, we obtain $\nabla$[Fe/H]$_\mathrm{L}$/log (R/R$\textsubscript{eff}$) = --0.20 $\pm$ 0.27 dex/dex and $\nabla$[Fe/H]$_\mathrm{L}$/log(R/R$\textsubscript{eff}$)= -- 0.17 $\pm$ 0.13 dex/dex for BC03 and E-MILES, respectively. Once again, none of the input SSP models used during the analysis make significant differences in the median gradients. In addition, luminosity- and mass-weighted gradients are remarkably similar.  \citet{Goddardetal2016} found  luminosity- and mass-weighted metallicities and their radial dependence to be indistinguishable with an average offset of $\sim$ 0.05 dex in agreement with our results. 
 
 The metallicity gradient distribution of previous studies (Fig. \ref{fig:11tab4}) show a wide range of values from --0.16 to --0.42 with a median of --0.25 dex/dex. The light-weighted metallicity gradient found in the present study (--0.17 dex/dex and --0.09 dex/R$\textsubscript{eff}$) is at the shallower side, but well within the distribution. Most notably, IFU studies of Table \ref{tab:grad} agree well with our metallicity gradients.

 \subsubsection{Extinction gradients}
Bottom panel of Fig. \ref{fig:8} shows a variety of extinction behaviors with a tendency of flat A$_\mathrm{v}$ gradients. MaNGA early-type galaxies also have a very small amount of dust and show shallow, relatively flat radial profiles \citep{Goddardetal2016}. \citet{GonzalezDelgadoetal2015} CALIFA sample of ellipticals also show negative A$_\mathrm{v}$ gradients with an almost dust-free behavior at distances larger than 1 R$_\mathrm{eff}$. Unfortunately, the lack of previous extinction gradient studies do not allow for further comparison.

\begin{figure} [h]
\begin{center}
\includegraphics[width=\columnwidth]{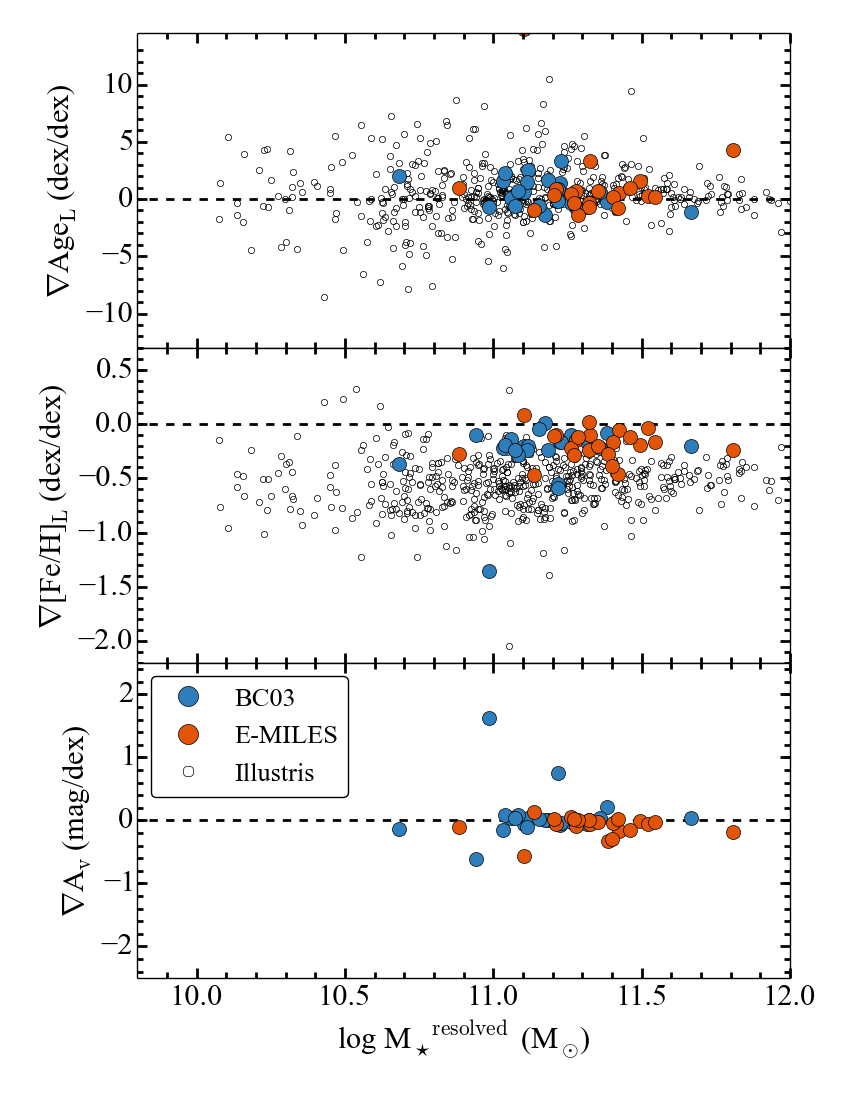}
\caption{Stellar population gradients Age$\textsubscript{L}$, [Fe/H]$\textsubscript{L}$ and A$_\textsubscript{v}$, in logarithmic space, as a function of stellar mass. Both, BC03 and E-MILES, SSP models have been included. Open circles correspond to Illustris simulation measurements in the outer (1 -- 2 R$_\mathrm{eff}$)  galaxy region \citep{Cooketal2016}. Dashed black lines show flat gradients. The 6 objects with peculiar profiles (Sect. \ref{Sec:peculiar})  have been removed from the analysis.} 
\label{fig:9}
\end{center}
\end{figure}

\subsubsection{Correlation with mass}
Figure \ref{fig:9} presents variations of stellar population gradients with the galaxy total mass. Total masses have been calculated by adding the estimated mass in each tessellation bin (see Sect. \ref{sec:integrated}). We have overplotted Illustris hydrodynamical simulation measurements in the outer galaxy region (1 -- 2 R$_\mathrm{eff}$) from \cite{Cooketal2016}. For comparison purposes, we have overplotted values obtained in logarithmic space and age gradients as $\nabla$Age$_\mathrm{L}$ rather than $\nabla$log Age$_\mathrm{L}$. No clear correlation between total stellar mass and stellar population gradients, for our mass range, has been found, in agreement with previous studies. \citet{GonzalezDelgadoetal2014a} also found no clear trend between age gradients with mass for very massive spheroidal galaxies, although it seems that larger inner gradients occur in less massive galaxies (log M$_{\star}$ < 11.2). Same behavior is found for metallicity gradients, where more massive galaxies tend to have weaker stellar metallicity gradients with no clear correlation.  MaNGA survey \citep{Goddardetal2016} also found no clear correlation between age and metallicity gradients with galaxy mass for massive early-type galaxies (log M$_{\star}$ > 10.5). Overall, our results are compatible with those of the Illustris simulation although our metallicity gradients are less steep than Illustris. Comparison with other studies shows that Illustris metallicity gradients ($\sim$ --0.5 dex/dex) are steeper than metallicity gradients from observations (see Table \ref{tab:grad} and Fig. \ref{fig:11tab4}) and previous simulations \citep[e.g.][]{Kobayashi2004, Hirschmannetal2015}.

\begin{table*}
\caption{Stellar population gradients of the master radial profiles in linear and logarithmic space.}
\label{tab:1}
\centering
\tiny
\begin{tabular}{ccccccccccc}

\hline 
\hline \\[-1ex]
 Models &  $\nabla$log Age$_\mathrm{L}$  &  log Age$_{0,}$$_\mathrm{L}$  & $\nabla$log Age$_\mathrm{M}$  &  log Age$_{0,}$$_\mathrm{M}$  &  $\nabla$[Fe/H]$_\mathrm{L}$   &  [Fe/H]$_{0,}$$_\mathrm{L}$ &  $\nabla$[Fe/H]$_\mathrm{M}$   &  [Fe/H]$_{0,}$$_\mathrm{M}$  & $\nabla$Av   &  Av$_{0}$\\[0.5ex]
 
              & (dex/R$\textsubscript{eff}$)      &   (Gyrs)    &   (dex/R$\textsubscript{eff}$)      &   (Gyrs)      &  (dex/R$\textsubscript{eff}$)  &  (dex)    &   (dex/R$\textsubscript{eff}$)  &  (dex)         &  (mag/R$\textsubscript{eff}$) &  (mag) \\[0.5ex]
\hline \\[-1ex]
         
       BC03    &    0.01 $\pm$ 0.01  &    0.73 $\pm$ 0.01    &      0.00 $\pm$ 0.01   &  0.83 $\pm$ 0.01  &   --0.11 $\pm$ 0.01 &  0.33 $\pm$ 0.02  & --0.10 $\pm$ 0.01 & 0.25 $\pm$ 0.02  &  0.03 $\pm$ 0.02 & 0.14 $\pm$ 0.04 \\
 E-MILES    &    0.00 $\pm$ 0.01  &    0.93 $\pm$ 0.01    &      0.01 $\pm$ 0.01   &  0.96 $\pm$ 0.01  &   --0.10 $\pm$ 0.01 &  0.07 $\pm$ 0.02  & --0.07 $\pm$ 0.01 & 0.12 $\pm$ 0.01  &  0.02 $\pm$ 0.01 & 0.22 $\pm$ 0.02 \\

 \hline \\[-1ex]
  
   &  $\nabla$log Age$_\mathrm{L}$  &  log Age$_{0,}$$_\mathrm{L}$  & $\nabla$log Age$_\mathrm{M}$  &  log Age$_{0,}$$_\mathrm{M}$  &  $\nabla$[Fe/H]$_\mathrm{L}$   &  [Fe/H]$_{0,}$$_\mathrm{L}$ &  $\nabla$[Fe/H]$_\mathrm{M}$   &  [Fe/H]$_{0,}$$_\mathrm{M}$  & $\nabla$Av   &  Av$_{0}$\\[0.5ex]
 
              & (dex/dex)      &   (Gyrs)    &   (dex/dex)      &   (Gyrs)      &  (dex/dex)  &  (dex)    &   (dex/dex)  &  (dex)         &  (mag/dex) &  (mag) \\[0.5ex]
  \hline \\[-1ex]

       BC03    &    0.03 $\pm$ 0.02  &    0.74 $\pm$ 0.01    &      0.02 $\pm$ 0.02   &  0.83 $\pm$ 0.01  &   --0.25 $\pm$ 0.04 &  0.16 $\pm$ 0.02  & --0.21 $\pm$ 0.04 & 0.11 $\pm$ 0.01  &  0.04 $\pm$ 0.05 & 0.20 $\pm$ 0.02 \\
 E-MILES    &    0.02 $\pm$ 0.01  &    0.93 $\pm$ 0.01    &      0.04 $\pm$ 0.01   &  0.98 $\pm$ 0.01  &   --0.21 $\pm$ 0.04 &  --                             0.07 $\pm$ 0.02  & --0.16 $\pm$ 0.03 & 0.01 $\pm$ 0.01  &  0.02 $\pm$ 0.03 & 0.25 $\pm$ 0.01 \\

\hline
\hline
\end{tabular}
\end{table*}

\subsection{Master radial profile}
\begin{figure*} [h]
\begin{center}
\includegraphics[width=0.9\textwidth]{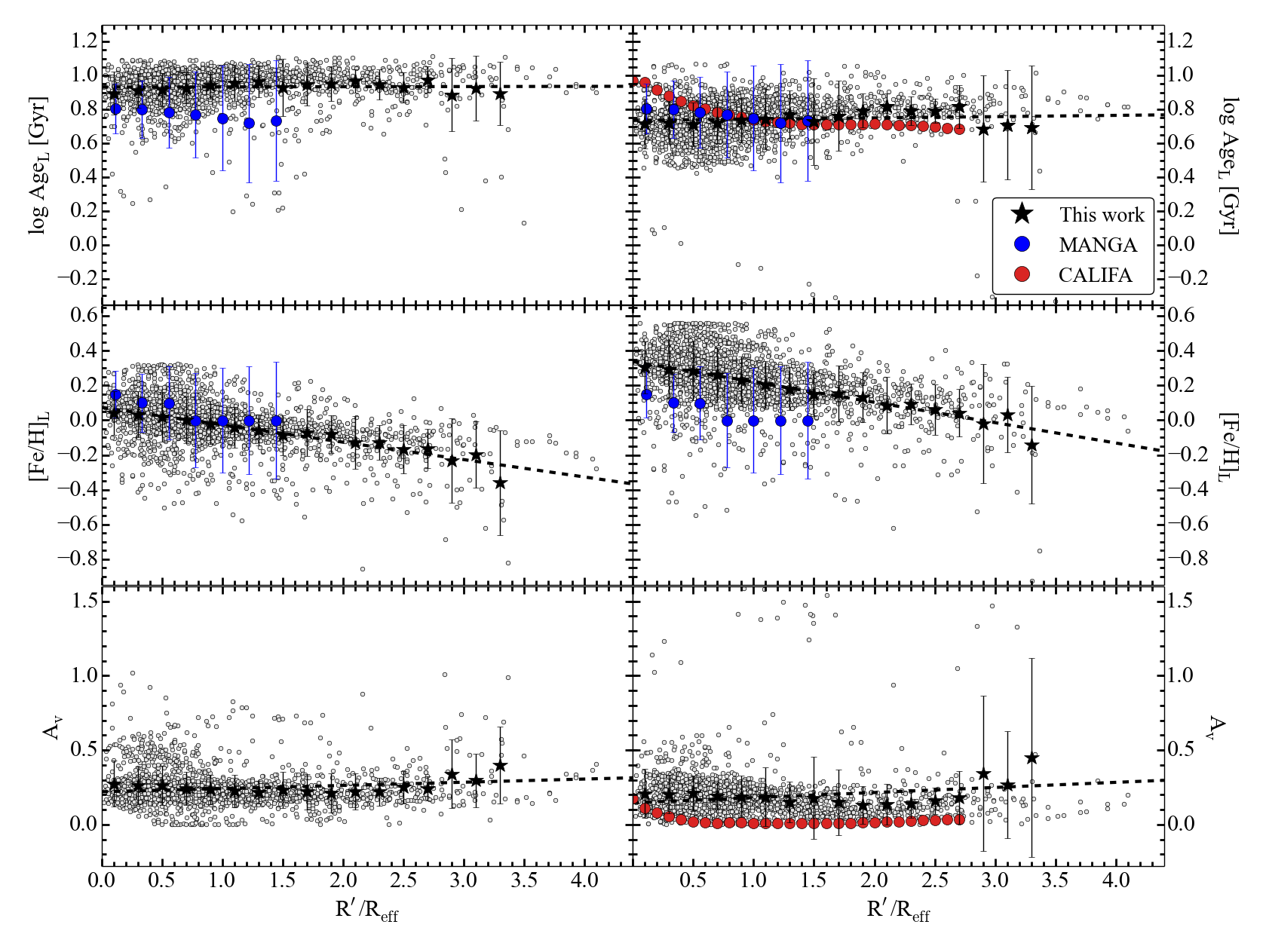}
\caption{Stacked radial profiles of log Age$\textsubscript{L}$, [Fe/H]$\textsubscript{L}$ and A$\textsubscript{v}$. Left panels present the case of E-MILES SSP models and right panels the case of BC03. Gray open symbols correspond to the stellar population values of each individual bin in each spatially resolved object. Star symbols have been obtained by averaging the stellar population values of the sample in constant bins of 0.2 R$\textsubscript{eff}$ between 0 $\leq$ R $\leq$ 3.6 R$\textsubscript{eff}$. The error bars are the standard deviation of the mean. When available, the profiles obtained by CALIFA \citep{GonzalezDelgadoetal2015} have been overplotted. MaNGA \citep{Goddardetal2016} for early-type galaxies with log M$_\star$ > 11.0 was also considered. The dashed black line corresponds to the best linear fit of the average stellar population values (star symbols). }
\label{fig:10}
\end{center}
\end{figure*}

\begin{figure*} [h]
\begin{center}
\includegraphics[width=0.9\textwidth]{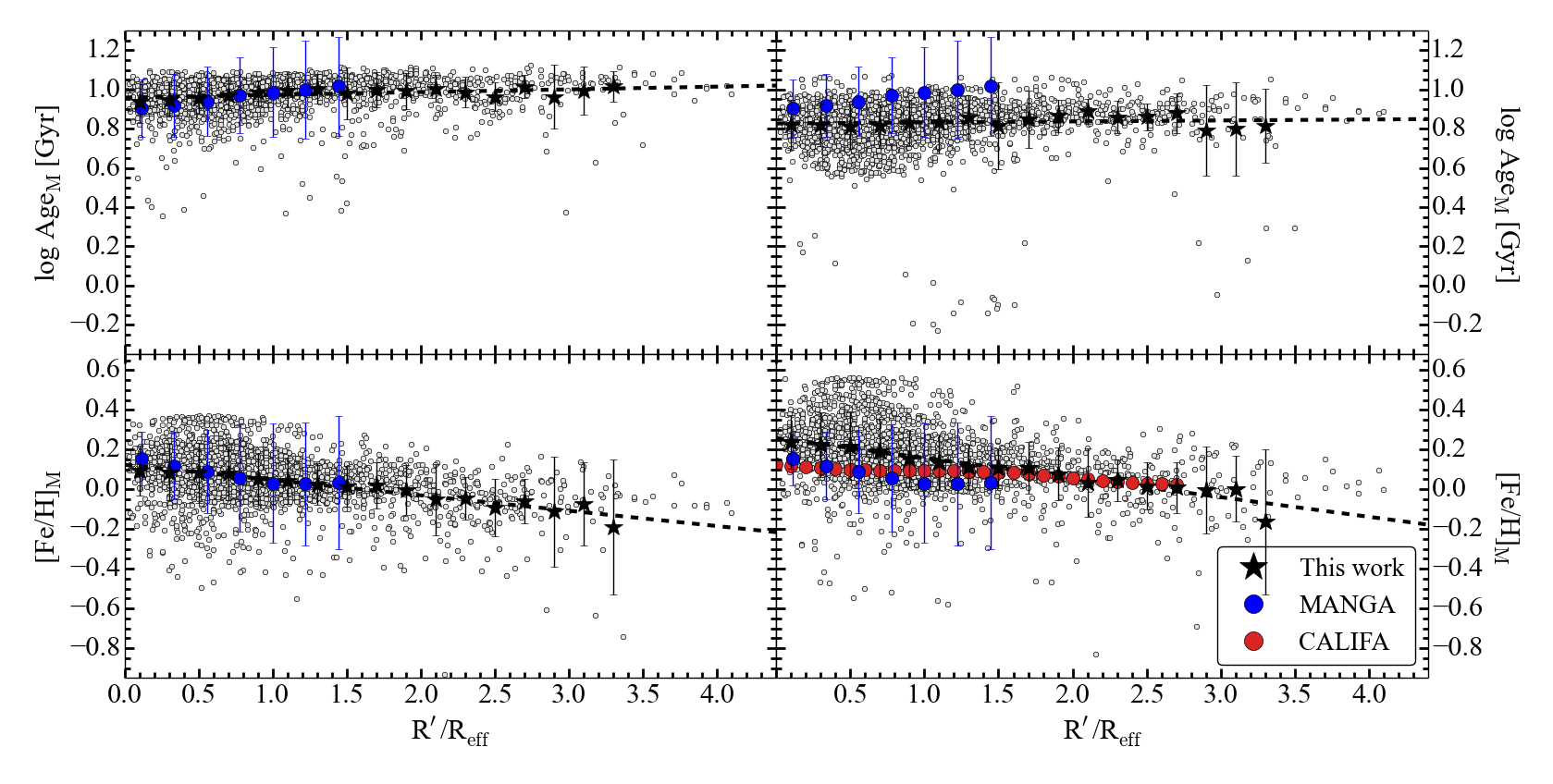}
\caption{Same as Fig. \ref{fig:10} but for the mass-weighted properties.} 
\label{fig:11}
\end{center}
\end{figure*}

Figures \ref{fig:10} and \ref{fig:11} show the age and metallicity (luminosity- and mass-weighted), and the extinction radial profiles  obtained by stacking all the galaxies obtaining a master profile. We have excluded from the analysis the peculiar cases (see Sect. \ref{Sec:peculiar}) for a final sub-sample of 23 objects. The individual gray open symbols in each panel correspond to the individual bin in each spatially resolved object. The master profiles (star symbols) have been obtained by averaging the stellar population properties of the sample in constant bins of 0.2 R$\textsubscript{eff}$ between 0 $\leq$ $\mathrm{R}^{\prime}$ $\leq$ 3.5 R$\textsubscript{eff}$. Left panels in both figures correspond to E-MILES SSP models and right panels correspond to BC03 SSP.  The error bars represent the standard deviation of the mean.  When available, the profiles obtained by CALIFA \citep{GonzalezDelgadoetal2014a} have been overplotted for the case of elliptical and the same BC03 SSP models than our study. Median stellar population parameters from MaNGA \citep{Goddardetal2016} have been also overplotted for the case of early-type galaxies with log M$_\star$ > 11.0. Linear relations have been fitted to the master profiles. The gradients and the values at $\mathrm{R}^{\prime}$=0 are summarized in Table \ref{tab:1}. Once again, the luminosity- and mass-weighted profiles show a similar behavior. We note again that while none of the input SSP models (BC03 versus E-MILES) make significant differences in the global profile, they do introduce a non-negligible offset between them.  

We first remark that our profiles extend to larger galactocentric distances than any previous spectroscopic study. The reason is certainly that photometry is more sensitive than standard spectroscopy although deeper photometry is needed to put stronger constrain on those outskirt areas. Overall, our stellar ages show flat profiles. Comparison with CALIFA shows remarkably similar radial gradients (flat ages) at R >  R$\textsubscript{eff}$. Larger differences are found in the inner regions (R < 0.7 R$\textsubscript{eff}$). While our sample show a continuous flat profile, CALIFA galaxies present a negative age gradient only present in these inner regions.  Since the same input SSP models (i.e. BC03) are used in both studies, the different methodology and techniques used are the main source to explain these discrepancies. The full spectral fitting of MaNGA use the stellar population models of \citet{Marastonetal2011} which use the MILES stellar library, so although Figs. \ref{fig:10} and \ref{fig:11} compare MaNGA results with our results in both SSP models, in principle, Maraston et al. models would be more similar to E-MILES than to BC03. Our results are remarkably similar to MaNGA when E-MILES and mass-weighted properties are considered and slightly differ for the luminosity-weighted age gradient. 

The stellar metallicity profiles show declining profiles at all radii although our study show slightly steeper gradients than the ones from CALIFA.  Once again, our results are remarkably similar to MaNGA when similar SSP models are used. 

A larger sample and deeper photometry is needed for further analysis at larger galactocentric distances. 

The stellar extinction behavior of the galaxies in our sample is consistent with a flat profile suggesting no significant changes in the dust content of early-type galaxies showing a constant A$_\mathrm{v}$= 0.2 mag at any given radius. MANGA early-type galaxies exhibit shallow, relatively flat radial profiles with reddening values of E(B--V) $\sim$ 0.05 for a similar mass range \citep{Goddardetal2016}. Considering R$_\mathrm{v}$= A$_\mathrm{v}$/E(B--V) and assuming a value of R$_\mathrm{v}$=3.1, MANGA extinction parameter is A$_\mathrm{v}$ $\sim$ 0.15, in good agreement with our results.  In contrast, CALIFA galaxies show negative A$_\mathrm{v}$ gradients out to R= 0.5 R$\textsubscript{eff}$ with a dust-free behavior at larger distances although absolute A$_\mathrm{v}$ estimation of CALIFA are smaller than MANGA and our estimations.

In spite of the different technique used, the behavior of the radial variation of the different stellar properties is comparable between our multi-filter ALHAMBRA study and recent IFU studies.  These results highlight  the scientific power of our 2D multi-filter methodology. From this analysis, we can conclude that an average massive ($\sim$ 10$^{11}$M$_{\sun}$), early-type galaxy at a redshift $z$ $\sim$ 0.2 has a flat age gradient with an inner age of log Age$_{0}$ $\sim$ 0.93, and a negative metallicity gradient of $\nabla$[Fe/H] $\sim$ --0.10 (dex/R$\textsubscript{eff}$) with an inner metallicity of [Fe/H]$_{0}$ $\sim$ 0.07.

\section{Discussion and Future Work} \label{Sec:discussion}
Although the goal of this study is to explore the potential of IFU-like studies with multi-filter surveys, the successful results and the reasonable sample size analyzed allow us to shed some light on the formation history of galaxies in our sample. In this section, we compare our results with cosmological simulations and include some discussion of the prospects for extending our analysis in future works.

The most recent cosmological simulations propose a two phase scenario for the formation of early-type galaxies: a first phase of dissipative processes (2 < $z$ < 6) dominated by in-situ star formation from  inflowing cold gas (the monolithic collapse scenario) and a second phase (0 < $z$ < 2) where the galaxies grow through external accretion (mergers) and ex-situ star formation \citep[e.g.][]{DominguezTenreiroetal2006, Naabetal2009, Oseretal2012}. The age and metallicity mean gradients can put constraint on the formation history of the galaxies as different formation models predict different gradients. In general, dissipative processes predict steep metallicity gradients,  while merging scenarios predict shallower gradients due to the dilution of any pre-existing gradient \citep[e.g.][]{Kobayashi2004, Pipinoetal2008}. However different physical processes like AGN feedback, SN feedback, young massive stars or galactic winds can strongly influence the fraction of in-situ formed stars versus accreted, and thus the overall gradients \citep{Hirschmannetal2013, Hirschmannetal2015}. Gas rich mergers with different degree of dissipation have also been proposed as enhanced metallicity processes that will produce secondary star-formation events leading to steeper gradients \citep{Kobayashi2004}. 

\citet{Kobayashi2004} successfully reproduces the formation and chemodynamical evolution of elliptical galaxies through SPH simulations. He found an average gradient of $\nabla$log Z/log R = --0.3 dex/dex with a dispersion of $\pm$ 0.2 in the local Universe. Metallicity gradients obtained in our study are considerably flatter than the values predicted by the dissipative collapse models, thus major and minor mergers are expected to play a significant role for the assembly of these massive galaxies. With a median metallicity gradient of $\nabla$[Fe/H]$\textsubscript{L}$ =  --0.17 $\pm$ 0.13 dex/dex, our study is in agreement with the predictions of \citet{Kobayashi2004} of galaxies that have undergone major mergers.

In a more recent study, \citet{Hirschmannetal2015} investigate the origin of stellar population gradients in cosmological, zoomed simulations of 10 massive galaxies at large radii. The simulations follow metal cooling and enrichment from SN and AGB winds as well as galactic winds. At $z$ $\sim$  0, galaxies have mean metallicity gradients of --0.35 dex/dex slightly steeper than our results. They explain the origin of these gradients based on two components: on one hand the accretion of metal-poorer stellar populations in major and minor mergers produces a dilution of a pre-existing gradient (in-situ gradient), on the other hand  re-infall of previously ejected metal-poor gas onto the galaxy, due to the galactic winds, could form metal-poor stars. Age gradients are in general mildly positive at $z$ = 0 (<$\nabla$logAge> = 0.04 dex/dex) due to older ages of the accreted stellar population than that of the in-situ formed stellar component. The younger in-situ formed stellar populations would be a consequence of the delayed and enhanced star formation in the wind model due to late infall of previously ejected gas.

According to \citet{Kobayashi2004} and \citet{Hirschmannetal2015}, our shallow negative metallicity and flat age gradients suggest that early-type galaxies in our sample have formed through major mergers where the gradients are driven by the higher metallicity and the older age of the accreted systems, together with the different mixing behavior. Using a novel approach, we have confirmed previous results  and extended them to larger radii consolidating the idea that the most massive early-type galaxies have formed primarily through major mergers, rather than in situ star formation processes.  However, although the general trends are established, the particular physical processes causing these trends and their relative role in galaxy formation are still unknown.

Illustris hydrodynamical simulations \citep{Cooketal2016} find significant differences  between the stellar population gradients in the inner galaxy regions (R < R$_\mathrm{eff}$) and the stellar halo (R > 2 R$_\mathrm{eff}$) in early-type galaxies. This study implies that information content of accretion history is retained in the stellar population profiles only at very large radii from the galaxy. Current limitations of spectroscopic studies at this low S/N regimes  suggest that deep photometric studies in galactic stellar halos are essential to uncover the formation and assembly of local galaxies. 

This paper describes a pilot study using a hybrid approach half-way between classical photometry and spectroscopy. We have demonstrate that our technique enables spatially resolved stellar population studies out to considerably fainter surface brightness than are not possible with current IFU surveys. This technique allows the analysis of galaxy profiles at larger galactocentric radii and at higher redshift than current surveys. The technique also allows the studies of very nearby galaxies (z < 0.01) so spatially extended that are not suitable for the small field of view of current IFU surveys.  This study opens the way to the analysis of larger early-type samples as well as to more general galaxy sample (i.e later-type galaxies). 

The upcoming multi-filter surveys J-PAS and J-PLUS will observe 8.500 deg$^{2}$ of the sky visible from the Northern Hemisphere using a set of  > 70 narrow and medium band filters that will provide low spectral resolution (R $\sim$50) multi-color information for every pixel on the sky, hence for hundred millions of galaxies in a large and contiguous area. The unique combination between these photometric data sets and our novel technique, will permit us spatially  resolved studies of passive and star-forming galaxies with redshift and environment in the largest sample up today.

\section{Summary and conclusions}
In this paper, we explore the potential of IFU-like studies with multi-filter surveys by using a method that combines a Centroidal Voronoi Tesselation and a multi-filter SED fitting method. This technique allows us to analyze unresolved stellar populations of spatially resolved galaxies based on photometric surveys. In order to validate and test the method,  we have analyzed the stellar population properties of 29 galaxies observed by ALHAMBRA survey using 23 medium filter bands at the 3.5m telescope of Calar Alto observatory. The sample includes elliptical galaxies covering a stellar mass range from log M$_\star$ = 10.5 to log M$_\star$ = 11.8  and with redshift 0.05 < z < 0.3. For each galaxy, 2D maps and radial profiles for (luminosity- and mass-weighted) age and metallicity, and extinction are presented. Radial gradients of these properties measured  out to 2 -- 3.5 R$_\mathrm{eff}$ are determined. Final stacked radial profiles are also obtained. Besides demonstrating the scientific potential of multi-filter photometry to explore the spatially resolved stellar populations of local galaxies, interesting results about the formation of early-type galaxies have been found:

\begin{enumerate}
\item \textit{Integrated galaxy properties vs. spatially averaged properties}: The total stellar mass, age and metallicity estimated from integrated photometry are remarkably robust  when compared with those from a spatially resolved analysis.\\ 

\item \textit{Luminosity-weighted properties vs. mass-weighted properties}: On average, the age and metallicity radial profiles and gradients are very similar when luminosity- or mass-weighted properties are used. This result suggests that early-type galaxies behave similarly on mass and on light. \\ 
 
\item \textit{Stellar ages}: Early-type galaxies have, on average, flat  age gradients in agreement with previous studies. No clear correlation between total stellar mass and stellar population age gradients has been found suggesting that, for this mass range, radial variations are not primary driven by mass. In addition, the more massive galaxies are also the older ones preserving the "downsizing" scenario. \\

\item \textit{Stellar metallicities}: All the  galaxies show [Fe/H](R) declining profiles with moderate radial dependence. The radial profiles do not scale with stellar mass. The mass-metallicity relation is also present in our sample\\

\item \textit{Stellar extinction}: The stacked A$_\mathrm{v}$ profile shows a flat behavior with no radial variation suggesting that early-type galaxies have a constant dust content of A$_\mathrm{v}$ $\sim$ 0.2 at any given galactocentric distance. 
\end{enumerate}

None of the input SSP models make significant differences in our results. From these results and comparison with theoretical predictions, we conclude that major mergers are likely the main processes assembling early-type galaxies where the gradients are driven by the higher metallicity and the older age of the accreted systems. 

Although more detailed investigations will require larger data sets, it is clear that photometric surveys as the upcoming J-PAS \citep{Benitezetal2014} will extend 2D multi-filter studies such as the one presented here to scientific cases not available to current IFU techniques. For example it will allow us the 2D analysis of nearby galaxies  at large galactocentric distance (R > 3 R$_\mathrm{eff}$) or the impact of environment in the 2D properties of nearby galaxies.

\begin{acknowledgements} 
We thank the anonymous referee, whose thoughtful comments greatly  improved  the  paper.  We thank all the CEFCA staff for useful and productive discussions. \\ 

This work has been mainly funded by the FITE (Fondos de Inversiones de Teruel) and the Spanish Ministry for Economy and Competitiveness and FEDER funds through grants AYA2012-30789 and AYA2015-66211-C2-1-P. 

 We also acknowledge financial support from the projects AYA2014-57490-P and AYA2016-77846-P, and from the Arag\'{o}n Government through the Research Group E103. BA has received funding from the European Union's Horizon 2020 research and innovation programme under the Marie Sklodowska-Curie grant agreement No 656354.\\

 This research made use of NASA's Astrophysics Data System Bibliographic Services, as well as the following software packages: Astropy \citep{Astropy2013}, Matplotlib \citep{Hunteretal2007}, IPython \citep{PerezGranger2007}, SciPy \citep{Jonesetal2001} and NumPy \citep{vanderWaltetal2011}.

\end{acknowledgements}


\begin{thebibliography}{108}
\expandafter\ifx\csname natexlab\endcsname\relax\def\natexlab#1{#1}\fi

\bibitem[{{Aparicio Villegas} {et~al}\mbox{.}(2010){Aparicio Villegas},
  {Alfaro}, {Cabrera-Ca{\~n}o}, {Moles}, {Ben{\'{\i}}tez}, {Perea}, {del Olmo},
  {Fern{\'a}ndez-Soto}, {Crist{\'o}bal-Hornillos}, {Husillos}, {Aguerri},
  {Broadhurst}, {Castander}, {Cepa}, {Cervi{\~n}o}, {Gonz{\'a}lez Delgado},
  {Infante}, {M{\'a}rquez}, {Masegosa}, {Mart{\'{\i}}nez}, {Prada}, {Quintana},
  \& {S{\'a}nchez}}]{AparicioVillegasetal2010}
{Aparicio Villegas} T. {et~al.}, 2010, \aj, 139, 1242

\bibitem[{{Astropy Collaboration} {et~al}\mbox{.}(2013){Astropy Collaboration},
  {Robitaille}, {Tollerud}, {Greenfield}, {Droettboom}, {Bray}, {Aldcroft},
  {Davis}, {Ginsburg}, {Price-Whelan}, {Kerzendorf}, {Conley}, {Crighton},
  {Barbary}, {Muna}, {Ferguson}, {Grollier}, {Parikh}, {Nair}, {Unther},
  {Deil}, {Woillez}, {Conseil}, {Kramer}, {Turner}, {Singer}, {Fox}, {Weaver},
  {Zabalza}, {Edwards}, {Azalee Bostroem}, {Burke}, {Casey}, {Crawford},
  {Dencheva}, {Ely}, {Jenness}, {Labrie}, {Lim}, {Pierfederici}, {Pontzen},
  {Ptak}, {Refsdal}, {Servillat}, \& {Streicher}}]{Astropy2013}
{Astropy Collaboration} {et~al.}, 2013, \aap, 558, A33

\bibitem[{{Benitez} {et~al}\mbox{.}(2014){Benitez}, {Dupke}, {Moles}, {Sodre},
  {Cenarro}, {Marin-Franch}, {Taylor}, {Cristobal}, {Fernandez-Soto}, {Mendes
  de Oliveira}, {Cepa-Nogue}, {Abramo}, {Alcaniz}, {Overzier},
  {Hernandez-Monteagudo}, {Alfaro}, {Kanaan}, {Carvano}, {Reis}, {Martinez
  Gonzalez}, {Ascaso}, {Ballesteros}, {Xavier}, {Varela}, {Ederoclite},
  {Vazquez Ramio}, {Broadhurst}, {Cypriano}, {Angulo}, {Diego}, {Zandivarez},
  {Diaz}, {Melchior}, {Umetsu}, {Spinelli}, {Zitrin}, {Coe}, {Yepes}, {Vielva},
  {Sahni}, {Marcos-Caballero}, {Shu Kitaura}, {Maroto}, {Masip}, {Tsujikawa},
  {Carneiro}, {Gonzalez Nuevo}, {Carvalho}, {Reboucas}, {Carvalho}, {Abdalla},
  {Bernui}, {Pigozzo}, {Ferreira}, {Chandrachani Devi}, {Bengaly}, {Campista},
  {Amorim}, {Asari}, {Bongiovanni}, {Bonoli}, {Bruzual}, {Cardiel}, {Cava},
  {Cid Fernandes}, {Coelho}, {Cortesi}, {Delgado}, {Diaz Garcia}, {Espinosa},
  {Galliano}, {Gonzalez-Serrano}, {Falcon-Barroso}, {Fritz}, {Fernandes},
  {Gorgas}, {Hoyos}, {Jimenez-Teja}, {Lopez-Aguerri}, {Lopez-San Juan},
  {Mateus}, {Molino}, {Novais}, {OMill}, {Oteo}, {Perez-Gonzalez}, {Poggianti},
  {Proctor}, {Ricciardelli}, {Sanchez-Blazquez}, {Storchi-Bergmann}, {Telles},
  {Schoennell}, {Trujillo}, {Vazdekis}, {Viironen}, {Daflon},
  {Aparicio-Villegas}, {Rocha}, {Ribeiro}, {Borges}, {Martins}, {Marcolino},
  {Martinez-Delgado}, {Perez-Torres}, {Siffert}, {Calvao}, {Sako}, {Kessler},
  {Alvarez-Candal}, {De Pra}, {Roig}, {Lazzaro}, {Gorosabel}, {Lopes de
  Oliveira}, {Lima-Neto}, {Irwin}, {Liu}, {Alvarez}, {Balmes}, {Chueca},
  {Costa-Duarte}, {da Costa}, {Dantas}, {Diaz}, {Fabregat}, {Ferrari},
  {Gavela}, {Gracia}, {Gruel}, {Gutierrez}, {Guzman}, {Hernandez-Fernandez},
  {Herranz}, {Hurtado-Gil}, {Jablonsky}, {Laporte}, {Le Tiran}, {Licandro},
  {Lima}, {Martin}, {Martinez}, {Montero}, {Penteado}, {Pereira}, {Peris},
  {Quilis}, {Sanchez-Portal}, {Soja}, {Solano}, {Torra}, \&
  {Valdivielso}}]{Benitezetal2014}
{Benitez} N. {et~al.}, 2014, ArXiv e-prints

\bibitem[{{Bershady} {et~al}\mbox{.}(2010){Bershady}, {Verheijen}, {Swaters},
  {Andersen}, {Westfall}, \& {Martinsson}}]{Bershadyetal2010}
{Bershady} M.~A., {Verheijen} M.~A.~W., {Swaters} R.~A., {Andersen} D.~R.,
  {Westfall} K.~B., {Martinsson} T., 2010, \apj, 716, 198

\bibitem[{{Bertin}(2011)}]{Bertin2011}
{Bertin} E., 2011, in Astronomical Society of the Pacific Conference Series,
  Vol. 442, Astronomical Data Analysis Software and Systems XX, {Evans} I.~N.,
  {Accomazzi} A., {Mink} D.~J., {Rots} A.~H., eds., p. 435

\bibitem[{{Bertin}(2013)}]{Bertinetal2013}
{Bertin} E., 2013, {PSFEx: Point Spread Function Extractor}. Astrophysics
  Source Code Library

\bibitem[{{Bertin} \& {Arnouts}(1996)}]{Bertinetal1996}
{Bertin} E., {Arnouts} S., 1996, \aaps, 117, 393

\bibitem[{{Blanc} {et~al}\mbox{.}(2010){Blanc}, {Gebhardt}, {Heiderman},
  {Evans}, {Jogee}, {van den Bosch}, {Marinova}, {Weinzirl}, {Yoachim},
  {Drory}, {Fabricius}, {Fisher}, {Hao}, {MacQueen}, {Shen}, {Hill}, \&
  {Kormendy}}]{Blancetal2010}
{Blanc} G.~A. {et~al.}, 2010, in Astronomical Society of the Pacific Conference
  Series, Vol. 432, New Horizons in Astronomy: Frank N. Bash Symposium 2009,
  {Stanford} L.~M., {Green} J.~D., {Hao} L., {Mao} Y., eds., p. 180

\bibitem[{{Bressan} {et~al}\mbox{.}(1993){Bressan}, {Fagotto}, {Bertelli}, \&
  {Chiosi}}]{Bressan1993}
{Bressan} A., {Fagotto} F., {Bertelli} G., {Chiosi} C., 1993, \aaps, 100, 647

\bibitem[{{Brough} {et~al}\mbox{.}(2007){Brough}, {Proctor}, {Forbes}, {Couch},
  {Collins}, {Burke}, \& {Mann}}]{Broughetal2007}
{Brough} S., {Proctor} R., {Forbes} D.~A., {Couch} W.~J., {Collins} C.~A.,
  {Burke} D.~J., {Mann} R.~G., 2007, \mnras, 378, 1507

\bibitem[{{Bruzual} \& {Charlot}(2003)}]{BruzualCharlot2003}
{Bruzual} G., {Charlot} S., 2003, \mnras, 344, 1000

\bibitem[{{Bryant} {et~al}\mbox{.}(2015){Bryant}, {Owers}, {Robotham}, {Croom},
  {Driver}, {Drinkwater}, {Lorente}, {Cortese}, {Scott}, {Colless}, {Schaefer},
  {Taylor}, {Konstantopoulos}, {Allen}, {Baldry}, {Barnes}, {Bauer},
  {Bland-Hawthorn}, {Bloom}, {Brooks}, {Brough}, {Cecil}, {Couch}, {Croton},
  {Davies}, {Ellis}, {Fogarty}, {Foster}, {Glazebrook}, {Goodwin}, {Green},
  {Gunawardhana}, {Hampton}, {Ho}, {Hopkins}, {Kewley}, {Lawrence},
  {Leon-Saval}, {Leslie}, {McElroy}, {Lewis}, {Liske}, {L{\'o}pez-S{\'a}nchez},
  {Mahajan}, {Medling}, {Metcalfe}, {Meyer}, {Mould}, {Obreschkow}, {O'Toole},
  {Pracy}, {Richards}, {Shanks}, {Sharp}, {Sweet}, {Thomas}, {Tonini}, \&
  {Walcher}}]{Bryantetal2015}
{Bryant} J.~J. {et~al.}, 2015, \mnras, 447, 2857

\bibitem[{{Bundy} {et~al}\mbox{.}(2015){Bundy}, {Bershady}, {Law}, {Yan},
  {Drory}, {MacDonald}, {Wake}, {Cherinka}, {S{\'a}nchez-Gallego}, {Weijmans},
  {Thomas}, {Tremonti}, {Masters}, {Coccato}, {Diamond-Stanic},
  {Arag{\'o}n-Salamanca}, {Avila-Reese}, {Badenes}, {Falc{\'o}n-Barroso},
  {Belfiore}, {Bizyaev}, {Blanc}, {Bland-Hawthorn}, {Blanton}, {Brownstein},
  {Byler}, {Cappellari}, {Conroy}, {Dutton}, {Emsellem}, {Etherington},
  {Frinchaboy}, {Fu}, {Gunn}, {Harding}, {Johnston}, {Kauffmann}, {Kinemuchi},
  {Klaene}, {Knapen}, {Leauthaud}, {Li}, {Lin}, {Maiolino}, {Malanushenko},
  {Malanushenko}, {Mao}, {Maraston}, {McDermid}, {Merrifield}, {Nichol},
  {Oravetz}, {Pan}, {Parejko}, {Sanchez}, {Schlegel}, {Simmons}, {Steele},
  {Steinmetz}, {Thanjavur}, {Thompson}, {Tinker}, {van den Bosch}, {Westfall},
  {Wilkinson}, {Wright}, {Xiao}, \& {Zhang}}]{Bundyetal2015}
{Bundy} K. {et~al.}, 2015, \apj, 798, 7

\bibitem[{{Cappellari} \& {Copin}(2003)}]{Cappellarietal2003}
{Cappellari} M., {Copin} Y., 2003, \mnras, 342, 345

\bibitem[{{Cappellari} {et~al}\mbox{.}(2011){Cappellari}, {Emsellem},
  {Krajnovi{\'c}}, {McDermid}, {Scott}, {Verdoes Kleijn}, {Young}, {Alatalo},
  {Bacon}, {Blitz}, {Bois}, {Bournaud}, {Bureau}, {Davies}, {Davis}, {de
  Zeeuw}, {Duc}, {Khochfar}, {Kuntschner}, {Lablanche}, {Morganti}, {Naab},
  {Oosterloo}, {Sarzi}, {Serra}, \& {Weijmans}}]{Cappellarietal2011}
{Cappellari} M. {et~al.}, 2011, \mnras, 413, 813

\bibitem[{{Castander} {et~al}\mbox{.}(2012){Castander}, {Ballester}, {Bauer},
  {Cardiel-Sas}, {Carretero}, {Casas}, {Castilla}, {Crocce}, {Delfino},
  {Eriksen}, {Fern{\'a}ndez}, {Fosalba}, {Garc{\'{\i}}a-Bellido},
  {Gazta{\~n}aga}, {Gra{\~n}ena}, {Hern{\'a}ndez}, {Jim{\'e}nez}, {L{\'o}pez},
  {Mart{\'{\i}}}, {Miquel}, {Neissner}, {Padilla}, {P{\'{\i}}o}, {Ponce},
  {Sanchez}, {Serrano}, {Sevilla}, {Tonello}, \& {de
  Vicente}}]{Castanderetal2012}
{Castander} F.~J. {et~al.}, 2012, in \procspie, Vol. 8446, Ground-based and
  Airborne Instrumentation for Astronomy IV, p. 84466D

\bibitem[{{Chabrier}(2003)}]{Chabrier2003}
{Chabrier} G., 2003, \pasp, 115, 763

\bibitem[{{Colless} {et~al}\mbox{.}(2001){Colless}, {Dalton}, {Maddox},
  {Sutherland}, {Norberg}, {Cole}, {Bland-Hawthorn}, {Bridges}, {Cannon},
  {Collins}, {Couch}, {Cross}, {Deeley}, {De Propris}, {Driver}, {Efstathiou},
  {Ellis}, {Frenk}, {Glazebrook}, {Jackson}, {Lahav}, {Lewis}, {Lumsden},
  {Madgwick}, {Peacock}, {Peterson}, {Price}, {Seaborne}, \&
  {Taylor}}]{Collessetal2001}
{Colless} M. {et~al.}, 2001, \mnras, 328, 1039

\bibitem[{{Cook} {et~al}\mbox{.}(2016){Cook}, {Conroy}, {Pillepich},
  {Rodriguez-Gomez}, \& {Hernquist}}]{Cooketal2016}
{Cook} B.~A., {Conroy} C., {Pillepich} A., {Rodriguez-Gomez} V., {Hernquist}
  L., 2016, ArXiv e-prints

\bibitem[{{Cowie} {et~al}\mbox{.}(1996){Cowie}, {Songaila}, {Hu}, \&
  {Cohen}}]{Cowieetal1996}
{Cowie} L.~L., {Songaila} A., {Hu} E.~M., {Cohen} J.~G., 1996, \aj, 112, 839

\bibitem[{{Crist{\'o}bal-Hornillos}
  {et~al}\mbox{.}(2009){Crist{\'o}bal-Hornillos}, {Aguerri}, {Moles}, {Perea},
  {Castander}, {Broadhurst}, {Alfaro}, {Ben{\'{\i}}tez}, {Cabrera-Ca{\~n}o},
  {Cepa}, {Cervi{\~n}o}, {Fern{\'a}ndez-Soto}, {Gonz{\'a}lez Delgado},
  {Husillos}, {Infante}, {M{\'a}rquez}, {Mart{\'{\i}}nez}, {Masegosa}, {del
  Olmo}, {Prada}, {Quintana}, \& {S{\'a}nchez}}]{CristobalHornillosetal2009}
{Crist{\'o}bal-Hornillos} D. {et~al.}, 2009, \apj, 696, 1554

\bibitem[{{Cutri} {et~al}\mbox{.}(2002){Cutri}, {Nelson}, {Francis}, \&
  {Smith}}]{cutri2002}
{Cutri} R.~M., {Nelson} B.~O., {Francis} P.~J., {Smith} P.~S., 2002, in
  Astronomical Society of the Pacific Conference Series, Vol. 284, IAU Colloq.
  184: AGN Surveys, {Green} R.~F., {Khachikian} E.~Y., {Sanders} D.~B., eds.,
  p. 127

\bibitem[{{Cutri} {et~al}\mbox{.}(2003){Cutri}, {Skrutskie}, {van Dyk},
  {Beichman}, {Carpenter}, {Chester}, {Cambresy}, {Evans}, {Fowler}, {Gizis},
  {Howard}, {Huchra}, {Jarrett}, {Kopan}, {Kirkpatrick}, {Light}, {Marsh},
  {McCallon}, {Schneider}, {Stiening}, {Sykes}, {Weinberg}, {Wheaton},
  {Wheelock}, \& {Zacarias}}]{cutri2003}
{Cutri} R.~M. {et~al.}, 2003, {2MASS All Sky Catalog of point sources.}

\bibitem[{{Davidge}(1992)}]{Davidge1992}
{Davidge} T.~J., 1992, \aj, 103, 1512

\bibitem[{{Davies}, {Sadler} \& {Peletier}(1993){Davies}, {Sadler}, \&
  {Peletier}}]{Daviesetal1993}
{Davies} R.~L., {Sadler} E.~M., {Peletier} R.~F., 1993, \mnras, 262, 650

\bibitem[{{de Vaucouleurs} {et~al}\mbox{.}(1991){de Vaucouleurs}, {de
  Vaucouleurs}, {Corwin}, {Buta}, {Paturel}, \&
  {Fouqu{\'e}}}]{Devaucouleursetal1991}
{de Vaucouleurs} G., {de Vaucouleurs} A., {Corwin}, Jr. H.~G., {Buta} R.~J.,
  {Paturel} G., {Fouqu{\'e}} P., 1991, {Third Reference Catalogue of Bright
  Galaxies. Volume I: Explanations and references. Volume II: Data for galaxies
  between 0$^{h}$ and 12$^{h}$. Volume III: Data for galaxies between 12$^{h}$
  and 24$^{h}$.}

\bibitem[{{de Zeeuw} {et~al}\mbox{.}(2002){de Zeeuw}, {Bureau}, {Emsellem},
  {Bacon}, {Carollo}, {Copin}, {Davies}, {Kuntschner}, {Miller}, {Monnet},
  {Peletier}, \& {Verolme}}]{deZeeuwetal2002}
{de Zeeuw} P.~T. {et~al.}, 2002, \mnras, 329, 513

\bibitem[{{D{\'{\i}}az-Garc{\'{\i}}a}
  {et~al}\mbox{.}(2015){D{\'{\i}}az-Garc{\'{\i}}a}, {Cenarro},
  {L{\'o}pez-Sanjuan}, {Ferreras}, {Varela}, {Viironen},
  {Crist{\'o}bal-Hornillos}, {Moles}, {Mar{\'{\i}}n-Franch}, {Arnalte-Mur},
  {Ascaso}, {Cervi{\~n}o}, {Gonz{\'a}lez Delgado}, {M{\'a}rquez}, {Masegosa},
  {Molino}, {Povi{\'c}}, {Alfaro}, {Aparicio-Villegas}, {Ben{\'{\i}}tez},
  {Broadhurst}, {Cabrera-Ca{\~n}o}, {Castander}, {Cepa}, {Fern{\'a}ndez-Soto},
  {Husillos}, {Infante}, {Aguerri}, {Mart{\'{\i}}nez}, {del Olmo}, {Perea},
  {Prada}, {Quintana}, \& {Gruel}}]{DiazGarciaetal2015}
{D{\'{\i}}az-Garc{\'{\i}}a} L.~A. {et~al.}, 2015, \aap, 582, A14

\bibitem[{{Dom{\'{\i}}nguez-Tenreiro}
  {et~al}\mbox{.}(2006){Dom{\'{\i}}nguez-Tenreiro}, {O{\~n}orbe}, {S{\'a}iz},
  {Artal}, \& {Serna}}]{DominguezTenreiroetal2006}
{Dom{\'{\i}}nguez-Tenreiro} R., {O{\~n}orbe} J., {S{\'a}iz} A., {Artal} H.,
  {Serna} A., 2006, \apjl, 636, L77

\bibitem[{{Driver} {et~al}\mbox{.}(2011){Driver}, {Hill}, {Kelvin}, {Robotham},
  {Liske}, {Norberg}, {Baldry}, {Bamford}, {Hopkins}, {Loveday}, {Peacock},
  {Andrae}, {Bland-Hawthorn}, {Brough}, {Brown}, {Cameron}, {Ching}, {Colless},
  {Conselice}, {Croom}, {Cross}, {de Propris}, {Dye}, {Drinkwater}, {Ellis},
  {Graham}, {Grootes}, {Gunawardhana}, {Jones}, {van Kampen}, {Maraston},
  {Nichol}, {Parkinson}, {Phillipps}, {Pimbblet}, {Popescu}, {Prescott},
  {Roseboom}, {Sadler}, {Sansom}, {Sharp}, {Smith}, {Taylor}, {Thomas},
  {Tuffs}, {Wijesinghe}, {Dunne}, {Frenk}, {Jarvis}, {Madore}, {Meyer},
  {Seibert}, {Staveley-Smith}, {Sutherland}, \& {Warren}}]{Driveretal2011}
{Driver} S.~P. {et~al.}, 2011, \mnras, 413, 971

\bibitem[{{Fagotto} {et~al}\mbox{.}(1994{\natexlab{a}}){Fagotto}, {Bressan},
  {Bertelli}, \& {Chiosi}}]{Fagotto1994a}
{Fagotto} F., {Bressan} A., {Bertelli} G., {Chiosi} C., 1994{\natexlab{a}},
  \aaps, 104

\bibitem[{{Fagotto} {et~al}\mbox{.}(1994{\natexlab{b}}){Fagotto}, {Bressan},
  {Bertelli}, \& {Chiosi}}]{Fagotto1994b}
{Fagotto} F., {Bressan} A., {Bertelli} G., {Chiosi} C., 1994{\natexlab{b}},
  \aaps, 105

\bibitem[{{Ferreras} \& {Silk}(2000)}]{Ferrerasetal2000}
{Ferreras} I., {Silk} J., 2000, \apjl, 541, L37

\bibitem[{{Fitzpatrick}(1999)}]{Fitzpatrick1999}
{Fitzpatrick} E.~L., 1999, \pasp, 111, 63

\bibitem[{{Flesch}(2016)}]{Flesch2016}
{Flesch} E.~W., 2016, \pasa, 33, e052

\bibitem[{{Gallazzi} {et~al}\mbox{.}(2005){Gallazzi}, {Charlot}, {Brinchmann},
  {White}, \& {Tremonti}}]{Gallazzietal2005}
{Gallazzi} A., {Charlot} S., {Brinchmann} J., {White} S.~D.~M., {Tremonti}
  C.~A., 2005, \mnras, 362, 41

\bibitem[{{Gibson} {et~al}\mbox{.}(2013){Gibson}, {Pilkington}, {Brook},
  {Stinson}, \& {Bailin}}]{Gibsonetal2013}
{Gibson} B.~K., {Pilkington} K., {Brook} C.~B., {Stinson} G.~S., {Bailin} J.,
  2013, \aap, 554, A47

\bibitem[{{Girardi} {et~al}\mbox{.}(1996){Girardi}, {Bressan}, {Chiosi},
  {Bertelli}, \& {Nasi}}]{Girardi1996}
{Girardi} L., {Bressan} A., {Chiosi} C., {Bertelli} G., {Nasi} E., 1996, \aaps,
  117, 113

\bibitem[{{Goddard} {et~al}\mbox{.}(2016){Goddard}, {Thomas}, {Maraston},
  {Westfall}, {Etherington}, {Riffel}, {Mallmann}, {Zheng}, {Argudo-Fernandez},
  {Lian}, {Bershady}, {Bundy}, {Drory}, {Law}, {Yan}, {Wake}, {Weijmans},
  {Bizyaev}, {Brownstein}, {Lane}, {Maiolino}, {Masters}, {Merrifield},
  {Nitschelm}, {Pan}, {Roman-Lopes}, {Storchi-Bergmann}, \&
  {Schneider}}]{Goddardetal2016}
{Goddard} D. {et~al.}, 2016, ArXiv e-prints

\bibitem[{{Gonzalez} \& {Gorgas}(1995)}]{Gonzalezetal1995}
{Gonzalez} J.~J., {Gorgas} J., 1995, in Astronomical Society of the Pacific
  Conference Series, Vol.~86, Fresh Views of Elliptical Galaxies, {Buzzoni} A.,
  {Renzini} A., {Serrano} A., eds., p. 225

\bibitem[{{Gonz{\'a}lez Delgado} {et~al}\mbox{.}(2015){Gonz{\'a}lez Delgado},
  {Garc{\'{\i}}a-Benito}, {P{\'e}rez}, {Cid Fernandes}, {de Amorim},
  {Cortijo-Ferrero}, {Lacerda}, {L{\'o}pez Fern{\'a}ndez}, {Vale-Asari},
  {S{\'a}nchez}, {Moll{\'a}}, {Ruiz-Lara}, {S{\'a}nchez-Bl{\'a}zquez},
  {Walcher}, {Alves}, {Aguerri}, {Bekerait{\'e}}, {Bland-Hawthorn}, {Galbany},
  {Gallazzi}, {Husemann}, {Iglesias-P{\'a}ramo}, {Kalinova},
  {L{\'o}pez-S{\'a}nchez}, {Marino}, {M{\'a}rquez}, {Masegosa}, {Mast},
  {M{\'e}ndez-Abreu}, {Mendoza}, {del Olmo}, {P{\'e}rez}, {Quirrenbach}, \&
  {Zibetti}}]{GonzalezDelgadoetal2015}
{Gonz{\'a}lez Delgado} R.~M. {et~al.}, 2015, \aap, 581, A103

\bibitem[{{Gonz{\'a}lez Delgado} {et~al}\mbox{.}(2014){Gonz{\'a}lez Delgado},
  {P{\'e}rez}, {Cid Fernandes}, {Garc{\'{\i}}a-Benito}, {de Amorim},
  {S{\'a}nchez}, {Husemann}, {Cortijo-Ferrero}, {L{\'o}pez Fern{\'a}ndez},
  {S{\'a}nchez-Bl{\'a}zquez}, {Bekeraite}, {Walcher}, {Falc{\'o}n-Barroso},
  {Gallazzi}, {van de Ven}, {Alves}, {Bland-Hawthorn}, {Kennicutt}, {Kupko},
  {Lyubenova}, {Mast}, {Moll{\'a}}, {Marino}, {Quirrenbach}, {V{\'{\i}}lchez},
  \& {Wisotzki}}]{GonzalezDelgadoetal2014a}
{Gonz{\'a}lez Delgado} R.~M. {et~al.}, 2014, \aap, 562, A47

\bibitem[{{Gorgas}, {Efstathiou} \& {Aragon Salamanca}(1990){Gorgas},
  {Efstathiou}, \& {Aragon Salamanca}}]{Gorgasetal1990}
{Gorgas} J., {Efstathiou} G., {Aragon Salamanca} A., 1990, \mnras, 245, 217

\bibitem[{{Gorgas} {et~al}\mbox{.}(1993){Gorgas}, {Faber}, {Burstein},
  {Gonzalez}, {Courteau}, \& {Prosser}}]{Gorgasetal1993}
{Gorgas} J., {Faber} S.~M., {Burstein} D., {Gonzalez} J.~J., {Courteau} S.,
  {Prosser} C., 1993, \apjs, 86, 153

\bibitem[{{Heap} \& {Lindler}(2007)}]{Heapetal2007}
{Heap} S.~R., {Lindler} D.~J., 2007, in Astronomical Society of the Pacific
  Conference Series, Vol. 374, From Stars to Galaxies: Building the Pieces to
  Build Up the Universe, {Vallenari} A., {Tantalo} R., {Portinari} L.,
  {Moretti} A., eds., p. 409

\bibitem[{{Hirschmann} {et~al}\mbox{.}(2013){Hirschmann}, {Naab}, {Dav{\'e}},
  {Oppenheimer}, {Ostriker}, {Somerville}, {Oser}, {Genzel}, {Tacconi},
  {F{\"o}rster-Schreiber}, {Burkert}, \& {Genel}}]{Hirschmannetal2013}
{Hirschmann} M. {et~al.}, 2013, \mnras, 436, 2929

\bibitem[{{Hirschmann} {et~al}\mbox{.}(2015){Hirschmann}, {Naab}, {Ostriker},
  {Forbes}, {Duc}, {Dav{\'e}}, {Oser}, \& {Karabal}}]{Hirschmannetal2015}
{Hirschmann} M., {Naab} T., {Ostriker} J.~P., {Forbes} D.~A., {Duc} P.-A.,
  {Dav{\'e}} R., {Oser} L., {Karabal} E., 2015, \mnras, 449, 528

\bibitem[{{Hopkins} {et~al}\mbox{.}(2013){Hopkins}, {Cox}, {Hernquist},
  {Narayanan}, {Hayward}, \& {Murray}}]{Hopkinsetal2013}
{Hopkins} P.~F., {Cox} T.~J., {Hernquist} L., {Narayanan} D., {Hayward} C.~C.,
  {Murray} N., 2013, \mnras, 430, 1901

\bibitem[{Hunter(2007)}]{Hunteretal2007}
Hunter J.~D., 2007, Computing In Science \& Engineering, 9, 90

\bibitem[{{Jim{\'e}nez-Teja} \& {Ben{\'{\i}}tez}(2012)}]{JimenezTejaetal2012}
{Jim{\'e}nez-Teja} Y., {Ben{\'{\i}}tez} N., 2012, \apj, 745, 150

\bibitem[{Jones {et~al}\mbox{.}(2001)Jones, Oliphant, Peterson,
  {et~al.}}]{Jonesetal2001}
Jones E., Oliphant T., Peterson P., {et~al.}, 2001, {SciPy}: Open source
  scientific tools for {Python}

\bibitem[{{Kaviraj} {et~al}\mbox{.}(2007){Kaviraj}, {Schawinski}, {Devriendt},
  {Ferreras}, {Khochfar}, {Yoon}, {Yi}, {Deharveng}, {Boselli}, {Barlow},
  {Conrow}, {Forster}, {Friedman}, {Martin}, {Morrissey}, {Neff},
  {Schiminovich}, {Seibert}, {Small}, {Wyder}, {Bianchi}, {Donas}, {Heckman},
  {Lee}, {Madore}, {Milliard}, {Rich}, \& {Szalay}}]{Kavirajetal2007}
{Kaviraj} S. {et~al.}, 2007, \apjs, 173, 619

\bibitem[{{Kobayashi}(2004)}]{Kobayashi2004}
{Kobayashi} C., 2004, \mnras, 347, 740

\bibitem[{{Koleva} {et~al}\mbox{.}(2011){Koleva}, {Prugniel}, {de Rijcke}, \&
  {Zeilinger}}]{Kolevaetal2011}
{Koleva} M., {Prugniel} P., {de Rijcke} S., {Zeilinger} W.~W., 2011, \mnras,
  417, 1643

\bibitem[{{Kroupa}(2001)}]{Kroupa2001}
{Kroupa} P., 2001, \mnras, 322, 231

\bibitem[{{Kuntschner} {et~al}\mbox{.}(2010){Kuntschner}, {Emsellem}, {Bacon},
  {Cappellari}, {Davies}, {de Zeeuw}, {Falc{\'o}n-Barroso}, {Krajnovi{\'c}},
  {McDermid}, {Peletier}, {Sarzi}, {Shapiro}, {van den Bosch}, \& {van de
  Ven}}]{Kuntschneretal2010}
{Kuntschner} H. {et~al.}, 2010, \mnras, 408, 97

\bibitem[{{La Barbera} {et~al}\mbox{.}(2010){La Barbera}, {De Carvalho}, {De La
  Rosa}, {Gal}, {Swindle}, \& {Lopes}}]{LaBarberaetal2010}
{La Barbera} F., {De Carvalho} R.~R., {De La Rosa} I.~G., {Gal} R.~R.,
  {Swindle} R., {Lopes} P.~A.~A., 2010, \aj, 140, 1528

\bibitem[{{La Barbera} {et~al}\mbox{.}(2005){La Barbera}, {de Carvalho}, {Gal},
  {Busarello}, {Merluzzi}, {Capaccioli}, \& {Djorgovski}}]{LaBarberaetal2005}
{La Barbera} F., {de Carvalho} R.~R., {Gal} R.~R., {Busarello} G., {Merluzzi}
  P., {Capaccioli} M., {Djorgovski} S.~G., 2005, \apjl, 626, L19

\bibitem[{{La Barbera} {et~al}\mbox{.}(2012){La Barbera}, {Ferreras}, {de
  Carvalho}, {Bruzual}, {Charlot}, {Pasquali}, \& {Merlin}}]{LaBarberaetal2012}
{La Barbera} F., {Ferreras} I., {de Carvalho} R.~R., {Bruzual} G., {Charlot}
  S., {Pasquali} A., {Merlin} E., 2012, \mnras, 426, 2300

\bibitem[{{Labb{\'e}} {et~al}\mbox{.}(2003){Labb{\'e}}, {Franx}, {Rudnick},
  {Schreiber}, {Rix}, {Moorwood}, {van Dokkum}, {van der Werf},
  {R{\"o}ttgering}, {van Starkenburg}, {van der Wel}, {Kuijken}, \&
  {Daddi}}]{Labbeetal2003}
{Labb{\'e}} I. {et~al.}, 2003, \aj, 125, 1107

\bibitem[{{Li} {et~al}\mbox{.}(2015){Li}, {Wang}, {Lin}, {Bershady}, {Bundy},
  {Tremonti}, {Xiao}, {Yan}, {Bizyaev}, {Blanton}, {Cales}, {Cherinka},
  {Cheung}, {Drory}, {Emsellem}, {Fu}, {Gelfand}, {Law}, {Lin}, {MacDonald},
  {Maraston}, {Masters}, {Merrifield}, {Pan}, {S{\'a}nchez}, {Schneider},
  {Thomas}, {Wake}, {Wang}, {Weijmans}, {Wilkinson}, {Yoachim}, {Zhang}, \&
  {Zheng}}]{Lietal2015}
{Li} C. {et~al.}, 2015, \apj, 804, 125

\bibitem[{{Lonoce} {et~al}\mbox{.}(2014){Lonoce}, {Longhetti}, {Saracco},
  {Gargiulo}, \& {Tamburri}}]{Lonoceetal2014}
{Lonoce} I., {Longhetti} M., {Saracco} P., {Gargiulo} A., {Tamburri} S., 2014,
  \mnras, 444, 2048

\bibitem[{{Lopez-Corredoira} {et~al}\mbox{.}(2017){Lopez-Corredoira},
  {Vazdekis}, {Gutierrez}, \& {Castro-Rodriguez}}]{LopezCorredoiraetal2017}
{Lopez-Corredoira} M., {Vazdekis} A., {Gutierrez} C.~M., {Castro-Rodriguez} N.,
  2017, ArXiv e-prints

\bibitem[{{MacArthur} {et~al}\mbox{.}(2004){MacArthur}, {Courteau}, {Bell}, \&
  {Holtzman}}]{MacArthuretal2004}
{MacArthur} L.~A., {Courteau} S., {Bell} E., {Holtzman} J.~A., 2004, \apjs,
  152, 175

\bibitem[{{MacArthur}, {Gonz{\'a}lez} \& {Courteau}(2009){MacArthur},
  {Gonz{\'a}lez}, \& {Courteau}}]{MacArthuretal2009}
{MacArthur} L.~A., {Gonz{\'a}lez} J.~J., {Courteau} S., 2009, \mnras, 395, 28

\bibitem[{{Maraston} \& {Str{\"o}mb{\"a}ck}(2011)}]{Marastonetal2011}
{Maraston} C., {Str{\"o}mb{\"a}ck} G., 2011, \mnras, 418, 2785

\bibitem[{{Mehlert} {et~al}\mbox{.}(2003){Mehlert}, {Thomas}, {Saglia},
  {Bender}, \& {Wegner}}]{Mehlertetal2003}
{Mehlert} D., {Thomas} D., {Saglia} R.~P., {Bender} R., {Wegner} G., 2003,
  \aap, 407, 423

\bibitem[{{Moles} {et~al}\mbox{.}(2008){Moles}, {Ben{\'{\i}}tez}, {Aguerri},
  {Alfaro}, {Broadhurst}, {Cabrera-Ca{\~n}o}, {Castander}, {Cepa},
  {Cervi{\~n}o}, {Crist{\'o}bal-Hornillos}, {Fern{\'a}ndez-Soto}, {Gonz{\'a}lez
  Delgado}, {Infante}, {M{\'a}rquez}, {Mart{\'{\i}}nez}, {Masegosa}, {del
  Olmo}, {Perea}, {Prada}, {Quintana}, \& {S{\'a}nchez}}]{Molesetal2008}
{Moles} M. {et~al.}, 2008, \aj, 136, 1325

\bibitem[{{Molino} {et~al}\mbox{.}(2014){Molino}, {Ben{\'{\i}}tez}, {Moles},
  {Fern{\'a}ndez-Soto}, {Crist{\'o}bal-Hornillos}, {Ascaso},
  {Jim{\'e}nez-Teja}, {Schoenell}, {Arnalte-Mur}, {Povi{\'c}}, {Coe},
  {L{\'o}pez-Sanjuan}, {D{\'{\i}}az-Garc{\'{\i}}a}, {Varela}, {Stefanon},
  {Cenarro}, {Matute}, {Masegosa}, {M{\'a}rquez}, {Perea}, {Del Olmo},
  {Husillos}, {Alfaro}, {Aparicio-Villegas}, {Cervi{\~n}o}, {Huertas-Company},
  {Aguerri}, {Broadhurst}, {Cabrera-Ca{\~n}o}, {Cepa}, {Gonz{\'a}lez},
  {Infante}, {Mart{\'{\i}}nez}, {Prada}, \& {Quintana}}]{Molinoetal2014}
{Molino} A. {et~al.}, 2014, \mnras, 441, 2891

\bibitem[{{Mu{\~n}oz-Mateos} {et~al}\mbox{.}(2011){Mu{\~n}oz-Mateos},
  {Boissier}, {Gil de Paz}, {Zamorano}, {Kennicutt}, {Moustakas}, {Prantzos},
  \& {Gallego}}]{MunozMateosetal2011}
{Mu{\~n}oz-Mateos} J.~C., {Boissier} S., {Gil de Paz} A., {Zamorano} J.,
  {Kennicutt}, Jr. R.~C., {Moustakas} J., {Prantzos} N., {Gallego} J., 2011,
  \apj, 731, 10

\bibitem[{{Naab}, {Johansson} \& {Ostriker}(2009){Naab}, {Johansson}, \&
  {Ostriker}}]{Naabetal2009}
{Naab} T., {Johansson} P.~H., {Ostriker} J.~P., 2009, \apjl, 699, L178

\bibitem[{{Oser} {et~al}\mbox{.}(2012){Oser}, {Naab}, {Ostriker}, \&
  {Johansson}}]{Oseretal2012}
{Oser} L., {Naab} T., {Ostriker} J.~P., {Johansson} P.~H., 2012, \apj, 744, 63

\bibitem[{{Page} {et~al}\mbox{.}(2012){Page}, {Brindle}, {Talavera}, {Still},
  {Rosen}, {Yershov}, {Ziaeepour}, {Mason}, {Cropper}, {Breeveld}, {Loiseau},
  {Mignani}, {Smith}, \& {Murdin}}]{Pageetal2012}
{Page} M.~J. {et~al.}, 2012, \mnras, 426, 903

\bibitem[{{Pan} {et~al}\mbox{.}(2015){Pan}, {Li}, {Lin}, {Wang}, {Fan}, \&
  {Kong}}]{Panetal2015}
{Pan} Z., {Li} J., {Lin} W., {Wang} J., {Fan} L., {Kong} X., 2015, \apjl, 804,
  L42

\bibitem[{{Peng}, {Maiolino} \& {Cochrane}(2015){Peng}, {Maiolino}, \&
  {Cochrane}}]{Pengetal2015}
{Peng} Y., {Maiolino} R., {Cochrane} R., 2015, \nat, 521, 192

\bibitem[{{P{\'e}rez} {et~al}\mbox{.}(2013){P{\'e}rez}, {Cid Fernandes},
  {Gonz{\'a}lez Delgado}, {Garc{\'{\i}}a-Benito}, {S{\'a}nchez}, {Husemann},
  {Mast}, {Rod{\'o}n}, {Kupko}, {Backsmann}, {de Amorim}, {van de Ven},
  {Walcher}, {Wisotzki}, {Cortijo-Ferrero}, \& {CALIFA
  Collaboration}}]{Perezetal2013}
{P{\'e}rez} E. {et~al.}, 2013, \apjl, 764, L1

\bibitem[{P\'erez \& Granger(2007)}]{PerezGranger2007}
P\'erez F., Granger B.~E., 2007, Computing in Science and Engineering, 9, 21

\bibitem[{{P{\'e}rez-Gonz{\'a}lez}
  {et~al}\mbox{.}(2013){P{\'e}rez-Gonz{\'a}lez}, {Cava}, {Barro}, {Villar},
  {Cardiel}, {Ferreras}, {Rodr{\'{\i}}guez-Espinosa}, {Alonso-Herrero},
  {Balcells}, {Cenarro}, {Cepa}, {Charlot}, {Cimatti}, {Conselice}, {Daddi},
  {Donley}, {Elbaz}, {Espino}, {Gallego}, {Gobat},
  {Gonz{\'a}lez-Mart{\'{\i}}n}, {Guzm{\'a}n}, {Hern{\'a}n-Caballero},
  {Mu{\~n}oz-Tu{\~n}{\'o}n}, {Renzini}, {Rodr{\'{\i}}guez-Zaur{\'{\i}}n},
  {Tresse}, {Trujillo}, \& {Zamorano}}]{PerezGonzalezetal2013}
{P{\'e}rez-Gonz{\'a}lez} P.~G. {et~al.}, 2013, \apj, 762, 46

\bibitem[{{Pietrinferni} {et~al}\mbox{.}(2004){Pietrinferni}, {Cassisi},
  {Salaris}, \& {Castelli}}]{Pietrinfernietal2004}
{Pietrinferni} A., {Cassisi} S., {Salaris} M., {Castelli} F., 2004, \apj, 612,
  168

\bibitem[{{Pipino}, {D'Ercole} \& {Matteucci}(2008){Pipino}, {D'Ercole}, \&
  {Matteucci}}]{Pipinoetal2008}
{Pipino} A., {D'Ercole} A., {Matteucci} F., 2008, \aap, 484, 679

\bibitem[{{Povi{\'c}} {et~al}\mbox{.}(2013){Povi{\'c}}, {Huertas-Company},
  {Aguerri}, {M{\'a}rquez}, {Masegosa}, {Husillos}, {Molino},
  {Crist{\'o}bal-Hornillos}, {Perea}, {Ben{\'{\i}}tez}, {Olmo},
  {Fern{\'a}ndez-Soto}, {Jim{\'e}nez-Teja}, {Moles}, {Alfaro},
  {Aparicio-Villegas}, {Ascaso}, {Broadhurst}, {Cabrera-Ca{\~n}o}, {Castander},
  {Cepa}, {Fernandez Lorenzo}, {Cervi{\~n}o}, {Delgado}, {Infante},
  {L{\'o}pez-Sanjuan}, {Mart{\'{\i}}nez}, {Matute}, {Oteo},
  {P{\'e}rez-Garc{\'{\i}}a}, {Prada}, \& {Quintana}}]{Povicetal2013}
{Povi{\'c}} M. {et~al.}, 2013, \mnras, 435, 3444

\bibitem[{{Rawle}, {Smith} \& {Lucey}(2010){Rawle}, {Smith}, \&
  {Lucey}}]{Rawleetal2010}
{Rawle} T.~D., {Smith} R.~J., {Lucey} J.~R., 2010, \mnras, 401, 852

\bibitem[{{Rawle} {et~al}\mbox{.}(2008){Rawle}, {Smith}, {Lucey}, \&
  {Swinbank}}]{Rawleetal2008}
{Rawle} T.~D., {Smith} R.~J., {Lucey} J.~R., {Swinbank} A.~M., 2008, \mnras,
  389, 1891

\bibitem[{{Reda} {et~al}\mbox{.}(2007){Reda}, {Proctor}, {Forbes}, {Hau}, \&
  {Larsen}}]{Redaetal2007}
{Reda} F.~M., {Proctor} R.~N., {Forbes} D.~A., {Hau} G.~K.~T., {Larsen} S.~S.,
  2007, \mnras, 377, 1772

\bibitem[{{R{\"o}ck} {et~al}\mbox{.}(2015){R{\"o}ck}, {Vazdekis}, {Peletier},
  {Knapen}, \& {Falc{\'o}n-Barroso}}]{Rocketal2015}
{R{\"o}ck} B., {Vazdekis} A., {Peletier} R.~F., {Knapen} J.~H.,
  {Falc{\'o}n-Barroso} J., 2015, \mnras, 449, 2853

\bibitem[{{Rogers} {et~al}\mbox{.}(2010){Rogers}, {Ferreras}, {Peletier}, \&
  {Silk}}]{Rogersetal2010}
{Rogers} B., {Ferreras} I., {Peletier} R., {Silk} J., 2010, \mnras, 402, 447

\bibitem[{{Rosales-Ortega} {et~al}\mbox{.}(2010){Rosales-Ortega}, {Kennicutt},
  {S{\'a}nchez}, {D{\'{\i}}az}, {Pasquali}, {Johnson}, \&
  {Hao}}]{RosalesOrtegaetal2010}
{Rosales-Ortega} F.~F., {Kennicutt} R.~C., {S{\'a}nchez} S.~F., {D{\'{\i}}az}
  A.~I., {Pasquali} A., {Johnson} B.~D., {Hao} C.~N., 2010, \mnras, 405, 735

\bibitem[{{S{\'a}nchez} {et~al}\mbox{.}(2012){S{\'a}nchez}, {Kennicutt}, {Gil
  de Paz}, {van de Ven}, {V{\'{\i}}lchez}, {Wisotzki}, {Walcher}, {Mast},
  {Aguerri}, {Albiol-P{\'e}rez}, {Alonso-Herrero}, {Alves}, {Bakos},
  {Bart{\'a}kov{\'a}}, {Bland-Hawthorn}, {Boselli}, {Bomans},
  {Castillo-Morales}, {Cortijo-Ferrero}, {de Lorenzo-C{\'a}ceres}, {Del Olmo},
  {Dettmar}, {D{\'{\i}}az}, {Ellis}, {Falc{\'o}n-Barroso}, {Flores},
  {Gallazzi}, {Garc{\'{\i}}a-Lorenzo}, {Gonz{\'a}lez Delgado}, {Gruel},
  {Haines}, {Hao}, {Husemann}, {Igl{\'e}sias-P{\'a}ramo}, {Jahnke}, {Johnson},
  {Jungwiert}, {Kalinova}, {Kehrig}, {Kupko}, {L{\'o}pez-S{\'a}nchez},
  {Lyubenova}, {Marino}, {M{\'a}rmol-Queralt{\'o}}, {M{\'a}rquez}, {Masegosa},
  {Meidt}, {Mendez-Abreu}, {Monreal-Ibero}, {Montijo}, {Mour{\~a}o},
  {Palacios-Navarro}, {Papaderos}, {Pasquali}, {Peletier}, {P{\'e}rez},
  {P{\'e}rez}, {Quirrenbach}, {Rela{\~n}o}, {Rosales-Ortega}, {Roth},
  {Ruiz-Lara}, {S{\'a}nchez-Bl{\'a}zquez}, {Sengupta}, {Singh}, {Stanishev},
  {Trager}, {Vazdekis}, {Viironen}, {Wild}, {Zibetti}, \&
  {Ziegler}}]{Sanchezetal2012}
{S{\'a}nchez} S.~F. {et~al.}, 2012, \aap, 538, A8

\bibitem[{{S{\'a}nchez} {et~al}\mbox{.}(2014){S{\'a}nchez}, {Rosales-Ortega},
  {Iglesias-P{\'a}ramo}, {Moll{\'a}}, {Barrera-Ballesteros}, {Marino},
  {P{\'e}rez}, {S{\'a}nchez-Blazquez}, {Gonz{\'a}lez Delgado}, {Cid Fernandes},
  {de Lorenzo-C{\'a}ceres}, {Mendez-Abreu}, {Galbany}, {Falcon-Barroso},
  {Miralles-Caballero}, {Husemann}, {Garc{\'{\i}}a-Benito}, {Mast}, {Walcher},
  {Gil de Paz}, {Garc{\'{\i}}a-Lorenzo}, {Jungwiert}, {V{\'{\i}}lchez},
  {J{\'{\i}}lkov{\'a}}, {Lyubenova}, {Cortijo-Ferrero}, {D{\'{\i}}az},
  {Wisotzki}, {M{\'a}rquez}, {Bland-Hawthorn}, {Ellis}, {van de Ven}, {Jahnke},
  {Papaderos}, {Gomes}, {Mendoza}, \&
  {L{\'o}pez-S{\'a}nchez}}]{Sanchezetal2014}
{S{\'a}nchez} S.~F. {et~al.}, 2014, \aap, 563, A49

\bibitem[{{S{\'a}nchez-Bl{\'a}zquez}
  {et~al}\mbox{.}(2007){S{\'a}nchez-Bl{\'a}zquez}, {Forbes}, {Strader},
  {Brodie}, \& {Proctor}}]{SanchezBlazquezetal2007}
{S{\'a}nchez-Bl{\'a}zquez} P., {Forbes} D.~A., {Strader} J., {Brodie} J.,
  {Proctor} R., 2007, \mnras, 377, 759

\bibitem[{{S{\'a}nchez-Bl{\'a}zquez}, {Gorgas} \&
  {Cardiel}(2006){S{\'a}nchez-Bl{\'a}zquez}, {Gorgas}, \&
  {Cardiel}}]{SanchezBlazquezetal2006}
{S{\'a}nchez-Bl{\'a}zquez} P., {Gorgas} J., {Cardiel} N., 2006, \aap, 457, 823

\bibitem[{{S{\'a}nchez-Bl{\'a}zquez}
  {et~al}\mbox{.}(2011){S{\'a}nchez-Bl{\'a}zquez}, {Ocvirk}, {Gibson},
  {P{\'e}rez}, \& {Peletier}}]{SanchezBlazquezetal2011}
{S{\'a}nchez-Bl{\'a}zquez} P., {Ocvirk} P., {Gibson} B.~K., {P{\'e}rez} I.,
  {Peletier} R.~F., 2011, \mnras, 415, 709

\bibitem[{{S{\'a}nchez-Bl{\'a}zquez}
  {et~al}\mbox{.}(2014){S{\'a}nchez-Bl{\'a}zquez}, {Rosales-Ortega},
  {M{\'e}ndez-Abreu}, {P{\'e}rez}, {S{\'a}nchez}, {Zibetti}, {Aguerri},
  {Bland-Hawthorn}, {Catal{\'a}n-Torrecilla}, {Cid Fernandes}, {de Amorim}, {de
  Lorenzo-Caceres}, {Falc{\'o}n-Barroso}, {Galazzi}, {Garc{\'{\i}}a Benito},
  {Gil de Paz}, {Gonz{\'a}lez Delgado}, {Husemann}, {Iglesias-P{\'a}ramo},
  {Jungwiert}, {Marino}, {M{\'a}rquez}, {Mast}, {Mendoza}, {Moll{\'a}},
  {Papaderos}, {Ruiz-Lara}, {van de Ven}, {Walcher}, \&
  {Wisotzki}}]{SanchezBlazquezetal2014}
{S{\'a}nchez-Bl{\'a}zquez} P. {et~al.}, 2014, \aap, 570, A6

\bibitem[{{Sorba} \& {Sawicki}(2015)}]{Sorbaetal2015}
{Sorba} R., {Sawicki} M., 2015, \mnras, 452, 235

\bibitem[{{Spolaor} {et~al}\mbox{.}(2008){Spolaor}, {Forbes}, {Proctor}, {Hau},
  \& {Brough}}]{Spolaoretal2008}
{Spolaor} M., {Forbes} D.~A., {Proctor} R.~N., {Hau} G.~K.~T., {Brough} S.,
  2008, \mnras, 385, 675

\bibitem[{{Spolaor} {et~al}\mbox{.}(2010){Spolaor}, {Kobayashi}, {Forbes},
  {Couch}, \& {Hau}}]{Spolaoretal2010}
{Spolaor} M., {Kobayashi} C., {Forbes} D.~A., {Couch} W.~J., {Hau} G.~K.~T.,
  2010, \mnras, 408, 272

\bibitem[{{Tantalo}, {Chiosi} \& {Bressan}(1998){Tantalo}, {Chiosi}, \&
  {Bressan}}]{Tantaloetal1998}
{Tantalo} R., {Chiosi} C., {Bressan} A., 1998, \aap, 333, 419

\bibitem[{{Tortora} {et~al}\mbox{.}(2010){Tortora}, {Napolitano}, {Cardone},
  {Capaccioli}, {Jetzer}, \& {Molinaro}}]{Tortoraetal2010}
{Tortora} C., {Napolitano} N.~R., {Cardone} V.~F., {Capaccioli} M., {Jetzer}
  P., {Molinaro} R., 2010, \mnras, 407, 144

\bibitem[{{van der Walt}, {Colbert} \& {Varoquaux}(2011){van der Walt},
  {Colbert}, \& {Varoquaux}}]{vanderWaltetal2011}
{van der Walt} S., {Colbert} S.~C., {Varoquaux} G., 2011, Computing in Science
  and Engineering, 13, 22

\bibitem[{{Vazdekis} {et~al}\mbox{.}(2016){Vazdekis}, {Koleva}, {Ricciardelli},
  {R{\"o}ck}, \& {Falc{\'o}n-Barroso}}]{Vazdekisetal2016}
{Vazdekis} A., {Koleva} M., {Ricciardelli} E., {R{\"o}ck} B.,
  {Falc{\'o}n-Barroso} J., 2016, \mnras, 463, 3409

\bibitem[{{Wilkinson} {et~al}\mbox{.}(2015){Wilkinson}, {Maraston}, {Thomas},
  {Coccato}, {Tojeiro}, {Cappellari}, {Belfiore}, {Bershady}, {Blanton},
  {Bundy}, {Cales}, {Cherinka}, {Drory}, {Emsellem}, {Fu}, {Law}, {Li},
  {Maiolino}, {Masters}, {Tremonti}, {Wake}, {Wang}, {Weijmans}, {Xiao}, {Yan},
  {Zhang}, {Bizyaev}, {Brinkmann}, {Kinemuchi}, {Malanushenko}, {Malanushenko},
  {Oravetz}, {Pan}, \& {Simmons}}]{Wilkinsonetal2015}
{Wilkinson} D.~M. {et~al.}, 2015, \mnras, 449, 328

\bibitem[{{Wolf} {et~al}\mbox{.}(2003){Wolf}, {Meisenheimer}, {Rix}, {Borch},
  {Dye}, \& {Kleinheinrich}}]{Wolfetal2003}
{Wolf} C., {Meisenheimer} K., {Rix} H.-W., {Borch} A., {Dye} S.,
  {Kleinheinrich} M., 2003, \aap, 401, 73

\bibitem[{{Worthey}, {Faber} \& {Gonzalez}(1992){Worthey}, {Faber}, \&
  {Gonzalez}}]{Wortheyetal1992}
{Worthey} G., {Faber} S.~M., {Gonzalez} J.~J., 1992, \apj, 398, 69

\bibitem[{{Worthey} {et~al}\mbox{.}(1994){Worthey}, {Faber}, {Gonzalez}, \&
  {Burstein}}]{Wortheyetal1994}
{Worthey} G., {Faber} S.~M., {Gonzalez} J.~J., {Burstein} D., 1994, \apjs, 94,
  687

\bibitem[{{Wu} {et~al}\mbox{.}(2005){Wu}, {Shao}, {Mo}, {Xia}, \&
  {Deng}}]{Wuetal2005}
{Wu} H., {Shao} Z., {Mo} H.~J., {Xia} X., {Deng} Z., 2005, \apj, 622, 244

\bibitem[{{York} {et~al}\mbox{.}(2000){York}, {Adelman}, {Anderson},
  {Anderson}, {Annis}, {Bahcall}, {Bakken}, {Barkhouser}, {Bastian}, {Berman},
  {Boroski}, {Bracker}, {Briegel}, {Briggs}, {Brinkmann}, {Brunner}, {Burles},
  {Carey}, {Carr}, {Castander}, {Chen}, {Colestock}, {Connolly}, {Crocker},
  {Csabai}, {Czarapata}, {Davis}, {Doi}, {Dombeck}, {Eisenstein}, {Ellman},
  {Elms}, {Evans}, {Fan}, {Federwitz}, {Fiscelli}, {Friedman}, {Frieman},
  {Fukugita}, {Gillespie}, {Gunn}, {Gurbani}, {de Haas}, {Haldeman}, {Harris},
  {Hayes}, {Heckman}, {Hennessy}, {Hindsley}, {Holm}, {Holmgren}, {Huang},
  {Hull}, {Husby}, {Ichikawa}, {Ichikawa}, {Ivezi{\'c}}, {Kent}, {Kim},
  {Kinney}, {Klaene}, {Kleinman}, {Kleinman}, {Knapp}, {Korienek}, {Kron},
  {Kunszt}, {Lamb}, {Lee}, {Leger}, {Limmongkol}, {Lindenmeyer}, {Long},
  {Loomis}, {Loveday}, {Lucinio}, {Lupton}, {MacKinnon}, {Mannery}, {Mantsch},
  {Margon}, {McGehee}, {McKay}, {Meiksin}, {Merelli}, {Monet}, {Munn},
  {Narayanan}, {Nash}, {Neilsen}, {Neswold}, {Newberg}, {Nichol}, {Nicinski},
  {Nonino}, {Okada}, {Okamura}, {Ostriker}, {Owen}, {Pauls}, {Peoples},
  {Peterson}, {Petravick}, {Pier}, {Pope}, {Pordes}, {Prosapio},
  {Rechenmacher}, {Quinn}, {Richards}, {Richmond}, {Rivetta}, {Rockosi},
  {Ruthmansdorfer}, {Sandford}, {Schlegel}, {Schneider}, {Sekiguchi}, {Sergey},
  {Shimasaku}, {Siegmund}, {Smee}, {Smith}, {Snedden}, {Stone}, {Stoughton},
  {Strauss}, {Stubbs}, {SubbaRao}, {Szalay}, {Szapudi}, {Szokoly}, {Thakar},
  {Tremonti}, {Tucker}, {Uomoto}, {Vanden Berk}, {Vogeley}, {Waddell}, {Wang},
  {Watanabe}, {Weinberg}, {Yanny}, {Yasuda}, \& {SDSS
  Collaboration}}]{Yorketal2000}
{York} D.~G. {et~al.}, 2000, \aj, 120, 1579

\bibitem[{{Zheng} {et~al}\mbox{.}(2016){Zheng}, {Wang}, {Ge}, {Mao}, {Li},
  {Li}, {Mo}, {Goddard}, {Bundy}, {Li}, {Nair}, {Lin}, {Long}, {Riffel},
  {Thomas}, {Masters}, {Bizyaev}, {Brownstein}, {Zhang}, {Law}, {Drory},
  {Lopes}, \& {Malanushenko}}]{Zhengetal2016}
{Zheng} Z. {et~al.}, 2016, ArXiv e-prints

\bibitem[{{Zibetti}, {Charlot} \& {Rix}(2009){Zibetti}, {Charlot}, \&
  {Rix}}]{Zibettietal2009}
{Zibetti} S., {Charlot} S., {Rix} H.-W., 2009, \mnras, 400, 1181

\end{thebibliography}

\appendix
\section{Stellar population properties}\label{ap:tables}
\onecolumn
\tiny
\begin{center}
\begin{landscape}
\begin{longtable}{cccclcccccc}
\caption{Stellar population properties.}\\
\hline\hline\\[-1ex]
ID & RA & Dec &  Redshift & Model & log M$_{\star}$  &  < Age >$_\mathrm{L}$$^\mathrm{resolved}$   &  < Age >$_\mathrm{M}$$^\mathrm{resolved}$   &  < [Fe/H] >$_\mathrm{L}$$^\mathrm{resolved}$ &  < [Fe/H] >$_\mathrm{M}$$^\mathrm{resolved}$ &  < $\chi^{2}$ > \\
    &           &   &  &     &  (M$_{\sun}$)       &   (Gyrs)      &   (Gyrs)      &  (dex) &  (dex)  &  \\[0.5ex]
\hline\\[-1ex]
\endfirsthead

\caption{continued.}\\
\hline\hline\\[-1ex]
ID &   RA & Dec & Redshift & Model  &  log M$_{\star}$ &  Age$_\mathrm{L}$  &  Age$_\mathrm{M}$   &  [Fe/H]$_\mathrm{L}$ &  [Fe/H]$_\mathrm{M}$ & < $\chi^{2}$ >\\
    &           &      &   &  &  (M$_{\sun}$)      &   (Gyrs)      &   (Gyrs)      &  (dex) &  (dex)  &  \\[0.5ex]

\hline\\[-1ex]
\endhead
\hline\\[-1ex]

\endfoot
\hline\\[-1ex]
\multicolumn{11}{l}{Notes. ID column corresponds to ALHAMBRA IDs. They are formed according to the following criteria: 3 digits (detection image) + 4 digits (field) + 2 digits (pointing) + 1 digit (CCD) + 5 digits (idenfification ID).}\\
\multicolumn{11}{l}{$\textsuperscript{a}$Peculiar galaxy: Identified object as a peculiar case that stand up from the analyzed sample (see Sect. \ref{Sec:peculiar} for further details).}\\
\multicolumn{11}{l}{$\textsuperscript{b}$Spectroscopic redshift.}
\endlastfoot

81461302615	&	14:15:20.376 & +52:20:45.24	&	0.07$\textsuperscript{b}$	&	BC03	&	11.03	&	6.36	$\pm$	0.09	&	7.93	$\pm$	0.12	&	0.20	$\pm$	0.01	&	0.12	$\pm$	0.01	&	0.61	$\pm$	0.15	\\
	&	&	&		&	E-MILES	&	11.28	&	9.76	$\pm$	0.08	&	10.83	$\pm$	0.10	&	-0.05	$\pm$	0.01	&	0.05	$\pm$	0.01	&	0.82	$\pm$	0.16	\\
81422100334	&	02:29:10.272 & +01:15:12.6	&	0.21$\textsuperscript{b}$	&	BC03	&	11.26	&	6.14	$\pm$	0.45	&	7.48	$\pm$	0.50	&	0.12	$\pm$	0.04	&	0.05	$\pm$	0.04	&	0.80	$\pm$	0.21	\\
	&	&	&		&	E-MILES	&	11.42	&	7.82	$\pm$	0.35	&	8.42	$\pm$	0.33	&	-0.11	$\pm$	0.03	&	-0.10	$\pm$	0.03	&	0.60	$\pm$	0.14	\\
81421207571	&	02:27:48.84 & +01:05:18.96	&	0.13$\textsuperscript{b}$	&	BC03	&	11.10	&	5.56	$\pm$	0.17	&	6.92	$\pm$	0.22	&	0.19	$\pm$	0.01	&	0.12	$\pm$	0.02	&	1.12	$\pm$	0.68	\\
	&	&	&		&	E-MILES	&	11.32	&	8.37	$\pm$	0.13	&	9.06	$\pm$	0.16	&	-0.06	$\pm$	0.01	&	0.02	$\pm$	0.01	&	1.19	$\pm$	0.46	\\
81422303905	&	02:26:53.928 & +00:39:44.28	&	0.13$\textsuperscript{b}$	&	BC03	&	10.68	&	5.11	$\pm$	0.32	&	6.36	$\pm$	0.39	&	0.04	$\pm$	0.03	&	0.00	$\pm$	0.03	&	0.90	$\pm$	0.21	\\
	&	&	&		&	E-MILES	&	10.88	&	7.66	$\pm$	0.28	&	8.61	$\pm$	0.30	&	-0.20	$\pm$	0.02	&	-0.12	$\pm$	0.03	&	0.92	$\pm$	0.18	\\
81422406945$\textsuperscript{a}$	&	02:29:20.928 & +00:35:38.04	&	0.13	&	BC03	&	11.24	&	7.27	$\pm$	0.12	&	8.50	$\pm$	0.12	&	-0.05	$\pm$	0.01	&	-0.14	$\pm$	0.01	&	0.99	$\pm$	0.41	\\
	&	&	&		&	E-MILES	&	11.40	&	9.61	$\pm$	0.10	&	10.23	$\pm$	0.09	&	-0.27	$\pm$	0.01	&	-0.24	$\pm$	0.01	&	1.13	$\pm$	0.40	\\
81432100744	&	09:17:41.448 & +46:30:00	&	0.19$\textsuperscript{b}$	&	BC03	&	11.12	&	6.48	$\pm$	0.29	&	7.67	$\pm$	0.35	&	0.18	$\pm$	0.02	&	0.11	$\pm$	0.03	&	0.90	$\pm$	0.38	\\
	&	&	&		&	E-MILES	&	11.33	&	9.81	$\pm$	0.19	&	10.39	$\pm$	0.22	&	-0.02	$\pm$	0.01	&	0.04	$\pm$	0.02	&	0.72	$\pm$	0.25	\\
81441106089	&	10:01:39.768 & +02:25:48.72	&	0.12$\textsuperscript{b}$	&	BC03	&	11.22	&	5.46	$\pm$	0.23	&	7.07	$\pm$	0.29	&	0.31	$\pm$	0.01	&	0.23	$\pm$	0.01	&	0.82	$\pm$	0.15	\\
	&	&	&		&	E-MILES	&	11.49	&	9.94	$\pm$	0.14	&	10.65	$\pm$	0.15	&	0.05	$\pm$	0.01	&	0.13	$\pm$	0.01	&	0.75	$\pm$	0.13	\\
81451305824	&	12:33:42.456 & +61:48:09.72	&	0.23$\textsuperscript{b}$	&	BC03	&	11.38	&	5.73	$\pm$	0.67	&	7.67	$\pm$	0.74	&	0.32	$\pm$	0.04	&	0.24	$\pm$	0.05	&	1.01	$\pm$	0.12	\\
	&	&	&		&	E-MILES	&	11.52	&	6.25	$\pm$	0.61	&	7.61	$\pm$	0.62	&	0.16	$\pm$	0.03	&	0.14	$\pm$	0.04	&	0.79	$\pm$	0.12	\\
81481104825	&	23:46:53.52 & +15:56:40.2	&	0.2$\textsuperscript{b}$	&	BC03	&	11.31	&	7.56	$\pm$	0.32	&	8.55	$\pm$	0.35	&	0.07	$\pm$	0.02	&	0.04	$\pm$	0.03	&	0.61	$\pm$	0.23	\\
	&	&	&		&	E-MILES	&	11.46	&	9.52	$\pm$	0.25	&	9.97	$\pm$	0.24	&	-0.12	$\pm$	0.02	&	-0.08	$\pm$	0.02	&	0.64	$\pm$	0.28	\\
81421106337	&	02:29:55.488 & +01:04:36.48	&	0.24	&	BC03	&	11.67	&	6.42	$\pm$	0.53	&	7.66	$\pm$	0.56	&	0.17	$\pm$	0.04	&	0.12	$\pm$	0.04	&	1.41	$\pm$	0.46	\\
	&	&	&		&	E-MILES	&	11.81	&	7.46	$\pm$	0.35	&	7.82	$\pm$	0.36	&	-0.05	$\pm$	0.03	&	-0.04	$\pm$	0.03	&	1.36	$\pm$	0.39	\\
81451206302$\textsuperscript{a}$	&	12:34:29.88 & +62:18:06.48	&	0.17	&	BC03	&	10.56	&	0.35	$\pm$	0.01	&	1.14	$\pm$	0.05	&	-0.50	$\pm$	0.01	&	0.12	$\pm$	0.01	&	9.55	$\pm$	7.96	\\
	&	&	&		&	E-MILES	&	10.85	&	1.94	$\pm$	0.02	&	5.85	$\pm$	0.07	&	0.01	$\pm$	0.01	&	0.22	$\pm$	0.01	&	18.32	$\pm$	16.99	\\
81461400772	&	14:18:59.952 & +52:23:29.4	&	0.24	&	BC03	&	11.06	&	5.09	$\pm$	0.48	&	6.27	$\pm$	0.62	&	0.22	$\pm$	0.04	&	0.15	$\pm$	0.04	&	0.57	$\pm$	0.17	\\
	&	&	&		&	E-MILES	&	11.21	&	6.08	$\pm$	0.33	&	6.55	$\pm$	0.39	&	-0.01	$\pm$	0.02	&	-0.01	$\pm$	0.03	&	0.88	$\pm$	0.18	\\
81473405681$\textsuperscript{a}$	&	16:14:57.984 & +54:08:53.16	&	0.21	&	BC03	&	10.67	&	0.88	$\pm$	0.04	&	1.65	$\pm$	0.09	&	0.02	$\pm$	0.02	&	0.32	$\pm$	0.02	&	0.71	$\pm$	0.30	\\
	&	&	&		&	E-MILES	&	10.94	&	2.26	$\pm$	0.06	&	5.49	$\pm$	0.17	&	-0.29	$\pm$	0.01	&	0.14	$\pm$	0.01	&	1.01	$\pm$	0.39	\\
81473406818	&	16:15:13.224 & +54:07:39	&	0.12	&	BC03	&	11.11	&	5.95	$\pm$	0.14	&	7.40	$\pm$	0.16	&	0.23	$\pm$	0.01	&	0.15	$\pm$	0.01	&	0.87	$\pm$	0.19	\\
	&	&	&		&	E-MILES	&	11.35	&	9.69	$\pm$	0.10	&	10.25	$\pm$	0.12	&	-0.03	$\pm$	0.01	&	0.04	$\pm$	0.01	&	0.95	$\pm$	0.16	\\
81474201317	&	16:09:46.56 & +54:42:08.28	&	0.23	&	BC03	&	10.94	&	3.46	$\pm$	0.38	&	4.52	$\pm$	0.63	&	0.02	$\pm$	0.05	&	0.18	$\pm$	0.05	&	0.74	$\pm$	0.33	\\
	&	&	&		&	E-MILES	&	11.10	&	4.53	$\pm$	0.43	&	5.78	$\pm$	0.56	&	-0.23	$\pm$	0.04	&	-0.03	$\pm$	0.05	&	0.83	$\pm$	0.23	\\
81474307526$\textsuperscript{a}$	&	16:09:02.112 & +54:02:36.24	&	0.26	&	BC03	&	10.62	&	0.86	$\pm$	0.06	&	1.84	$\pm$	0.19	&	0.08	$\pm$	0.03	&	0.23	$\pm$	0.04	&	0.77	$\pm$	0.16	\\
	&	&	&		&	E-MILES	&	10.91	&	2.42	$\pm$	0.09	&	6.67	$\pm$	0.23	&	-0.41	$\pm$	0.02	&	0.12	$\pm$	0.02	&	1.20	$\pm$	0.45	\\
81474404152	&	16:13:31.848 & +54:09:29.16	&	0.19	&	BC03	&	11.36	&	5.81	$\pm$	0.30	&	7.05	$\pm$	0.39	&	0.21	$\pm$	0.02	&	0.13	$\pm$	0.03	&	0.61	$\pm$	0.18	\\
	&	&	&		&	E-MILES	&	11.54	&	7.90	$\pm$	0.24	&	8.42	$\pm$	0.27	&	-0.02	$\pm$	0.01	&	0.00	$\pm$	0.02	&	0.63	$\pm$	0.15	\\
81474404716	&	16:12:49.44 & +54:08:37.32	&	0.15	&	BC03	&	11.18	&	5.60	$\pm$	0.26	&	6.80	$\pm$	0.32	&	0.18	$\pm$	0.02	&	0.09	$\pm$	0.02	&	0.72	$\pm$	0.15	\\
	&	&	&		&	E-MILES	&	11.40	&	8.44	$\pm$	0.19	&	8.88	$\pm$	0.22	&	-0.06	$\pm$	0.01	&	-0.01	$\pm$	0.02	&	0.81	$\pm$	0.15	\\
81481201181	&	23:45:38.472 & +16:02:24.36	&	0.23	&	BC03	&	10.98	&	4.79	$\pm$	0.65	&	6.35	$\pm$	0.87	&	-0.03	$\pm$	0.06	&	-0.09	$\pm$	0.09	&	1.70	$\pm$	1.59	\\
	&	&	&		&	E-MILES	&	11.14	&	6.57	$\pm$	0.45	&	7.38	$\pm$	0.52	&	-0.28	$\pm$	0.04	&	-0.23	$\pm$	0.06	&	2.42	$\pm$	2.85	\\
81481402928	&	23:47:39.936 & +15:30:35.28	&	0.19	&	BC03	&	11.08	&	6.86	$\pm$	0.41	&	7.90	$\pm$	0.46	&	0.08	$\pm$	0.03	&	0.03	$\pm$	0.03	&	0.88	$\pm$	0.21	\\
	&	&	&		&	E-MILES	&	11.26	&	9.77	$\pm$	0.28	&	10.16	$\pm$	0.28	&	-0.14	$\pm$	0.02	&	-0.10	$\pm$	0.02	&	0.70	$\pm$	0.14	\\
81421103140	&	02:29:50.448 & +01:10:17.76	&	0.24	&	BC03	&	11.04	&	5.09	$\pm$	0.47	&	6.16	$\pm$	0.64	&	0.23	$\pm$	0.04	&	0.17	$\pm$	0.05	&	1.35	$\pm$	0.51	\\
	&	&	&		&	E-MILES	&	11.20	&	7.12	$\pm$	0.38	&	7.79	$\pm$	0.45	&	-0.03	$\pm$	0.03	&	-0.03	$\pm$	0.03	&	1.29	$\pm$	0.47	\\
81422407693$\textsuperscript{a}$	&	02:29:13.296 & +00:34:30.72	&	0.19	&	BC03	&	10.67	&	1.95	$\pm$	0.09	&	4.32	$\pm$	0.18	&	-0.34	$\pm$	0.02	&	-0.46	$\pm$	0.03	&	3.39	$\pm$	1.96	\\
	&	&	&		&	E-MILES	&	10.97	&	3.29	$\pm$	0.07	&	5.68	$\pm$	0.14	&	-0.33	$\pm$	0.01	&	-0.12	$\pm$	0.02	&	11.00	$\pm$	12.10	\\
81431100706	&	09:17:41.472 & +46:30:00.36	&	0.17	&	BC03	&	11.18	&	8.06	$\pm$	0.21	&	8.94	$\pm$	0.23	&	0.02	$\pm$	0.02	&	-0.04	$\pm$	0.02	&	1.33	$\pm$	0.36	\\
	&	&	&		&	E-MILES	&	11.32	&	9.99	$\pm$	0.18	&	10.59	$\pm$	0.22	&	-0.11	$\pm$	0.01	&	-0.06	$\pm$	0.01	&	1.42	$\pm$	0.35	\\
81473103857$\textsuperscript{a}$	&	16:14:33.84 & +54:40:50.52	&	0.24	&	BC03	&	10.62	&	0.61	$\pm$	0.04	&	1.74	$\pm$	0.14	&	-0.01	$\pm$	0.01	&	0.35	$\pm$	0.01	&	5.14	$\pm$	4.09	\\
	&	&	&		&	E-MILES	&	10.95	&	1.94	$\pm$	0.04	&	4.57	$\pm$	0.13	&	0.00	$\pm$	0.01	&	0.28	$\pm$	0.01	&	18.05	$\pm$	17.56	\\
81473111498	&	16:14:22.08 & +54:32:18.6	&	0.27	&	BC03	&	11.22	&	5.36	$\pm$	0.68	&	6.96	$\pm$	0.79	&	0.20	$\pm$	0.04	&	0.20	$\pm$	0.04	&	1.38	$\pm$	0.87	\\
	&	&	&		&	E-MILES	&	11.39	&	6.38	$\pm$	0.63	&	7.49	$\pm$	0.77	&	-0.02	$\pm$	0.03	&	0.16	$\pm$	0.03	&	1.69	$\pm$	0.87	\\
81474103381	&	16:12:13.2 & +54:39:38.88	&	0.22	&	BC03	&	11.22	&	4.42	$\pm$	0.27	&	5.92	$\pm$	0.37	&	0.07	$\pm$	0.03	&	0.07	$\pm$	0.03	&	1.45	$\pm$	0.63	\\
	&	&	&		&	E-MILES	&	11.42	&	5.61	$\pm$	0.28	&	6.85	$\pm$	0.39	&	-0.17	$\pm$	0.02	&	0.02	$\pm$	0.03	&	1.83	$\pm$	0.65	\\
81482205540	&	23:44:51.096 & +15:56:55.32	&	0.17	&	BC03	&	11.15	&	7.83	$\pm$	0.29	&	8.57	$\pm$	0.30	&	-0.01	$\pm$	0.02	&	-0.07	$\pm$	0.03	&	0.81	$\pm$	0.17	\\
	&	&	&		&	E-MILES	&	11.29	&	9.21	$\pm$	0.25	&	9.85	$\pm$	0.24	&	-0.19	$\pm$	0.02	&	-0.19	$\pm$	0.02	&	0.92	$\pm$	0.21	\\
81481401355	&	23:47:04.656 & +15:32:59.28	&	0.19	&	BC03	&	11.07	&	6.11	$\pm$	0.34	&	7.46	$\pm$	0.39	&	0.21	$\pm$	0.02	&	0.14	$\pm$	0.02	&	0.80	$\pm$	0.20	\\
	&	&	&		&	E-MILES	&	11.27	&	8.70	$\pm$	0.31	&	9.22	$\pm$	0.35	&	-0.02	$\pm$	0.02	&	0.02	$\pm$	0.02	&	0.68	$\pm$	0.09	\\
81422404933	&	02:28:55.752 & +00:39:10.44	&	0.26	&	BC03	&	11.23	&	5.66	$\pm$	0.57	&	7.18	$\pm$	0.69	&	0.17	$\pm$	0.04	&	0.08	$\pm$	0.05	&	1.24	$\pm$	0.59	\\
	&	&	&		&	E-MILES	&	11.40	&	7.74	$\pm$	0.41	&	8.40	$\pm$	0.44	&	-0.05	$\pm$	0.03	&	-0.04	$\pm$	0.03	&	1.23	$\pm$	0.50	\\

\label{tab:A1}
\end{longtable}
\end{landscape}
\end{center}
\normalsize

\twocolumn

\newpage
\onecolumn

\tiny
\begin{landscape}
\begin{longtable}{clcccccccccc}
\caption{Light-weighted and mass-weighted age, metallicity annd extinction gradients.}\\
\hline\hline\\[-1ex]
ID & Model &  $\nabla$log Age$_\mathrm{L}$  &  log Age$_{0,}$$_\mathrm{L}$  & $\nabla$log Age$_\mathrm{M}$  &  log Age$_{0,}$$_\mathrm{M}$  &  $\nabla$[Fe/H]$_\mathrm{L}$   &  [Fe/H]$_{0,}$$_\mathrm{L}$ &  $\nabla$[Fe/H]$_\mathrm{M}$   &  [Fe/H]$_{0,}$$_\mathrm{M}$  & $\nabla$A$_\mathrm{V}$   &  A$_\mathrm{V0}$ \\
    &           & (dex/R$\textsubscript{eff}$)      &   (Gyrs)    &   (dex/R$\textsubscript{eff}$)      &   (Gyrs)      &  (dex/R$\textsubscript{eff}$)  &  (dex)    &   (dex/R$\textsubscript{eff}$)  &  (dex)         &  (mag/R$\textsubscript{eff}$) &  (mag) \\[0.5ex]
\hline\\[-1ex]
\endfirsthead

\caption{continued.}\\
\hline\hline\\[-1ex]
ID & Model &  $\nabla$log Age$_\mathrm{L}$  &  log Age$_{0,}$$_\mathrm{L}$  & $\nabla$log Age$_\mathrm{M}$  &  log Age$_{0,}$$_\mathrm{M}$  &  $\nabla$[Fe/H]$_\mathrm{L}$   &  [Fe/H]$_{0,}$$_\mathrm{L}$ &  $\nabla$[Fe/H]$_\mathrm{M}$   &  [Fe/H]$_{0,}$$_\mathrm{M}$  & $\nabla$A$_\mathrm{V}$   &  A$_\mathrm{V0}$ \\
    &           & (dex/R$\textsubscript{eff}$)      &   (Gyrs)    &   (dex/R$\textsubscript{eff}$)      &   (Gyrs)      &  (dex/R$\textsubscript{eff}$)  &  (dex)    &   (dex/R$\textsubscript{eff}$)  &  (dex)         &  (mag/R$\textsubscript{eff}$) &  (mag)  \\[0.5ex]
\hline\\[-1ex]
\endhead
\hline\\[-1ex]

\endfoot
\hline\\[-1ex]
\multicolumn{12}{l}{Notes. ID column corresponds to ALHAMBRA IDs. They are formed according to the following criteria: 3 digits (detection image) + 4 digits (field) + 2 digits (pointing) + 1 digit (CCD) + 5 digits (idenfification ID).}\\
\multicolumn{12}{l}{$\textsuperscript{a}$Peculiar galaxy: Identified object as a peculiar case that stand up from the analyzed sample. Stellar population gradients of these objects should be used with caution (see Sect. \ref{Sec:peculiar} for further details).}
\endlastfoot

81461302615	&	BC03	&	0.02	$\pm$	0.00	&	0.76	$\pm$	0.01	&	0.00	$\pm$	0.00	&	0.88	$\pm$	0.00	&	-0.08	$\pm$	0.00	&	0.30	$\pm$	0.01	&	-0.04	$\pm$	0.00	&	0.17	$\pm$	0.01	&	-0.07	$\pm$	0.00	&	0.23	$\pm$	0.00	\\
	&	E-MILES	&	0.01	$\pm$	0.00	&	0.97	$\pm$	0.00	&	0.02	$\pm$	0.00	&	1.01	$\pm$	0.00	&	-0.05	$\pm$	0.00	&	0.03	$\pm$	0.00	&	-0.04	$\pm$	0.00	&	0.10	$\pm$	0.01	&	-0.04	$\pm$	0.00	&	0.28	$\pm$	0.00	\\
81422100334	&	BC03	&	-0.01	$\pm$	0.01	&	0.79	$\pm$	0.02	&	-0.00	$\pm$	0.01	&	0.87	$\pm$	0.01	&	-0.08	$\pm$	0.03	&	0.23	$\pm$	0.03	&	-0.05	$\pm$	0.03	&	0.13	$\pm$	0.04	&	-0.01	$\pm$	0.02	&	0.30	$\pm$	0.02	\\
	&	E-MILES	&	0.02	$\pm$	0.02	&	0.85	$\pm$	0.02	&	0.02	$\pm$	0.01	&	0.88	$\pm$	0.01	&	-0.05	$\pm$	0.02	&	-0.04	$\pm$	0.02	&	-0.04	$\pm$	0.02	&	-0.05	$\pm$	0.03	&	-0.09	$\pm$	0.01	&	0.50	$\pm$	0.01	\\
81421207571	&	BC03	&	0.05	$\pm$	0.01	&	0.66	$\pm$	0.02	&	0.03	$\pm$	0.01	&	0.76	$\pm$	0.02	&	-0.12	$\pm$	0.01	&	0.34	$\pm$	0.02	&	-0.11	$\pm$	0.01	&	0.26	$\pm$	0.01	&	-0.01	$\pm$	0.01	&	0.16	$\pm$	0.01	\\
	&	E-MILES	&	0.01	$\pm$	0.01	&	0.90	$\pm$	0.01	&	0.00	$\pm$	0.01	&	0.95	$\pm$	0.01	&	-0.10	$\pm$	0.02	&	0.05	$\pm$	0.02	&	-0.10	$\pm$	0.01	&	0.15	$\pm$	0.02	&	0.01	$\pm$	0.02	&	0.19	$\pm$	0.02	\\
81422303905	&	BC03	&	0.07	$\pm$	0.02	&	0.60	$\pm$	0.02	&	0.05	$\pm$	0.02	&	0.71	$\pm$	0.02	&	-0.17	$\pm$	0.04	&	0.25	$\pm$	0.04	&	-0.12	$\pm$	0.03	&	0.21	$\pm$	0.03	&	-0.08	$\pm$	0.02	&	0.49	$\pm$	0.02	\\
	&	E-MILES	&	0.04	$\pm$	0.01	&	0.81	$\pm$	0.01	&	0.02	$\pm$	0.01	&	0.89	$\pm$	0.01	&	-0.11	$\pm$	0.03	&	-0.06	$\pm$	0.03	&	-0.14	$\pm$	0.04	&	0.05	$\pm$	0.04	&	-0.08	$\pm$	0.02	&	0.56	$\pm$	0.02	\\
81422406945$\textsuperscript{a}$	&	BC03	&	0.19	$\pm$	0.01	&	0.62	$\pm$	0.02	&	0.14	$\pm$	0.01	&	0.75	$\pm$	0.01	&	-0.35	$\pm$	0.02	&	0.34	$\pm$	0.02	&	-0.33	$\pm$	0.02	&	0.27	$\pm$	0.02	&	-0.07	$\pm$	0.01	&	0.32	$\pm$	0.01	\\
	&	E-MILES	&	0.14	$\pm$	0.01	&	0.82	$\pm$	0.01	&	0.09	$\pm$	0.01	&	0.90	$\pm$	0.01	&	-0.28	$\pm$	0.01	&	0.05	$\pm$	0.02	&	-0.28	$\pm$	0.01	&	0.09	$\pm$	0.02	&	-0.05	$\pm$	0.01	&	0.32	$\pm$	0.01	\\
81432100744	&	BC03	&	0.08	$\pm$	0.02	&	0.70	$\pm$	0.02	&	0.06	$\pm$	0.02	&	0.80	$\pm$	0.02	&	-0.14	$\pm$	0.04	&	0.35	$\pm$	0.03	&	-0.14	$\pm$	0.04	&	0.29	$\pm$	0.04	&	-0.0	$\pm$	0.01	&	0.07	$\pm$	0.01	\\
	&	E-MILES	&	0.05	$\pm$	0.01	&	0.92	$\pm$	0.01	&	0.03	$\pm$	0.01	&	0.97	$\pm$	0.01	&	-0.08	$\pm$	0.03	&	0.08	$\pm$	0.03	&	-0.10	$\pm$	0.04	&	0.16	$\pm$	0.03	&	-0.05	$\pm$	0.01	&	0.14	$\pm$	0.01	\\
81441106089	&	BC03	&	0.07	$\pm$	0.02	&	0.61	$\pm$	0.02	&	0.05	$\pm$	0.01	&	0.75	$\pm$	0.01	&	-0.09	$\pm$	0.02	&	0.45	$\pm$	0.02	&	-0.09	$\pm$	0.02	&	0.39	$\pm$	0.02	&	-0.03	$\pm$	0.00	&	0.24	$\pm$	0.00	\\
	&	E-MILES	&	0.03	$\pm$	0.01	&	0.95	$\pm$	0.01	&	0.02	$\pm$	0.00	&	1.00	$\pm$	0.00	&	-0.10	$\pm$	0.01	&	0.19	$\pm$	0.02	&	-0.09	$\pm$	0.02	&	0.25	$\pm$	0.02	&	-0.01	$\pm$	0.00	&	0.19	$\pm$	0.00	\\
81451305824	&	BC03	&	-0.02	$\pm$	0.03	&	0.77	$\pm$	0.03	&	-0.01	$\pm$	0.02	&	0.89	$\pm$	0.02	&	-0.07	$\pm$	0.02	&	0.42	$\pm$	0.02	&	-0.11	$\pm$	0.03	&	0.39	$\pm$	0.03	&	0.12	$\pm$	0.06	&	0.14	$\pm$	0.07	\\
	&	E-MILES	&	0.02	$\pm$	0.01	&	0.76	$\pm$	0.01	&	-0.00	$\pm$	0.01	&	0.87	$\pm$	0.01	&	-0.04	$\pm$	0.02	&	0.21	$\pm$	0.02	&	-0.07	$\pm$	0.02	&	0.24	$\pm$	0.02	&	-0.08	$\pm$	0.02	&	0.43	$\pm$	0.02	\\
81481104825	&	BC03	&	0.01	$\pm$	0.01	&	0.86	$\pm$	0.01	&	0.00	$\pm$	0.01	&	0.92	$\pm$	0.01	&	-0.07	$\pm$	0.03	&	0.21	$\pm$	0.03	&	-0.05	$\pm$	0.03	&	0.14	$\pm$	0.03	&	-0.05	$\pm$	0.02	&	0.23	$\pm$	0.02	\\
	&	E-MILES	&	0.01	$\pm$	0.01	&	0.94	$\pm$	0.01	&	0.00	$\pm$	0.01	&	0.97	$\pm$	0.01	&	-0.04	$\pm$	0.02	&	-0.04	$\pm$	0.03	&	-0.04	$\pm$	0.02	&	-0.01	$\pm$	0.03	&	-0.09	$\pm$	0.02	&	0.39	$\pm$	0.02	\\
81421106337	&	BC03	&	-0.03	$\pm$	0.04	&	0.85	$\pm$	0.05	&	-0.03	$\pm$	0.04	&	0.90	$\pm$	0.05	&	-0.11	$\pm$	0.05	&	0.34	$\pm$	0.06	&	-0.12	$\pm$	0.05	&	0.31	$\pm$	0.06	&	-0.03	$\pm$	0.05	&	0.32	$\pm$	0.07	\\
	&	E-MILES	&	0.02	$\pm$	0.04	&	0.82	$\pm$	0.05	&	0.02	$\pm$	0.03	&	0.85	$\pm$	0.04	&	-0.08	$\pm$	0.05	&	0.08	$\pm$	0.06	&	-0.08	$\pm$	0.05	&	0.09	$\pm$	0.06	&	-0.15	$\pm$	0.06	&	0.62	$\pm$	0.08	\\
81451206302$\textsuperscript{a}$	&	BC03	&	-0.06	$\pm$	0.01	&	-0.30	$\pm$	0.02	&	0.51	$\pm$	0.04	&	-0.72	$\pm$	0.06	&	0.07	$\pm$	0.01	&	-0.59	$\pm$	0.02	&	-0.09	$\pm$	0.03	&	0.17	$\pm$	0.04	&	-0.39	$\pm$	0.02	&	1.63	$\pm$	0.03	\\
	&	E-MILES	&	0.17	$\pm$	0.01	&	0.10	$\pm$	0.01	&	0.12	$\pm$	0.01	&	0.63	$\pm$	0.02	&	-0.58	$\pm$	0.02	&	0.61	$\pm$	0.03	&	-0.21	$\pm$	0.01	&	0.54	$\pm$	0.01	&	0.15	$\pm$	0.01	&	-0.08	$\pm$	0.01	\\
81461400772	&	BC03	&	0.02	$\pm$	0.04	&	0.67	$\pm$	0.05	&	0.02	$\pm$	0.04	&	0.74	$\pm$	0.05	&	-0.07	$\pm$	0.04	&	0.33	$\pm$	0.05	&	-0.06	$\pm$	0.04	&	0.26	$\pm$	0.05	&	0.01	$\pm$	0.01	&	0.05	$\pm$	0.02	\\
	&	E-MILES	&	0.04	$\pm$	0.04	&	0.72	$\pm$	0.05	&	0.04	$\pm$	0.03	&	0.74	$\pm$	0.05	&	-0.06	$\pm$	0.03	&	0.08	$\pm$	0.04	&	-0.05	$\pm$	0.03	&	0.08	$\pm$	0.04	&	-0.03	$\pm$	0.01	&	0.22	$\pm$	0.01	\\
81473405681$\textsuperscript{a}$	&	BC03	&	-0.00	$\pm$	0.01	&	-0.10	$\pm$	0.01	&	-0.02	$\pm$	0.01	&	0.21	$\pm$	0.01	&	-0.11	$\pm$	0.01	&	0.16	$\pm$	0.02	&	-0.11	$\pm$	0.01	&	0.50	$\pm$	0.01	&	0.0	$\pm$	0.01	&	0.89	$\pm$	0.01	\\
	&	E-MILES	&	-0.05	$\pm$	0.01	&	0.41	$\pm$	0.01	&	-0.12	$\pm$	0.01	&	0.87	$\pm$	0.02	&	-0.15	$\pm$	0.01	&	-0.11	$\pm$	0.02	&	-0.21	$\pm$	0.01	&	0.41	$\pm$	0.02	&	-0.09	$\pm$	0.01	&	0.69	$\pm$	0.02	\\
81473406818	&	BC03	&	0.07	$\pm$	0.01	&	0.67	$\pm$	0.01	&	0.04	$\pm$	0.01	&	0.80	$\pm$	0.01	&	-0.13	$\pm$	0.01	&	0.39	$\pm$	0.01	&	-0.12	$\pm$	0.01	&	0.30	$\pm$	0.01	&	-0.05	$\pm$	0.00	&	0.30	$\pm$	0.00	\\
	&	E-MILES	&	0.01	$\pm$	0.01	&	0.97	$\pm$	0.01	&	0.01	$\pm$	0.00	&	1.00	$\pm$	0.00	&	-0.09	$\pm$	0.01	&	0.08	$\pm$	0.01	&	-0.09	$\pm$	0.01	&	0.16	$\pm$	0.01	&	-0.01	$\pm$	0.00	&	0.25	$\pm$	0.00	\\
81474201317	&	BC03	&	0.14	$\pm$	0.07	&	0.35	$\pm$	0.06	&	0.15	$\pm$	0.06	&	0.45	$\pm$	0.06	&	-0.01	$\pm$	0.06	&	0.03	$\pm$	0.06	&	-0.14	$\pm$	0.06	&	0.35	$\pm$	0.06	&	-0.41	$\pm$	0.11	&	0.70	$\pm$	0.11	\\
	&	E-MILES	&	0.15	$\pm$	0.05	&	0.46	$\pm$	0.05	&	0.12	$\pm$	0.06	&	0.60	$\pm$	0.05	&	0.05	$\pm$	0.04	&	-0.29	$\pm$	0.04	&	-0.14	$\pm$	0.06	&	0.12	$\pm$	0.06	&	-0.4	$\pm$	0.07	&	0.77	$\pm$	0.07	\\
81474307526$\textsuperscript{a}$	&	BC03	&	0.09	$\pm$	0.01	&	-0.19	$\pm$	0.01	&	0.03	$\pm$	0.01	&	0.21	$\pm$	0.01	&	-0.13	$\pm$	0.01	&	0.23	$\pm$	0.01	&	-0.21	$\pm$	0.01	&	0.49	$\pm$	0.01	&	-0.07	$\pm$	0.01	&	0.91	$\pm$	0.01	\\
	&	E-MILES	&	-0.01	$\pm$	0.01	&	0.43	$\pm$	0.01	&	-0.08	$\pm$	0.01	&	0.93	$\pm$	0.01	&	-0.09	$\pm$	0.02	&	-0.32	$\pm$	0.02	&	-0.12	$\pm$	0.01	&	0.34	$\pm$	0.01	&	-0.04	$\pm$	0.01	&	0.54	$\pm$	0.01	\\
81474404152	&	BC03	&	0.04	$\pm$	0.01	&	0.70	$\pm$	0.01	&	0.02	$\pm$	0.01	&	0.79	$\pm$	0.01	&	-0.14	$\pm$	0.01	&	0.41	$\pm$	0.01	&	-0.14	$\pm$	0.01	&	0.33	$\pm$	0.01	&	0.03	$\pm$	0.00	&	0.06	$\pm$	0.00	\\
	&	E-MILES	&	0.03	$\pm$	0.01	&	0.84	$\pm$	0.01	&	0.03	$\pm$	0.01	&	0.88	$\pm$	0.01	&	-0.11	$\pm$	0.01	&	0.13	$\pm$	0.01	&	-0.11	$\pm$	0.01	&	0.16	$\pm$	0.01	&	-0.02	$\pm$	0.00	&	0.20	$\pm$	0.00	\\
81474404716	&	BC03	&	0.06	$\pm$	0.01	&	0.66	$\pm$	0.01	&	0.05	$\pm$	0.01	&	0.76	$\pm$	0.01	&	-0.14	$\pm$	0.01	&	0.38	$\pm$	0.01	&	-0.14	$\pm$	0.01	&	0.28	$\pm$	0.01	&	-0.01	$\pm$	0.00	&	0.14	$\pm$	0.00	\\
	&	E-MILES	&	0.03	$\pm$	0.01	&	0.88	$\pm$	0.01	&	0.02	$\pm$	0.00	&	0.91	$\pm$	0.00	&	-0.10	$\pm$	0.01	&	0.07	$\pm$	0.01	&	-0.10	$\pm$	0.01	&	0.10	$\pm$	0.01	&	-0.03	$\pm$	0.00	&	0.21	$\pm$	0.00	\\
81481201181	&	BC03	&	-0.18	$\pm$	0.07	&	0.74	$\pm$	0.12	&	-0.05	$\pm$	0.04	&	0.77	$\pm$	0.07	&	-0.32	$\pm$	0.06	&	0.46	$\pm$	0.10	&	-0.33	$\pm$	0.06	&	0.36	$\pm$	0.12	&	0.48	$\pm$	0.12	&	-0.24	$\pm$	0.23	\\
	&	E-MILES	&	-0.11	$\pm$	0.04	&	0.87	$\pm$	0.08	&	-0.01	$\pm$	0.02	&	0.84	$\pm$	0.04	&	-0.26	$\pm$	0.04	&	0.08	$\pm$	0.08	&	-0.28	$\pm$	0.06	&	0.14	$\pm$	0.10	&	0.08	$\pm$	0.03	&	0.08	$\pm$	0.06	\\
81481402928	&	BC03	&	0.03	$\pm$	0.01	&	0.79	$\pm$	0.01	&	0.02	$\pm$	0.01	&	0.86	$\pm$	0.01	&	-0.14	$\pm$	0.01	&	0.31	$\pm$	0.01	&	-0.11	$\pm$	0.01	&	0.22	$\pm$	0.01	&	0.04	$\pm$	0.01	&	0.08	$\pm$	0.01	\\
	&	E-MILES	&	0.02	$\pm$	0.01	&	0.95	$\pm$	0.01	&	0.02	$\pm$	0.01	&	0.98	$\pm$	0.01	&	-0.11	$\pm$	0.01	&	0.04	$\pm$	0.01	&	-0.10	$\pm$	0.00	&	0.06	$\pm$	0.01	&	0.02	$\pm$	0.01	&	0.15	$\pm$	0.01	\\
81421103140	&	BC03	&	0.01	$\pm$	0.06	&	0.68	$\pm$	0.09	&	0.02	$\pm$	0.05	&	0.73	$\pm$	0.07	&	-0.06	$\pm$	0.11	&	0.33	$\pm$	0.15	&	-0.05	$\pm$	0.11	&	0.28	$\pm$	0.16	&	0.03	$\pm$	0.02	&	0.03	$\pm$	0.03	\\
	&	E-MILES	&	0.03	$\pm$	0.08	&	0.78	$\pm$	0.12	&	0.03	$\pm$	0.07	&	0.82	$\pm$	0.10	&	-0.05	$\pm$	0.09	&	0.05	$\pm$	0.13	&	-0.05	$\pm$	0.10	&	0.06	$\pm$	0.14	&	-0.02	$\pm$	0.05	&	0.18	$\pm$	0.07	\\
81422407693$\textsuperscript{a}$	&	BC03	&	0.84	$\pm$	0.05	&	-1.29	$\pm$	0.07	&	0.49	$\pm$	0.04	&	-0.34	$\pm$	0.05	&	-0.41	$\pm$	0.04	&	0.16	$\pm$	0.05	&	-0.42	$\pm$	0.04	&	0.30	$\pm$	0.06	&	-0.83	$\pm$	0.05	&	2.61	$\pm$	0.06	\\
	&	E-MILES	&	0.25	$\pm$	0.02	&	0.14	$\pm$	0.02	&	0.12	$\pm$	0.02	&	0.57	$\pm$	0.02	&	-0.56	$\pm$	0.05	&	0.34	$\pm$	0.06	&	-0.56	$\pm$	0.05	&	0.58	$\pm$	0.07	&	-0.33	$\pm$	0.03	&	1.21	$\pm$	0.03	\\
81431100706	&	BC03	&	-0.01	$\pm$	0.04	&	0.87	$\pm$	0.03	&	-0.01	$\pm$	0.03	&	0.91	$\pm$	0.03	&	-0.03	$\pm$	0.08	&	0.09	$\pm$	0.07	&	-0.02	$\pm$	0.08	&	0.07	$\pm$	0.08	&	0.0	$\pm$	0.02	&	0.09	$\pm$	0.02	\\
	&	E-MILES	&	-0.00	$\pm$	0.02	&	0.98	$\pm$	0.02	&	0.00	$\pm$	0.01	&	1.01	$\pm$	0.01	&	-0.01	$\pm$	0.06	&	-0.07	$\pm$	0.05	&	-0.00	$\pm$	0.06	&	-0.04	$\pm$	0.06	&	-0.01	$\pm$	0.01	&	0.17	$\pm$	0.01	\\
81473103857$\textsuperscript{a}$	&	BC03	&	0.21	$\pm$	0.02	&	-0.60	$\pm$	0.02	&	-0.04	$\pm$	0.02	&	0.12	$\pm$	0.03	&	-0.06	$\pm$	0.03	&	0.02	$\pm$	0.03	&	-0.08	$\pm$	0.01	&	0.46	$\pm$	0.01	&	-0.12	$\pm$	0.03	&	1.58	$\pm$	0.04	\\
	&	E-MILES	&	0.15	$\pm$	0.01	&	0.13	$\pm$	0.02	&	0.13	$\pm$	0.02	&	0.50	$\pm$	0.03	&	-0.44	$\pm$	0.03	&	0.46	$\pm$	0.04	&	-0.13	$\pm$	0.01	&	0.45	$\pm$	0.01	&	0.07	$\pm$	0.01	&	0.66	$\pm$	0.01	\\
81473111498	&	BC03	&	0.04	$\pm$	0.08	&	0.62	$\pm$	0.11	&	0.04	$\pm$	0.09	&	0.68	$\pm$	0.12	&	-0.17	$\pm$	0.10	&	0.37	$\pm$	0.13	&	-0.21	$\pm$	0.07	&	0.47	$\pm$	0.09	&	-0.02	$\pm$	0.03	&	0.14	$\pm$	0.04	\\
	&	E-MILES	&	0.11	$\pm$	0.05	&	0.61	$\pm$	0.07	&	0.12	$\pm$	0.04	&	0.69	$\pm$	0.06	&	-0.12	$\pm$	0.12	&	0.08	$\pm$	0.17	&	-0.05	$\pm$	0.01	&	0.24	$\pm$	0.01	&	-0.13	$\pm$	0.10	&	0.50	$\pm$	0.14	\\
81474103381	&	BC03	&	-0.23	$\pm$	0.08	&	0.63	$\pm$	0.11	&	-0.16	$\pm$	0.06	&	0.67	$\pm$	0.09	&	-0.19	$\pm$	0.03	&	0.25	$\pm$	0.04	&	-0.10	$\pm$	0.02	&	0.21	$\pm$	0.02	&	0.28	$\pm$	0.10	&	0.28	$\pm$	0.13	\\
	&	E-MILES	&	-0.08	$\pm$	0.03	&	0.73	$\pm$	0.04	&	-0.01	$\pm$	0.02	&	0.78	$\pm$	0.03	&	-0.15	$\pm$	0.03	&	-0.05	$\pm$	0.04	&	-0.05	$\pm$	0.02	&	0.07	$\pm$	0.02	&	0.02	$\pm$	0.03	&	0.49	$\pm$	0.04	\\
81482205540	&	BC03	&	-0.02	$\pm$	0.01	&	0.89	$\pm$	0.02	&	-0.01	$\pm$	0.01	&	0.93	$\pm$	0.01	&	-0.03	$\pm$	0.02	&	0.08	$\pm$	0.03	&	-0.02	$\pm$	0.02	&	0.03	$\pm$	0.03	&	0.0	$\pm$	0.01	&	0.13	$\pm$	0.02	\\
	&	E-MILES	&	-0.03	$\pm$	0.02	&	0.98	$\pm$	0.02	&	-0.02	$\pm$	0.01	&	0.99	$\pm$	0.02	&	-0.04	$\pm$	0.02	&	-0.10	$\pm$	0.02	&	-0.02	$\pm$	0.02	&	-0.12	$\pm$	0.02	&	0.01	$\pm$	0.01	&	0.25	$\pm$	0.01	\\
81481401355	&	BC03	&	-0.01	$\pm$	0.02	&	0.79	$\pm$	0.02	&	-0.02	$\pm$	0.01	&	0.89	$\pm$	0.02	&	-0.10	$\pm$	0.02	&	0.37	$\pm$	0.03	&	-0.09	$\pm$	0.02	&	0.29	$\pm$	0.03	&	0.02	$\pm$	0.00	&	0.09	$\pm$	0.01	\\
	&	E-MILES	&	-0.00	$\pm$	0.01	&	0.94	$\pm$	0.02	&	-0.00	$\pm$	0.01	&	0.96	$\pm$	0.01	&	-0.11	$\pm$	0.02	&	0.15	$\pm$	0.03	&	-0.10	$\pm$	0.02	&	0.16	$\pm$	0.03	&	0.01	$\pm$	0.01	&	0.15	$\pm$	0.01	\\
81422404933	&	BC03	&	0.13	$\pm$	0.07	&	0.53	$\pm$	0.09	&	0.10	$\pm$	0.05	&	0.67	$\pm$	0.07	&	-0.27	$\pm$	0.02	&	0.65	$\pm$	0.03	&	-0.31	$\pm$	0.03	&	0.66	$\pm$	0.03	&	-0.03	$\pm$	0.03	&	0.23	$\pm$	0.04	\\
	&	E-MILES	&	0.16	$\pm$	0.05	&	0.62	$\pm$	0.07	&	0.07	$\pm$	0.05	&	0.80	$\pm$	0.06	&	-0.22	$\pm$	0.04	&	0.36	$\pm$	0.05	&	-0.26	$\pm$	0.08	&	0.37	$\pm$	0.10	&	-0.18	$\pm$	0.04	&	0.48	$\pm$	0.06	\\

\label{tab:A2}

\end{longtable}
\end{landscape}
\normalsize

\newpage
\twocolumn

\section{Mass-weighted properties}\label{ap:resolved}
Mass-weighted version of equations 1 and 2: 

\begin{equation}\label{eq:3}
 < Age >_\textsubscript{M} ^\textsubscript{resolved} = \frac{{\sum_{i} M_{i}\cdot Age_{M,i}}}{{\sum_{i} M_{i}}}, \text{and}
\end{equation}

\begin{equation}\label{eq:4}
< [Fe/H] >_\textsubscript{M} ^\textsubscript{resolved} = \frac{{\sum_{i} M_{i}\cdot [Fe/H]_{M,i}}}{{\sum_{i} M_{i}}}
\end{equation}

\begin{figure} [h]
\begin{center}
\includegraphics[width=\columnwidth]{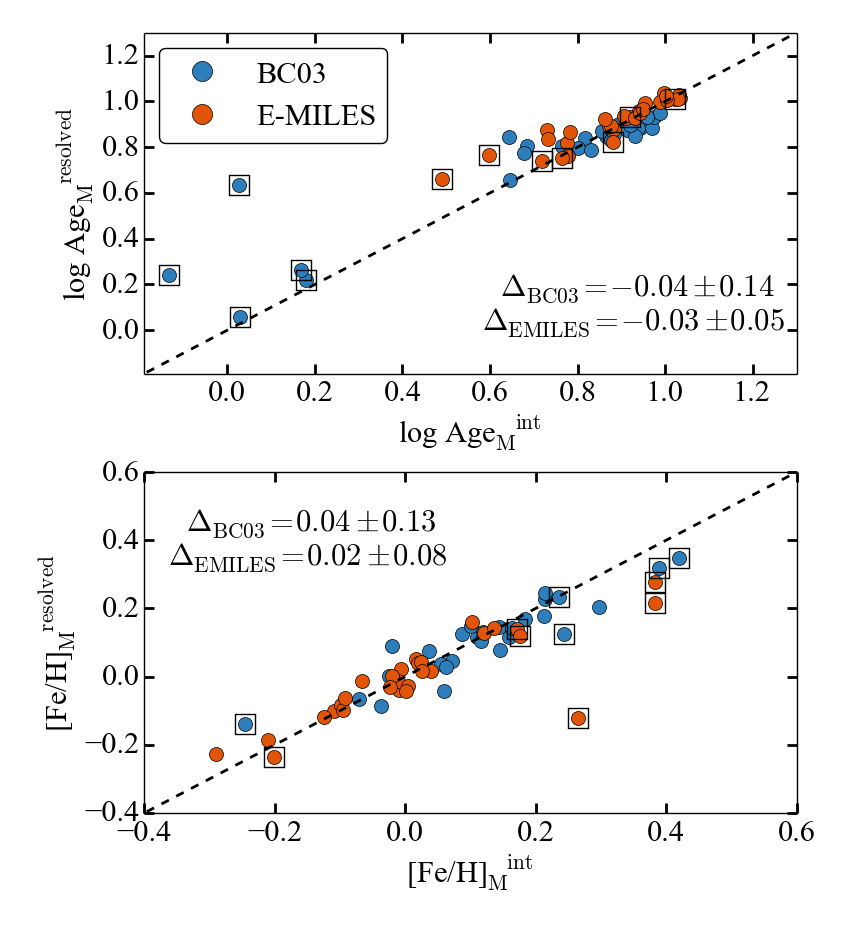}
\caption{Same as Fig. \ref{fig:4} but for mass-weighted properties.} 
\label{fig:4a}
\end{center}
\end{figure}

\begin{figure} [h]
\begin{center}
\includegraphics[width=\columnwidth]{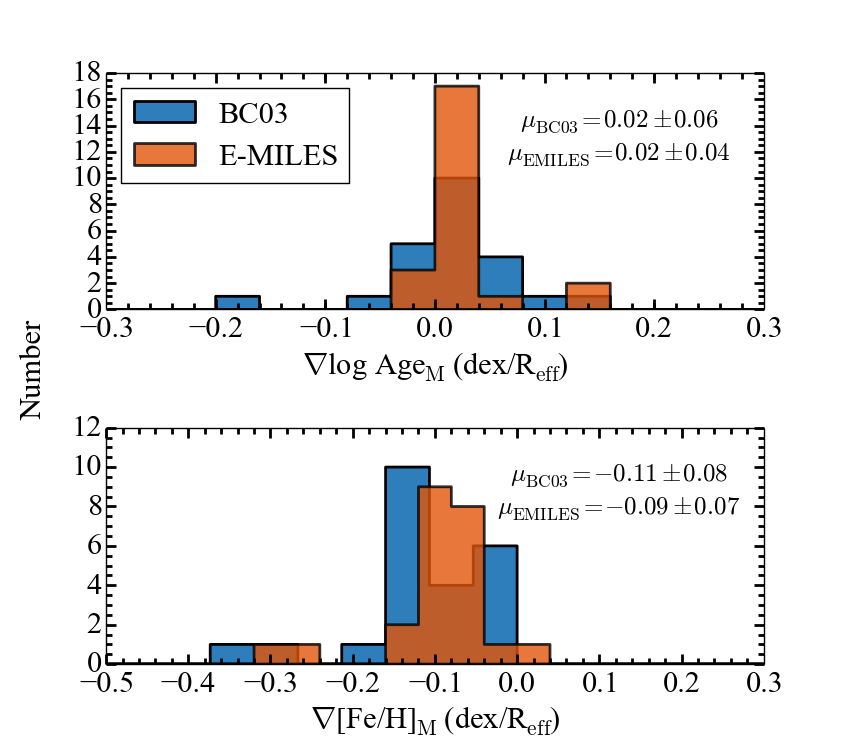}
\caption{Same as Fig. \ref{fig:8} but for mass-weighted stellar properties.} 
\label{fig:8b}
\end{center}
\end{figure}

\begin{figure} [h]
\begin{center}
\includegraphics[width=\columnwidth]{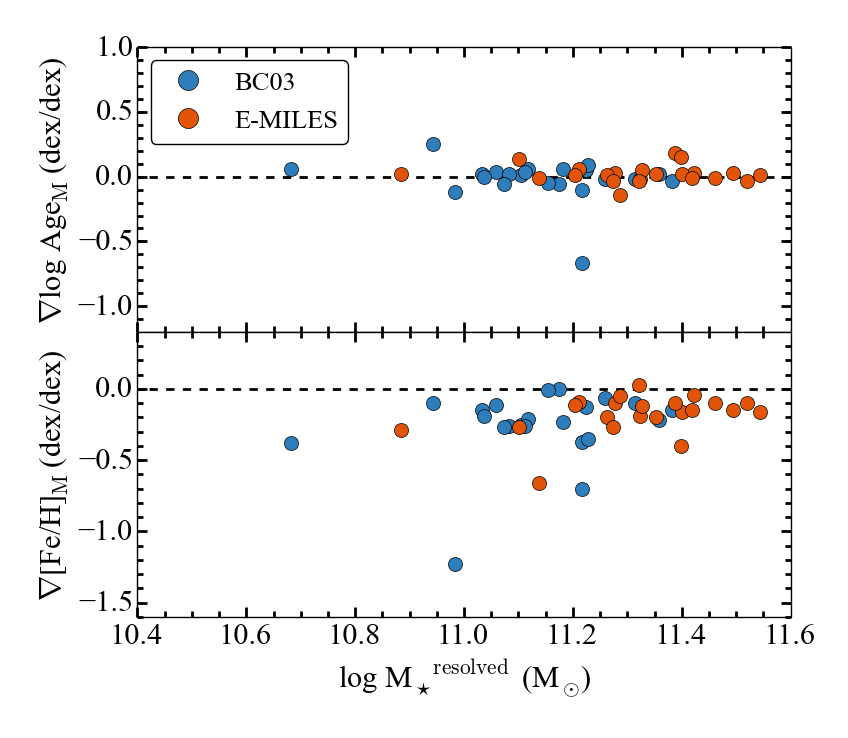}
\caption{Same as Fig. \ref{fig:9} but for mass-weighted stellar properties.} 
\label{fig:9b}
\end{center}
\end{figure}


\end{document}